\documentclass[11pt]{article}
\usepackage{amsfonts}
\usepackage{amssymb}
\usepackage{amsmath,amssymb}
\usepackage{bbm}

\def\slasha#1{\setbox0=\hbox{$#1$}#1\hskip-\wd0\hbox to\wd0{\hss\sl/\/\hss}}

\def\periodb#1{\setbox0=\hbox{$#1$}#1\hskip-\wd0\hbox to\wd0{-}}

\def\sfrac#1#2{{\textstyle\frac{#1}{#2}}}
\newcommand{\unit}{\mathbbm{1}}   
\newcommand{\frg}{\mathfrak{g}}
\newcommand{\CN}{\mathcal{N}}    
\newcommand{\CA}{\mathcal{A}}    
\newcommand{\CG}{\mathcal{G}} 
\newcommand{\CB}{\mathcal{B}} 
\newcommand{\CO}{\mathcal{O}}    
\newcommand{\CL}{\mathcal{L}}    
\newcommand{\CF}{\mathcal{F}}   
\newcommand{\CE}{\mathcal{E}}  
\newcommand{\CP}{\mathcal{P}}
\newcommand{\CH}{\mathcal{H}}
\newcommand{\CV}{\mathcal{V}}
\newcommand{\CW}{\mathcal{W}} 
\newcommand{\Z}{\mathbb{Z}}   
\newcommand{\W}{\mathbb{W}}   
\newcommand{\R}{\mathbb{R}}     
\newcommand{\C}{\mathbb{C}}     
\newcommand{\Hbb}{\mathbb{H}}     
\newcommand{\Abb}{\mathbb{A}}     
\newcommand{\CPP}{{\mathbb{C}P}}    
\newcommand{\Nbb}{{\mathbb{N}}}    

\newcommand{\im}{\mathrm{i}} 
 
\newcommand{\dd}{\mathrm{d}}     
\newcommand{\dpar}{\partial}     
\newcommand{\diag}{{\mathrm{diag}}}

\newcommand{\+}{\dagger}
\newcommand{\wt}{\widetilde}

\newcommand{\sU}{\mathrm{U}}     
\newcommand{\sSU}{\mathrm{SU}}     
     
\newcommand{\sSO}{\mathrm{SO}}     
\newcommand{\sGL}{\mathrm{GL}}  
\newcommand{\rsu}{\mathrm{su}} 
\newcommand{\ru}{\mathrm{u}}     
\newcommand{\Spin}{\mathrm{Spin}} 
\newcommand{\rCl}{\mathrm{Cl}}

\newcommand{\al}{{{\alpha}}} 
 \newcommand{\ga}{{{\gamma}}}

\newcommand{\vph}{{{\varphi}}}

\newcommand{\ah}{{\hat{a}}} 
\newcommand{\bh}{{\hat{b}}} 
\newcommand{\xh}{\hat{x}} 
\newcommand{\ph}{\hat{p}}
\newcommand{\fh}{{\hat{f}}}

\newcommand{\yb}{{\bar{y}}}
\newcommand{\zb}{{\bar{z}}}

\makeatletter

\@addtoreset{equation}{section}
\makeatother

\begin{document}
\begin{titlepage}
\setcounter{page}{0}
.
\vskip 15mm
\begin{center}
{\LARGE \bf Quantum connection, charges and virtual particles}\\
\vskip 2cm
{\Large Alexander D. Popov}
\vskip 1cm
{\em Institut f\"{u}r Theoretische Physik,
Leibniz Universit\"{a}t Hannover\\
Appelstra{\ss}e 2, 30167 Hannover, Germany}\\
{Email: alexander.popov@itp.uni-hannover.de}
\vskip 1.1cm
\end{center}
\begin{center}
{\bf Abstract}
\end{center}
Geometrically, quantum mechanics is defined by a complex line bundle $L_\hbar$ over the classical particle phase space $T^*\R^3\cong\R^6$
with coordinates $x^a$ and momenta $p_a$, $a,...=1,2,3$. This quantum bundle $L_\hbar$ is endowed with a connection $A_\hbar$, and its sections are standard wave functions $\psi$ obeying the Schr\"odinger equation. The components of covariant derivatives 
$\nabla_{A_\hbar}^{}$ in $L_\hbar$ are equivalent to operators $\xh^a$ and $\ph_a$. The bundle $L_\hbar=: L_\C^+$ is associated with 
symmetry group U(1)$_\hbar$ and describes particles with quantum charge $q=1$ which is eigenvalue of the generator of the group 
U(1)$_\hbar$. The complex conjugate bundle $L^-_\C:={\overline{L_\C^+}}$ describes antiparticles with quantum charge $q=-1$. We will 
lift the bundles $L_\C^\pm$ and connection $A_\hbar$ on them to the relativistic phase space $T^*\R^{3,1}$ and couple them to the Dirac spinor bundle describing both particles and antiparticles. Free relativistic quarks and leptons are described by the Dirac equation on Minkowski space $\R^{3,1}$. This equation does not contain interaction with the quantum connection $A_\hbar$ on bundles $L^\pm_\C\to T^*\R^{3,1}$  because $A_\hbar$ has non-vanishing components only along $p_a$-directions in $T^*\R^{3,1}$. To enable the interaction of elementary fermions $\Psi$ with quantum connection $A_\hbar$ on $L_\C^\pm$, we will extend the Dirac equation to the phase space while maintaining the condition that $\Psi$ depends only on $t$ and $x^a$.
The extended equation has an infinite number of oscillator-type solutions with discrete energy values as well as wave packets of coherent states. We argue that all these normalized solutions describe virtual particles and antiparticles living outside the mass shell hyperboloid. The transition to free particles is possible through squeezed coherent states. In the limit of infinite width of squeezed solutions, the interaction with quantum connection $A_\hbar$ disappears and the squeezed wave packets become wave solutions describing free particles and antiparticles.
\end{titlepage}

\newpage
\setcounter{page}{1}

\tableofcontents


\section{Introduction}

\noindent  The main idea discussed in this paper is to consider quantization as adding internal degrees of freedom to classical particles. We start from Newtonian particles in $\R^3$ with phase space $T^*\R^3$ and canonical quantization. As explained in geometric quantization approach (see e.g. \cite{Sni, Wood}) one should introduce a complex line bundle $L_\hbar =: L^+_\C$ over $T^*\R^3$ and complex Schr\"odinger wave functions $\psi$ are sections of this bundle depending only on coordinates and time $t\in\R$. This bundle describes quantum particles, i.e. particles with internal space $\C$ which is the typical fibre of the bundle $L^+_\C$. This bundle has the structure group $\sU(1)_\hbar$ and a connection $A_\hbar\in \ru(1)_\hbar$, and with $\psi$-function one naturally associates the quantum charge $q=1$ in the same way as is done for electromagnetic bundles with the structure group $\sU(1)_{\mathrm {em}}$. We will discuss in detail the geometry of the bundle $L^+_\C$ and introduce the complex conjugate bundle $L^-_\C :=\overline{L^+_\C}$ whose sections have the quantum charge $q=-1$ and describe nonrelativistic antiparticles. We will show that connection $A_\hbar$ on these bundles defines a vacuum gauge field acting on particles and antiparticles, that leads to bound states.

Moving on to relativistic quantum mechanics in Minkowski space $\R^{3,1}$, we discuss Dirac spinors $\Psi$ with values in complex vector bundles $E^{\pm}_{\C^N}$, $N=1,2,3$, as particles with internal degrees of freedom $\C^N$ in addition to the degrees of freedom of quantum bundles $L^{\pm}_{\C}$ lifted to $T^*\R^{3,1}$. Considering the nonrelativistic limits of the Dirac and Klein-Gordon equations to the Pauli and Schr\"odinger equations, we show that  positive frequency $\Psi^+$ and negative frequency $\Psi^-$ solutions of the relativistic equations are reduced to solutions of the nonrelativistic equations taking values in the bundles $L^{+}_{\C}$ and $L^{-}_{\C}$, respectively. Due to the fact that spinors $\Psi^+$ and $\Psi^-$ belong to orthogonal subbundles $L^{\pm}_{\C}$ of the bundle $L_{\C^2}=L^{+}_{\C}\oplus L^{-}_{\C}$,
the Dirac spinor $\Psi$ must have an additional index parametrizing these two subbundles. We introduce and study a reformulation of relativistic quantum mechanics with fields taking values in the bundle $L_{\C^2}$ or its subbundles.

A view of quantum mechanics as a gauge theory with a gauge field $A_\hbar$ on the bundle $L_\hbar$ leads to the statement that in addition to the four known interactions, there is also an interaction with the field $A_\hbar$ that has no sources and can be considered as a vacuum gauge field fixed by the canonical commutation relations. To see clearly the interactions with these vacuum fields $A_\hbar$, we lift the Dirac and Klein-Gordon equations to phase space, since $A_\hbar$ has components only along momenta. We describe an infinite number of oscillatory type solutions to these equations that have finite energy, do not depend on momenta, and are localized in space and time. These solutions describe virtual particles (both fermions and gauge bosons) which we define as particles interacting with vacuum fields $A_\hbar$.


\section{Internal degrees of freedom and quantum mechanics}
 
{\bf\large 2.1. Quantum bundle}

Consider the phase space $T^*\R^3=\R^6$ of a nonrelativistic particle of mass $m$ without internal degrees of freedom. It has coordinates $x^a\in\R^3$, conjugate momenta $p_a\in\R^3$ and moves along a trajectory in $\R^6$ defined by a Hamiltonian vector field $V_H$ generated by a function $H(x, p)$ (Hamiltonian) with evolution parameter $t\in\R$ (time). According the canonical quantization program, the coordinate functions $(x^a, p_a)$ on $\R^6$ are replaced by operators $\xh^a$ and $\ph_a$, acting on a complex wave function $\psi (x,t)$ as
\begin{equation}\label{2.1}
\ph_a\psi (x,t)=-\im\hbar\frac{\partial\psi (x,t)}{\partial x^a}\ ,\quad \xh^a\psi (x,t)= x^a\psi (x,t)
\end{equation}
and the Hamiltonian equations of motion are replaced by the Schr\"odinger equation
\begin{equation}\label{2.2}
\im\hbar\frac{\partial\psi (x,t)}{\partial t}= \hat H (\xh , \ph ) \psi (x,t)\ ,
\end{equation}
where $\hbar$ is the Planck's constant \cite{Dirac}.

In the geometric quantization approach (see e.g. \cite{Sni, Wood}) it is explained that $\psi$ in \eqref{2.1}, \eqref{2.2} is a section of the complex line bundle
\begin{equation}\label{2.3}
\pi\ :\quad L_\hbar\stackrel{\C}{\longrightarrow} T^*\R^3
\end{equation}
over the phase space $T^*\R^3$ with the real polarization condition
\begin{equation}\label{2.4}
\partial^a\psi := \frac{\partial\psi}{\partial p_a} =0\quad \Rightarrow\quad \psi =\psi (x, t)\ .
\end{equation}
Hence, the operators in \eqref{2.1} are equivalent to the covariant derivatives
\begin{equation}\label{2.5}
\nabla_a:= \frac{\im}{\hbar}\,\ph_a =\dpar_a \quad\mathrm{and}\quad \nabla^{a+3}:=-\frac{\im}{\hbar}\,\xh^a =
\dpar^a-\frac{\im}{\hbar}\,x^a\ \mathrm{with}\ \dpar_a{=}\frac{\partial}{\partial x^a}\ \mathrm{and}\  \dpar^a{=}\frac{\partial}{\partial p_a}\ ,
\end{equation}
acting on sections $\psi$ of the bundle $L_\hbar$ subject to \eqref{2.4}. For greater rigor, one can introduce the tensor product of the bundle $L_\hbar$ and the line bundle of half-form  \cite{Sni, Wood} but we will not complicate the discussion.

\bigskip

\noindent{\bf\large 2.2. Internal degrees of freedom}

Starting with Dirac, Weyl and von Neumann, quantization is understood as a linear mapping $f\to \fh$ of the Poisson algebra $\CP (X, \omega_X)$ of  functions $f\in C^\infty (X)$ on a symplectic manifold $(X, \omega_X)$ into the set of operators $\fh$ on some Hilbert space in such a way that some axioms are satisfied \cite{Dirac, Sni, Wood}. By now, it has long been known that this does not work, not only for canonical quantization but also for all other existing quantization schemes (see discussion e.g. in \cite{Ali}). Everything works only for a very narrow class of functions on $X$, including coordinates of the phase space $X$. Therefore, it seems more rational to speak not about quantization, but about the introduction of {\it internal degrees of freedom}, which implies the transition from phase space $X$ to complex vector bundles over it and gauge fields $A$ that specify the forces acting on internal degrees of freedom of particles:
\begin{tabbing}
nnn\=\=\kill
manifold $(X, \omega_X)$\=$\quad\longrightarrow\quad$\= vector bundle $E(X,V)$ over $X$ with fibres $V_x\cong\C^N$ (internal spaces)\\[3pt]
points  $x$ of $X$\>$\quad\longrightarrow\quad$\= sections $(x,\psi_x)$ of $E$, $\psi_x\in V_x$\\[3pt]
forces $-\frac{\dpar H(x, p)}{\dpar x}$\>$\quad\longrightarrow\quad$\= covariant derivatives $\nabla_A$ on $E$ with a connection $A$
\end{tabbing}
Discussion of internal degrees of freedom $V\cong\C^N$ of particles and carriers of interactions $A$ corresponding to them is the main topic of this paper.

From \eqref{2.3}-\eqref{2.5} we conclude that when creating quantum mechanics in 1925-1926, it was actually discovered that particles have an additional internal degree of freedom, parametrized by the complex space $\C_u$, attached to each point $u=(x,p)\in \R^6$ of the phase space $\R^6$ of a Newtonian particle. These spaces $\C_u$ are combined into an {\it extended phase space}
\begin{equation}\label{2.6}
L_\hbar = \mathop{\bigsqcup}_{u\in\R^6}\C_u = \mathop{\bigcup}_{u\in\R^6}\left\{(u, \psi_u) \mid \psi_u\in\C_u\right\}\ ,
\end{equation}
which is the complex line bundle \eqref{2.3} over $T^*\R^3 = \R^6$. In this ``quantum" bundle $L_\hbar$, a ``quantum" connection $A_\hbar$ is given, which defines the parallel transport of sections $\psi$ of $L_\hbar$, i.e. a way to identify fibres $\C_u$ (quantum degrees of freedom) over nearby points. A little bit later we will describe in detail the geometry of this bundle, but first we will describe the complex structure $J$ on the fibres of $L_\hbar$ in differential geometric terms \cite{KobNom}, since this seemingly simple structure will be constantly used in this paper.

\bigskip
\noindent{\bf\large 2.3. Internal space $\C$}

Let us consider the Euclidean vector space $\R^2$ with the metric
\begin{equation}\label{2.7}
g_{\R^2}^{}=\delta_{AB}\dd y^A\dd y^B\qquad\mathrm{for}\qquad A,B=1,2\ .
\end{equation}
We introduce the standard complex structure $J\in\,$End$(T\R^2)$ acting on tangent vector bundle $T\R^2$ of $\R^2$. It is defined as a tensor $J=(J^A_B)$ such that $J^A_CJ^C_B = -\delta^A_B\ \Leftrightarrow  \ J^2=-\unit_2$. We choose $J$ as the matrix
\begin{equation}\label{2.8}
J=(J^A_B)=\begin{pmatrix} 0&-1\\1&0\end{pmatrix} \qquad\mathrm{with}\qquad J^1_2=-1, J^2_1=1
\end{equation}
in coordinates $y^A$ on $\R^2$, i.e. $J=J^A_B\dd y^B\dpar_A$ with $\dpar_A=\dpar /\dpar y^A$.

For any real vector field $\psi\in\Gamma (T\R^2)$,
\begin{equation}\label{2.9}
\psi =\psi^1\dpar_1 + \psi^2\dpar_2
\end{equation}
we define $J(\dpar_1)=\dpar_2$ and $J(\dpar_2)=-\dpar_1$. Since $J^2=-\unit_2$, eigenvalues of $J$ on vector fields are $\pm\im$, and cannot be realized on real vector fields \eqref{2.9}. However, we can consider complexified tangent bundle $T^{\C}\R^2$ with sections \eqref{2.9} having complex component $\psi^1_\C$ and $\psi^2_\C$. Then
\begin{equation}\label{2.10}
\psi_\C = \psi_\C^1\dpar_1 + \psi_\C^2\dpar_2=(\psi_\C^1 + \im\psi_\C^2)\dpar_y + (\psi_\C^1 - \im\psi_\C^2)\dpar_{\bar y}
=: \psi^y\dpar_y   +  \psi^{\bar y}\dpar_{\bar y}      \ ,
\end{equation}
where
\begin{equation}\label{2.11}
\dpar_y = \sfrac12\, (\dpar_1 - \im\dpar_2)\qquad\mathrm{and}\qquad \dpar_{\bar y}=\sfrac12\, (\dpar_1 + \im\dpar_2)
\end{equation}
are the basis vector fields in the subbundles
\begin{equation}\label{2.12}
T^{1,0}\R^2{\cong}\R^2{\times}\C\subset  T^{\C}\R^2{\cong}\R^2{\times} (\C{\oplus}\bar\C)\ \ \mathrm{and}\ \ T^{0,1}\R^2{\cong}
\R^2{\times}\bar\C\subset  T^{\C}\R^2{\cong}\R^2{\times}(\C{\oplus}\bar\C)\ .
\end{equation}
This means that $T^{1,0}\R^2\cong\C\times\C$ and $T^{0,1}\R^2\cong\bar\C\times\bar\C$, i.e. $(\R^2, J)\cong\C$ and $(\R^2, -J)\cong\bar\C$.
We have
\begin{equation}\label{2.13}
J(\psi^y\dpar_y) = \psi^y J(\dpar_y) = \im\psi^y\dpar_y\qquad\mathrm{and}\qquad J(\psi^{\bar y}\dpar_{\bar y}) = \psi^{\bar y} J(\dpar_{\bar y}) = - \im\psi^{\bar y}\dpar_{\bar y}
\end{equation}
where $\psi^{\bar y}$ is not complex conjugate to $\psi^y$. If $\psi^{\bar y}=\overline{\psi^y}$ then $\psi_\C$ becomes the real vector field \eqref{2.9}. 

\bigskip
\noindent{\bf\large 2.4. Pairing and Hermitian metric}

We introduced the splitting $T^{\C}\R^2=T^{1,0}\R^2\oplus T^{0,1}\R^2$ of complexified tangent bundles on $\R^2$ into (1,0) and (0,1) parts. Similarly, the complexified cotangent bundle $T^*_{\C}\R^2$ can be splitted as
\begin{equation}\label{2.14}
T^*_{\C}\R^2=T_{1,0}\R^2\oplus T_{0,1}\R^2\quad\ni\quad \psi_y\dd y+ \psi_{\bar y}\dd{\bar y}\ ,
\end{equation}
where $\dd y$ and $\dd\bar y$ are one-forms dual to the vector fields $\dpar_y$ and $\dpar_{\bar y}$, and formulae
\begin{equation}\label{2.15}
\psi^y\dpar_y\lrcorner \psi_y\dd y = \psi_y\psi^y\qquad\mathrm{and}\qquad \psi^{\bar y}\dpar_{\bar y}\lrcorner \psi_{\bar y}\dd {\bar y} = \psi_{\bar y}\psi^{\bar y}
\end{equation}
define pairing of the dual vector spaces.

On the space $\R^2$ we have the metric \eqref{2.7} which in complex coordinates  $y=y^1+\im y^2$ and $\bar y=y^1-\im y^2$ can be written as
\begin{equation}\label{2.16}
g_{\R^2} = \delta_{y\bar y}\, \dd y\, \dd\bar y =\delta_{\bar y y}\, \dd\bar y\,\dd y\ ,
\end{equation}
i.e. we have a Hermitian metric on $\C$ (and on $\bar\C$). This metric allows to define the isomorphisms
\begin{equation}\label{2.17}
\C\ni\psi^y\ \to\ \psi_{\yb}=\delta_{\bar y y}\psi^y\in\bar\C^{\vee}\quad\mathrm{and}\quad \bar\C\ni \psi^\yb\ \to\ \psi_{y}=\delta_{ y\yb}\psi^\yb\in\C^{\vee}
\end{equation}
which are used for definition of scalar product (sesquilinear form) of two vectors $\psi_1, \psi_2\in T^{1,0}\R^2$ as
\begin{equation}\label{2.18}
\langle\psi_1, \psi_2\rangle = \bar\psi_1^\yb\delta_{\yb y}\psi^y_2\in\C\ ,
\end{equation}
i.e. one maps $\psi^y_1\ \to \ \overline{\psi^y_1}=\bar\psi_1^\yb\ \to \delta_{y\yb}\bar\psi^\yb_1\in\C^{\vee}$ and then use the pairing  \eqref{2.15}. Similarly, for $\phi_1, \phi_2\in T^{0,1}\R^2$ we have 
\begin{equation}\label{2.19}
\langle\phi_1, \phi_2\rangle = \phi_1^y\delta_{y\yb}\phi^\yb_2\in\C\ ,
\end{equation}
and for $\psi_1=\psi_2$, $\phi_1=\phi_2$ these formulae define real-valued Hermitian metric on the spaces of (1,0) and (0,1) vector fields.

In calculations with vector fields \eqref{2.10} and one-forms \eqref{2.14}, it is often more convenient to use columns and rows. To do this, one needs to make substitutions
\begin{equation}\label{2.20}
\dpar_1\to\begin{pmatrix}1\\0\end{pmatrix} ,\quad \dpar_2\to\begin{pmatrix}0\\1\end{pmatrix}\quad\mathrm{and}\quad
\psi =\psi^1\dpar_1 + \psi^2\dpar_2\to\begin{pmatrix}\psi^1\\\psi^2\end{pmatrix}
\end{equation}
\begin{equation}\label{2.21}
\dpar_y\to v_+:=\frac{1}{\sqrt{2}}\begin{pmatrix}1\\-\im\end{pmatrix} ,\quad \psi^y\to\sqrt{2}\psi_+\quad\mathrm{and}\quad
\psi^y\dpar_y\to\psi_+v_+
\end{equation}
\begin{equation}\label{2.22}
\dpar_\yb\to v_-:=\frac{1}{\sqrt{2}}\begin{pmatrix}1\\\im\end{pmatrix} ,\quad \phi^\yb\to\sqrt{2}\psi_-\quad\mathrm{and}\quad
\phi^\yb\dpar_\yb\to\psi_-v_-
\end{equation}
\begin{equation}\label{2.23}
\psi_y\dd y\to\psi_+^{\vee}v_+^\dagger \quad\mathrm{and}\quad\phi_\yb\dd \yb\to\psi_-^{\vee}v_-^\dagger \ .
\end{equation}
In matrix notation we have
\begin{equation}\label{2.24}
\langle\psi_1, \psi_2\rangle = \psi^\dagger_{1}\psi_{2}\quad\mathrm{and}\quad\langle\phi_1, \phi_2\rangle = \phi^\dagger_{1}\phi_{2}\ ,
\end{equation}
and $Jv_\pm =\pm\im v_\pm$ for $J$ given in \eqref{2.8}.

\section{Geometry of quantum bundle $L_\hbar$}

{\bf\large 3.1. Phase space $T^*\R^3$}

Let us consider the phase space $T^*\R^3=\R^6$ with the symplectic structure
\begin{equation}\label{3.1}
\omega_{\R^6}=\omega_a^{\ \,b+3}\,\dd x^a\wedge\dd p_b=\omega^{b+3}_{\quad\ a}\,\dd p_b\wedge\dd x^a=\dd x^a\wedge\dd p_a\ ,
\end{equation}
where $(x^a, p_b)\in\R^6$ are coordinates and momenta with $a,b =1,2,3$. Classical particle is characterized by a point $(x^a, p_b)$ in $\R^6$ moving along a trajectory $(x^a(t),p_b(t))$ defined by a Hamiltonian vector field
\begin{equation}\label{3.2}
V_H=\omega^a_{\ \,b+3}\,\dpar_a H\dpar^b + \omega_{b+3}^{\quad\ a}\,\dpar^bH\dpar_a\quad\mathrm{for}\quad \dpar_a=\frac{\dpar}{\dpar x^a}\quad\mathrm{and}\quad \dpar^b=\frac{\dpar}{\dpar p_b}\ .
\end{equation}
Here we used the bivector field
\begin{equation}\label{3.3}
\omega^{-1}_{\R^6}=\omega^a_{\ \,b+3}\,\dpar_a\wedge\dpar^b= \omega_{b+3}^{\quad\ a}\,\dpar^b\wedge\dpar_a
\end{equation}
inverse to $\omega_{\R^6}$ in \eqref{3.1}. We consider time independent Hamiltonians $H(x,p)$. Equations of motion are
\begin{equation}\label{3.4}
\dot{x}^a=\frac{\dd x^a}{\dd t}=V_H(x^a)\quad\mathrm{and}\quad\dot{p}_a=V_H(p_a)\ ,
\end{equation}
where $t\in\R$ is an evolution parameter. Classical particles have no internal degrees of freedom - they are points in phase space. 

On $\R^6$ we introduce the metric
\begin{equation}\label{3.5}
g_{\R^6}=\delta_{ab}\dd x^a\dd x^b + w^4\delta^{ab}\dd p_a\dd p_b
\end{equation}
with the inverse
\begin{equation}\label{3.6}
g_{\R^6}^{-1}=\delta^{ab}\dpar_a\otimes\dpar_b + w^{-4}\delta_{ab}\dpar^a\otimes\dpar^b\ ,
\end{equation}
i.e.
\begin{equation}\label{3.7}
g^{{a+3}\ {b+3}}= w^4\delta^{ab}\quad\mathrm{and}\quad g_{{a+3}\ {b+3}}= w^{-4}\delta_{ab}\ .
\end{equation}
Here $w\in\R^+$ is a parameter such that $[w^2p_a]=[$length$]=[x^a]$. Note that definitions of $\omega_{\R^6}$ and $g_{\R^6}$
are independent.

\bigskip
\noindent{\bf\large 3.2. Extended phase space $T^*\R^3\times \R^2$}

From a geometric point of view adopted in this paper, in quantum mechanics particles are not points in the phase space $T^*\R^3$, but extended objects with internal degrees of freedom parametrized by the space $(\R^2, J)\cong\C$, where $J$ is a complex structure on $\R^2$. A quantum particle is a point $u=(x,p)\in\R^6$ and the internal space $\C_u$ of the particle, attached to the point $u$. The union of all internal spaces give the space
\begin{equation}\nonumber
L_\hbar = \mathop{\bigsqcup}_{u\in\R^6}\C_u = \mathop{\bigcup}_{u\in\R^6}\left\{(u, \psi_u) \mid \psi_u\in\C_u\right\}
\end{equation}
briefly discussed in Section 2. Topologically, this space is the direct product
\begin{equation}\label{3.8}
L_\hbar\cong T^*\R^3\times \C
\end{equation}
but this is not so from the differential geometry point of view, since the space \eqref{3.8} is endowed not with the metric $g_{\R^6}+g_{\R^2}$ of the direct product, but the twisted metric of the fibre bundle \eqref{2.3}, determined by a connection $A_\hbar$, which we now describe. We will see that the introduction of canonical commutation relations is nothing else than the introduction of covariant derivative on $L_\hbar$ defined by ``quantum" connection $A_\hbar$.

\bigskip
\noindent{\bf\large 3.3. Frame and coframe on $L_\hbar$}

On the space \eqref{3.8} we introduce vector fields
\begin{equation}\label{3.9}
\nabla_a:=\frac{\im}{\hbar}\,\ph_a=\dpar_a\ ,\quad \nabla^{a+3}:=-\frac{\im}{\hbar}\,\xh^a=\dpar^a + A^{a+3}J^v\ , \quad \nabla_y=\dpar_y\ ,
\end{equation}
where $y$ is a complex coordinate on $\C$ and the vector field
\begin{equation}\label{3.10}
J^v = -\im (y\dpar_y - \bar y\dpar_{\bar y})
\end{equation}
is the generator of the group SO(2)$\,\cong\,$U(1)  acting on the space $\R^2\cong\C$. In \eqref{3.9} we choose
\begin{equation}\label{3.11}
A_a=0\quad\mathrm{and}\quad A^{a+3}=-\frac{1}{\hbar}\,x^a
\end{equation}
as components of the u(1)-valued connection
\begin{equation}\label{3.12}
A_\hbar =(A_a\dd x^a + A^{a+3}\dd p_a) J^v = -\frac{1}{\hbar}\,x^a \dd p_a J^v 
\end{equation}
on the bundle $L_\hbar\to \R^6$. One-forms
\begin{equation}\label{3.13}
\theta^a=\dd x^a\ ,\quad \theta_{a+3}=\dd p_a\quad\mathrm{and}\quad\theta^y=\dd y +\im y A^{a+3}\dd p_a
\end{equation}
are dual to the vector fields \eqref{3.9}.

\bigskip
\noindent{\bf\large 3.4. Metric on $L_\hbar$}

Vector fields \eqref{3.9} form a frame on the tangent bundle to $L_\hbar$ and one-forms \eqref{3.13} form a coframe. Hence, on $L_\hbar$ we can introduce the metric
\begin{equation}\label{3.14}
g^{}_{L_\hbar}=g^{}_{\R^6} + \delta_{y\bar y}\,\theta^y\bar\theta^{\bar y}
\end{equation}
which is not flat due to the connection one-form \eqref{3.12}. For sections of the bundle $L_\hbar$,
\begin{equation}\label{3.15}
\psi = \psi^y(u)\dpar_y
\end{equation}
parametrizing internal spaces $\C_u$ of particles, we define the scalar product
\begin{equation}\label{3.16}
\langle\psi , \psi\rangle = \delta_{\bar y y}\bar\psi^{\bar y}\psi^y\ ,
\end{equation}
introduced in \eqref{2.18}.

\bigskip
\noindent{\bf\large 3.5. Curvature $F_\hbar$}

For commutators of the covariant derivatives \eqref{3.9} in $L_\hbar$ we obtain
\begin{equation}\label{3.17}
[\nabla_a, \nabla^{b+3}]=-\frac{1}{\hbar}\,\delta_a^b J^v=F_a^{\ \,b+3} J^v\quad\mathrm{and}\quad
[\nabla_y, \nabla^{b+3}]=\frac{\im}{\hbar}\,x^b\nabla_y\ .
\end{equation}
The two-form
\begin{equation}\label{3.18}
F_\hbar =F_a^{\ \,b+3}J^v\dd x^a\wedge\dd p_b
\end{equation}
is the curvature of the connection \eqref{3.12} on $L_\hbar$ taking values in the Lie algebra u(1) with the generator $J^v$ from \eqref{3.10}. On sections \eqref{3.15} of the complex line bundle $L_\hbar$ we have
\begin{equation}\label{3.19}
[\nabla_a, \psi]= (\dpar_a\psi^y)\dpar_y\quad\mathrm{and}\quad
[\nabla^{a+3}, \psi]= \left (\dpar^a\psi^y - \frac{\im}{\hbar}\,x^a\psi^y\right )\dpar_y\ ,
\end{equation}
i.e. in the covariant derivatives for $\psi^y\in\C$ we can substitute $J^v\to \im$.

For a transition to matrix-valued objects, instead of vector fields \eqref{3.15} and dual one-forms, we must use correspondences \eqref{2.20}-\eqref{2.23}. Then $\psi , A_\hbar$ and $F_\hbar$ will take the form
\begin{equation}\label{3.20}
\psi=\psi_+ v_+ \ ,\quad A_{\hbar}=-\frac{1}{\hbar}\,x^a\dd p_a J\quad\mathrm{and}\quad
F_\hbar = \dd A_\hbar =-\frac{1}{\hbar}\, \dd x^a\wedge\dd p_a J\  .
\end{equation}
For the covariant derivatives we obtain
\begin{equation}\label{3.21}
\nabla_a\psi = (\dpar_a\psi_+)v_+ \quad\mathrm{and}\quad\nabla^{a+3}\psi = \left(\dpar^a\psi_+ - \frac{\im}{\hbar}\,x^a\psi_+\right )v_+\ .
\end{equation}
If we consider only the bundle $L_\hbar$, then we can further simplify these formulae by replacing $J$ with $\im$ in \eqref{3.20}, removing the basis column $v_+$ from \eqref{3.21} and deleting the index ``+" in $\psi_+$. Note that when passing to matrix formulae \eqref{3.20} and \eqref{3.21}, the commutation relations \eqref{3.17} will take the form
\begin{equation}\label{3.22}
[\nabla_a, \nabla^{b+3}]=-\frac{1}{\hbar}\,\delta_a^b J\quad\mathrm{for}\quad
\nabla^{b+3}=\dpar^b-\frac{1}{\hbar}\,x^b J
\end{equation}
and when passing to the (1,0)-component $\psi_+$, we obtain the standard formulae
\begin{equation}\label{3.23}
\nabla_a = \dpar_a\ ,\quad \nabla^{b+3}=\dpar^b-\frac{\im}{\hbar}\,x^b\quad\Rightarrow\quad
[\nabla_a, \nabla^{b+3}]=-\frac{\im}{\hbar}\,\delta_a^b 
\end{equation}
for the operators
\begin{equation}\label{3.24}
\ph_a = -\im\hbar\,\nabla_a=-\im\hbar\,\dpar_a\ ,\quad \xh^a =\im\hbar\,\nabla^{a+3}=\im\hbar\left (\dpar^a - \frac{\im}{\hbar}\,x^a\right )=
x^a + \im\hbar\,\dpar^a
\end{equation}
and commutators
\begin{equation}\label{3.25}
[\ph_a, \xh^b]=-\im\hbar\,\delta_a^b\ .
\end{equation}
However, when passing to formulae \eqref{3.23}-\eqref{3.25}, one should always remember that $\psi_+$ is not a function, but a (1,0)-component of a vector $\psi\in\C^2$. Therefore, it is reasonable to keep the vector notation, in particular, because then the description of complex conjugate and dual bundles becomes clearer.

\bigskip
\noindent{\bf\large 3.6. Four quantum bundles}

In quantum mechanics, polarized sections $\psi (x,t)=\psi_+(x,t)v_+$ of the bundle $L_\hbar$ are considered to be wave functions describing particles, and it is usually stated that antiparticles appear only in relativistic theory. From a mathematical point of view, this is incorrect. In Section 2 we discussed the complex structure $J$ on the space $\R^2$, complex vector fields and complex one-forms dual to vector fields. Consequently, if $L_\hbar$ is a complex line bundle whose sections are vector fields $\psi$ (columns $\psi_+ v_+$), then there will also be a dual bundle $L_\hbar^\vee$ of one-forms $\psi^\vee$ (rows $\psi_+^\vee v_+^\dagger$) and their complex  conjugate bundles $\bar L_\hbar$ and 
$\bar L_\hbar^\vee$ with sections $\psi\in L_\hbar$, $\psi^\vee\in  L_\hbar^\vee$, $\phi\in\bar L_\hbar$ and $\phi^\vee\in\bar L_\hbar^\vee$. In the general case, these sections are independent, but the Hermitian metric \eqref{3.14} on fibres (see also \eqref{2.16}-\eqref{2.24}) defines isomorphisms of two pairs of these bundles,
\begin{equation}\label{3.26}
L^+_\C :=L_\hbar \quad\cong\quad \bar L_\hbar^\vee =\bar L_\hbar^\dagger = L_\hbar^\top
\end{equation}
\begin{equation}\label{3.27}
L^-_\C :=\bar L_\hbar \quad\cong\quad  L_\hbar^\vee = L_\hbar^\dagger \ .
\end{equation}
In terms of sections, the isomorphisms \eqref{3.26}, \eqref{3.27} mean  that
\begin{equation}\label{3.28}
L_\C^+\ni\psi_+=\frac{1}{\sqrt 2}\psi^y\leftrightarrow \frac{1}{\sqrt 2}\delta_{\yb y}\psi^y =:\phi^\vee_-\in\bar L_\hbar^\vee
\end{equation}
\begin{equation}\label{3.29}
L_\C^-\ni\psi_-=\frac{1}{\sqrt 2}\psi^\yb\leftrightarrow \frac{1}{\sqrt 2}\delta_{y\yb}\psi^\yb =:\phi^\vee_+\in L_\hbar^\vee\ .
\end{equation}
That is why, we leave only $L_\C^+$ and $L_\C^-$ as independent bundles and denote their sections as
\begin{equation}\label{3.30}
\Psi_+ = \psi_+ v_+\quad\mathrm{and}\quad\Psi_- = \psi_-v_-\ ,
\end{equation}
where $v_+$ and $v_-$ are given in \eqref{2.21} and \eqref{2.22}. The Hermitian scalar product for sections of $L_\C^+$ and $L_\C^-$ is defined by formulae
\begin{equation}\label{3.31}
\Psi_+^\+\Psi_+ = \psi_+^*\psi_+ v_+^\+ v_+=\psi_+^*\psi_+\quad\mathrm{and}\quad
\Psi_-^\+\Psi_- = \psi_-^*\psi_- v_-^\+ v_-=\psi_-^*\psi_-\ .
\end{equation}
Note that $\Psi_+^\+$ and $\Psi_-^\+$ are sections of the bundles $(L_\C^+)^\+ = L_\hbar^\vee$ and $(L_\C^-)^\+ = \bar L_\hbar^\vee$, respectively.

\bigskip
\noindent{\bf\large 3.7. Particles and antiparticles}

Consider now the direct sum
\begin{equation}\label{3.32}
L_{\C^2}=L_\C^+\oplus L_\C^-
\end{equation}
of complex line bundles $L_\C^+$ and $L_\C^-$ and a section
\begin{equation}\label{3.33}
\Psi = \Psi_++\Psi_-=\psi_+ v_+ + \psi_-v_- =\frac{1}{\sqrt 2}\begin{pmatrix}\psi_++\psi_-\\-\im (\psi_+-\psi_-)\end{pmatrix}
\end{equation}
of $L_{\C^2}$. It is easy to see that the structure group U(1)$_\hbar$ of the bundles $L^\pm_\C$ acts on $\Psi_\pm$ in the opposite way,
\begin{equation}\label{3.34}
\sU(1)\ni e^{\theta J} = \cos\theta + J\sin\theta\ : \quad e^{\theta J}\Psi = e^{\im\theta}\Psi_+ +  e^{-\im\theta}\Psi_-\ ,
\end{equation}
where $\theta$ is an angle variable. This means that sections $\Psi_\pm$ of $L^\pm_\C$ have opposite quantum charge $q=\pm 1$, and covariant derivatives in $L^\pm_\C$ have the form
\begin{equation}\label{3.35}
\nabla_a\Psi_\pm = \dpar_a\Psi_\pm\quad\mathrm{and}\quad\nabla^{a+3}\Psi_\pm = (\dpar^a+ A^{a+3}J)\Psi_\pm =\pm\im A^{a+3}\Psi_\pm\ ,
\end{equation}
where $\Psi_\pm = \Psi_\pm (x,t)$. Thus, sections $\Psi_+$ of $L^+_\C$ describe quantum particles with $q=+1$ and sections $\Psi_-$ of the bundle $L^-_\C$ describe antiparticles with $q=-1$.

Quantum mechanics considers $\Psi_+\in L_\C^+$ (particles), time evolution of which is described by the Schr\"odinger equation
\begin{equation}\label{3.36}
\im\hbar\frac{\dpar\Psi_+(x,t)}{\dpar t} = \hat H\Psi_+(x,t)\ \Rightarrow\ \im\hbar\frac{\dpar\psi_+(x,t)}{\dpar t} = \hat H_+\psi_+(x,t)
\end{equation}
\begin{equation}\label{3.37}
\Rightarrow\hat H_+\psi_+(x)=\hbar\omega\,\psi_+(x)\quad\mathrm{for}\quad\psi_+(x,t) = e^{-\im\omega t}\psi_+(x)\ .
\end{equation}
Recall that we consider time independent Hamiltonians $\hat H$. Sections $\Psi_-$ of the bundle $L_\C^-$ describe antiparticles ($q=-1$) satisfying the conjugate equation
\begin{equation}\label{3.38}
-\im\hbar\frac{\dpar\Psi_-(x,t)}{\dpar t} = \hat H\Psi_-(x,t)\ \Rightarrow\ -\im\hbar\frac{\dpar\psi_-(x,t)}{\dpar t} = \hat H_-\psi_-(x,t)
\end{equation}
\begin{equation}\label{3.39}
\Rightarrow\hat H_-\psi_-(x)=\hbar\omega\,\psi_-(x)\quad\mathrm{for}\quad\psi_-(x,t) = e^{\im\omega t}\psi_-(x)\ ,
\end{equation}
where 
\begin{equation}\label{3.40}
\hat H= \frac{\hat p^2}{2m} + V(\hat x)\ ,\quad \hat H_\pm= \frac{\hat p^2}{2m} + V_\pm( x)\quad\mathrm{and}\quad
V(\hat x)v_\pm = V_\pm(x)v_\pm\ .
\end{equation}
Note that energy $E=\hbar\omega$ for both particle $\Psi_+$ and antiparticle $\Psi_-$ is positive.

An important fact is that $\Psi_+$ and $\Psi_-$ are sections of different vector bundles $L_\C^+$ and $L^-_\C$ and therefore they are orthogonal:
\begin{equation}\label{3.41}
\Psi_+^\+\Psi_- = \psi_+^*\psi_- v_+^\+ v_-=0=
\Psi_-^\+\Psi_+\ .
\end{equation}
For the direct sum of these two vector-valued functions we have
\begin{equation}\nonumber
\Psi=\Psi_+ +\Psi_- =e^{-\im\omega t}\psi_+(x)v_+ + e^{\im\omega t}\psi_-(x)v_-
\end{equation}
\begin{equation}\nonumber
=\sfrac12 e^{-\im\omega t}(\psi_1+\im\psi_2)\begin{pmatrix}1\\-\im\end{pmatrix} + 
\sfrac12 e^{\im\omega t}(\psi_1-\im\psi_2)\begin{pmatrix}1\\\im\end{pmatrix}
\end{equation}
\begin{equation}\label{3.42}
=\begin{pmatrix}\cos\omega t&\sin\omega t\\-\sin\omega t&\cos\omega t\end{pmatrix}\begin{pmatrix}\psi_1\\\psi_2\end{pmatrix}=
e^{-\omega tJ}\begin{pmatrix}\psi_1\\\psi_2\end{pmatrix}\sim \begin{pmatrix} e^{-\im\omega t}&0\\0& e^{\im\omega t}\end{pmatrix}\begin{pmatrix}\psi_+\\\psi_-\end{pmatrix}\ ,
\end{equation}
where $e^{\mp\im\omega t}\in\sU(1)_\hbar$, $e^{-\omega tJ}\in\sSO(2)_\hbar$, $\psi_\pm=\frac{1}{\sqrt{2}}(\psi_1\pm\im\psi_2)$ and $\psi_1$, $\psi_2$ are complex functions of $x\in\R^3$. I believe that formulae \eqref{3.42} of non-relativistic quantum mechanics show quite clearly why particles and antiparticles in relativistic theories are associated with positive and negative frequencies. The energies of particles and antiparticles in \eqref{3.36}-\eqref{3.40} are non-negative and the correspondence principle requires that they be non-negative in relativistic theories as well.

\section{Quantum charge}

\noindent{\bf\large 4.1. Non-relativistic limit}

For the Hamiltonian
\begin{equation}\label{4.1}
\hat H_0=-\frac{\hbar^2}{2m}\, \delta^{ab}\dpar_a\dpar_b
\end{equation}
of a free particle or antiparticle, both of the equations \eqref{3.36} and \eqref{3.38} can be obtained from the Klein-Gordon equation
\begin{equation}\label{4.2}
\left (\frac{1}{c^2}\frac{\dpar^2}{\dpar t^2} - \delta^{ab}\dpar_a\dpar_b + \frac{m^2c^2}{\hbar^2}\right )\Phi =0
\end{equation}
when considering the non-relativistic limit $c\to\infty$ (see e.g. \cite{Greiner1}). Namely, by decomposing the complex scalar field $\Phi =\Phi_+ +\Phi_-$ into positive and negative frequency parts $\Phi_+$, $\Phi_-$ and redefining them as
\begin{equation}\label{4.3}
\Phi_\pm (x, t) = \psi_\pm(x,t)\exp (\mp\frac{\im}{\hbar}mc^2t)\ ,
\end{equation}
in the non-relativistic limit we obtain equations \eqref{3.36} for $\psi_+(x,t)$ and \eqref{3.38} for $\psi_-(x,t)$ with $\hat H_0$ in the right side. This confirm our assertion that \eqref{3.36} and \eqref{3.38} describe particles $\psi_+$ and antiparticles $\psi_-$. Returning to the columns $\Psi_+, \Psi_-$ in \eqref{3.36}, \eqref{3.38}, we note that
\begin{equation}\label{4.4}
\Psi = \Psi_++\Psi_-=\frac{1}{\sqrt{2}}\begin{pmatrix}\psi_++\psi_-\\-\im(\psi_+-\psi_-)\end{pmatrix}
\end{equation}
i.e. when in relativistic quantum mechanics (QM) they speak of a complex scalar function $\Phi$ in \eqref{4.2}, they mean only the first component of the vector \eqref{4.4}. However, the correspondence principle requires that when formulating the Klein-Gordon equation \eqref{4.2}, we must take into account that the states with particles and antiparticles in the non-relativistic limit are orthogonal. In other words, one should consider the Klein-Gordon field $\Phi$ not as a scalar, but as a section of the quantum bundle \eqref{3.32} and similarly with the Dirac field.

Recall that in the Klein-Gordon (KG) theory the scalar product of two complex functions $\phi$ and $\chi$ is introduced with formula \cite{Greiner2}
\begin{equation}\label{4.5}
(\phi , \chi)=\im\int\dd^3x\,\bigl(\phi^*(x,t)\dot\chi (x,t)-\dot\phi^*(x,t)\chi (x,t)\bigr)\ ,
\end{equation}
where dot means $\dpar/\dpar t$. Also, the KG Lagrangian is invariant under the transformation  $\phi\to e^{\im\theta}\phi$  which leads to the conserved charge
\begin{equation}\label{4.6}
q=\int\dd^3x\,j^0(x)   =  \frac{\im\hbar}{2mc^2} \int\dd^3x\,  (\phi^*\dot\phi-\dot\phi^*\phi) = \frac{\hbar}{2mc^2}\,(\phi , \phi)\ ,
\end{equation}
which, in nonrelativistic approximation \eqref{4.3}, reduce for $c\to\infty$ to
\begin{equation}\label{4.7}
q=\frac{\hbar}{2mc^2}\,(\phi , \phi)\quad\to\quad q= \int\dd^3x\,  (\psi^*_+\psi_+ -\psi^*_-\psi_-)\ .
\end{equation}
Thus, we conclude that $\psi_+^*\psi_+$ is the probability function for particles described by \eqref{3.36}, and $\psi_-^*\psi_-$ is the probability density function for antiparticles described by \eqref{3.38}, and for normalized functions $\psi_\pm$ we get the quantum charge $q=+1$ for particles $\psi_+$ and $q=-1$ for antiparticles $\psi_-$.

In terms of sections  \eqref{4.4} of $\C^2$-bundle \eqref{3.32} the scalar product \eqref{4.5} in nonrelativistic limit reduces to the scalar product
\begin{equation}\label{4.8}
\Psi_1^\+Q\Psi_2=\psi^*_{1+}\psi_{2+} - \psi^*_{1-}\psi_{2-}\quad\Rightarrow\quad\Psi^\+Q\Psi = \psi^*_{+}\psi_{+} - \psi^*_{-}\psi_{-}\ ,
\end{equation}
where 
\begin{equation}\label{4.9}
Q:=-\im J=\begin{pmatrix}0&\im\\-\im&0\end{pmatrix}=-\sigma_2\ ,\quad Qv_\pm=\pm v_\pm\ ,
\end{equation}
is the quantum charge operator. Note that when lifting to the KG equation for $\Phi\in L_{\C^2}$ from \eqref{3.32}, the matrix $Q$ goes over to $\im\dpar_t$, and in the case of the Dirac field $\Psi$ with values in $L_{\C^2}$, the scalar product $\Psi^\+\gamma^0\Psi$ does not contain a time derivative so that the metric \eqref{4.8} on the bundle $L_{\C^2}$ is included in the definition of the scalar product of spinors $\Psi$ with values in $L_{\C^2}$ as ${\overline\Psi}^q\Psi :=\Psi^\+\gamma^0\otimes Q\Psi$.

Returning to the formula \eqref{4.7} we see that for the bundle $L_\C^+$  we have $\psi_-=0$ and $q=+1$, for $L_\C^-$  we have $\psi_+=0$ and $q=-1$, and $q=0$ for three cases
\begin{equation}\label{4.10}
\psi_- = \psi_+^*\ \Rightarrow\     \Psi=\frac{1}{\sqrt{2}}\begin{pmatrix}\psi_++\psi_+^*\\-\im(\psi_+-\psi_+^*)\end{pmatrix}=
\begin{pmatrix}\psi_1\\\psi_2\end{pmatrix}\in L_{\R^2}\subset L_{\C^2}=L^+_\C\oplus L^-_\C
\end{equation}
\begin{equation}\label{4.11}
\psi_- = \psi_+\ \Rightarrow\   
 \Psi=\sqrt{2}\begin{pmatrix}\psi_+\\0\end{pmatrix}=
\begin{pmatrix}\psi_1\\0\end{pmatrix}\in L_\C^{\uparrow}=\diag (L^+_\C\oplus L^-_\C)\subset L^+_\C\oplus L^-_\C
\end{equation}
\begin{equation}\label{4.12}
\psi_- = -\psi_+\ \Rightarrow\   
 \Psi=\im\sqrt{2}\begin{pmatrix}0\\\psi_-\end{pmatrix}=
\begin{pmatrix}0\\\psi_2\end{pmatrix}\in L_\C^{\downarrow}=\mbox{adiag} (L^+_\C\oplus L^-_\C)\subset L^+_\C\oplus L^-_\C\ .
\end{equation}
Case \eqref{4.10} is realized for real scalar fields of spin zero.

\bigskip
\noindent{\bf\large  4.2. Vacuum forces}

From a geometric point of view, canonical quantization of Newtonian particles is the introduction of a new type of Abelian gauge fields $(A_\hbar , F_\hbar )$ given on the phase space $(\R^6, \{\cdot , \cdot\})$ of particle, where $\{\cdot , \cdot\}$ is the Poisson bracket on the space of functions $C^\infty (\R^6)$. The fields $A_\hbar$ and $F_\hbar$, connection and curvature of the quantum bundle $L^+_\C\to\R^6$, are not dynamical, they are given by formulae \eqref{3.20}-\eqref{3.25} encoding the symplectic geometry of the phase space $(\R^6, \{\cdot , \cdot\})$. The components of the curvature $F_\hbar$ are proportional to the components \eqref{3.1} of the ``flat" symplectic 2-form $\omega_{\R^6}$. Hence, the fields $(A_\hbar , F_\hbar )$ can be considered as vacuum fields since the 2-form $\omega_{\R^6}$ has no sources, it is a flat geometric structure on $\R^6$, similar to how Minkowski metric defines a flat space $\R^{3,1}$ which is a vacuum in the theory of gravity.
Thereby, $(A_\hbar , F_\hbar )$ are background fields, they have no sources and are massive for a finite parameter $w^2$ in \eqref{3.5}. Thus, nonrelativistic QM introduces the internal space $(\R^2, J)\cong\C$ of particles and gauge fields $A_\hbar$ acting on it, describing interaction of first quantized particles with vacuum fields.

Antiparticles have the phase space $(\R^6, -\{\cdot , \cdot\})$ with the sign of the Poisson bracket reversed, which is equivalent to replacing $t\to -t$ in the Hamiltonian equations of motion. Their internal space is $(\R^2, -J)\cong\bar\C$ which leads to the replacement of the bundle $L_\C^+$ by the complex line bundle $L_\C^-$ with the opposite quantum charge $q$. Later we will show that the vacuum fields $(A_\hbar , F_\hbar )$ lead to the attraction of particles with opposite signs of the charge $q$.

Analogous fields $(\tilde A_\hbar , \tilde F_\hbar )$ can also be introduced in quantum field theory (QFT) as vacuum fields coming from a symplectic structure on the space of fields. Interaction of these vacuum fields with QFT particles and antiparticles can be explicitly described. In fact, interaction with vacuum fields $(A_\hbar , F_\hbar )$ in first quantized theories leads to the appearance of bound states in one particle-antiparticle sector, while interaction with vacuum fields $(\tilde A_\hbar , \tilde F_\hbar )$ in QFT leads to bound states in a sector of many particles and antiparticles.

\bigskip
\noindent{\bf\large 4.3. Electric charge}

Particles with electric charges $\pm e$ are introduced similarly to particles with quantum charge $q=\pm 1$ through the introduction of additional internal degrees of freedom of classical particles. They again are parametrized by the space $(\R^2, J)_{\rm{em}}\cong\C$, where $J$ is the complex structure on $\R^2$, but they are different from the quantum degrees of freedom $(\R^2, J)_\hbar$ already considered.

In quantum mechanics, we can introduce electromagnetic fields as connections in complex line bundles $E_\C^\pm$ over Galilean space-time $\R\times\R^3$, and not over the space $\R\times T^*\R^3$ as it was for bundles $L_\C^\pm$. We pull-back bundles $E_\C^\pm$ to $\R\times T^*\R^3$ using the projection $T^*\R^3\to\R^3$. After that we should consider the tensor product of the bundles,
\begin{equation}\label{4.13}
(L^+_\C\oplus L^-_\C)\otimes (E^+_\C\oplus E^-_\C) = L^+_\C\otimes E^+_\C \oplus L^+_\C\otimes E^-_\C\oplus L^-_\C\otimes E^+_\C\oplus L^-_\C\otimes E^-_\C\ ,
\end{equation}
whose sections are the vectors
\begin{equation}\label{4.14}
\Psi = \psi_{++}\,v_+\otimes v_+ + \psi_{+-}\,v_+\otimes v_- +\psi_{-+}\,v_-\otimes v_+ +\psi_{--}\,v_-\otimes v_-
\end{equation}
taking values in the space $\C^4$ of fibres of the vector bundle \eqref{4.13}. Note that the tensor products of basis vectors can be constructed as follows:
\begin{equation}\label{4.15}
v_\pm{=}\frac{1}{\sqrt 2}\begin{pmatrix}1\\\mp\im\end{pmatrix}\ \Rightarrow\ v_+\otimes v_+{=}\frac{1}{\sqrt 2}
\begin{pmatrix}v_+\\-\im v_+\end{pmatrix}{=}\frac{1}{ 2}\begin{pmatrix}1\\-\im\\-\im\\-1\end{pmatrix},
\ v_+\otimes v_- {=}\frac{1}{\sqrt 2}\begin{pmatrix}v_+\\\im v_+\end{pmatrix}{=}\frac{1}{2}\begin{pmatrix}1\\-\im\\\im\\1\end{pmatrix}\ \mathrm{etc.}
\end{equation}
and we have $Jv_{\pm} =\pm\im v_{\pm}$.

On vectors \eqref{4.14}, the action of the group $\sU(1)_\hbar$ (first index) and the group $\sU(1)_{\rm{em}}$ (second index)  is given, and the generators of the group U(1)$_\hbar\times\,$ U(1)$_{\rm{em}}$ are $4\times 4$ matrices
\begin{equation}\label{4.16}
I_\hbar = J\otimes\unit_2\quad\mathrm{and}\quad I_e = \unit_2\otimes J\ .
\end{equation}
We have pairs of complex conjugate bundles:
\begin{equation}\label{4.17}
L^+_\C\otimes E^+_\C \quad\mathrm{and}\quad L^-_\C\otimes E^-_\C\ ,\quad
L^+_\C\otimes E^-_\C \quad\mathrm{and}\quad L^-_\C\otimes E^+_\C\ .
\end{equation}
The signs $++, +-$ etc. for components $\psi_{++}, \psi_{+-}$ etc. of the vector \eqref{4.14} indicate combinations of charges $q=\pm 1$ and $\pm e$, where $e>0$ is the modulus of the electric charge of electron. Therefore, particles and antiparticles always belong to either the first or second pair of bundles in \eqref{4.17}. From \eqref{4.16} we see that operators of quantum and electric charges are
\begin{equation}\label{4.18}
Q_\hbar =-\im J\otimes\unit_2\quad\mathrm{and}\quad Q_e =-\im e \unit_2\otimes J\ ,
\end{equation}
where $J=-\im\sigma_2$.

\bigskip
\noindent{\bf\large 4.4. Hydrogen atom}

So, for the transition to a quantum particle, the phase space $T^*\R^3$ of Newtonian particle is extended to the space $T^*\R^3\times\R^2_\hbar$. With this extension by $(\R^2, J)_\hbar =\C$ and $(\R^2, -J)_\hbar =\bar\C$, we get two quantum bundles $L^+_\C$ and $L^-_\C$ and the 
Schr\"odinger equations \eqref{3.36} and \eqref{3.38} for particles and antiparticles. The introduction of electromagnetic internal degrees of freedom requires further extension of the phase space,
\begin{equation}\label{4.19}
T^*\R^3\quad\to\quad T^*\R^3\times (\R^2, \pm J)_\hbar\quad \to\quad  T^*\R^3\times (\R^2, \pm J)_\hbar\times (\R^2, \pm J)_{\rm em}\ ,
\end{equation}
and we get four quantum bundles \eqref{4.13} that allows us to discuss the hydrogen atom by using a spinless charged Schr\"odinger field.

The main task of quantum mechanics was to explain the radiation energy spectrum of the hydrogen atom, the electron-proton system. 
We will not derive this spectrum, but only make some remarks. We consider the wave function of the electron as a section of the bundle 
$L_\C^+\otimes E^-_\C$, which means that, in accordance with the general logic, the wave function of the proton must be a section of the bundle $L_\C^-\otimes E^+_\C$, so that the total electric and quantum charges are equal to zero. The system is described by the potential energy of interaction $V(r)=-e^2/r$ of $e^-$ and $p^+$, $r^2 = \delta_{ab}x^ax^b$. Due to the presence of conserved quantities (energy, angular momentum and the Runge-Lenz vector), the phase space $T^*\R^3$ is reduced to a manifold $\CPP^1\times\CPP^1\subset T^*\R^3$, and the bundle $L_\C^+\otimes E^-_\C$ is reduced to the holomorphic line bundle \cite{Hurt}
\begin{equation}\label{4.20}
\CO (n-1, n-1)\ \stackrel{\C}{\longrightarrow}\ \CPP^1\times\CPP^1\quad\mathrm{for}\quad n=1,2,...
\end{equation}
with a topological index $n$ parametrizing the energy level $E_n\sim -1/n^2$. However, the complete resolving of all equations and reality conditions reduces bundles \eqref{4.20} to smooth complex line bundles
\begin{equation}\label{4.21}
\CB (0, n-1)\ \stackrel{\C}{\longrightarrow}\ S^2=\diag(\CPP^1\times\overline{\CPP}^1)\ ,
\end{equation}
where $\overline{\CPP}^1$ denotes $\CPP^1$ with the opposite complex structure. Solutions of the Schr\"odinger equations are now described by free sections of the bundle \eqref{4.21} with $n=1,2,...$. The space of sections of the line bundle \eqref{4.21} is a reducible representation of the group SU(2) of the form
\begin{equation}\label{4.22}
\C^n\otimes\bar\C^n = \mathop{\sum}_{\ell =0}^{n-1}\C^{2\ell + 1}
\end{equation}
and corresponds to the eigenfunctions $\psi_{n\ell m}(x)$, $\ell =0,...,n-1$, $m=-\ell ,..., +\ell$. The bundles \eqref{4.21}, described in detail in the Appendix in \cite{Popov1}, are not topological in contrast to \eqref{4.20}, which explains the possibility of transitions between levels. Interestingly, the curvature $F_\hbar$ of the bundle $L^+_\C$ vanishes when reduced to the free bundle \eqref{4.21}, $F_{\hbar\mid S^2}=0$. Perhaps this is what ensures the stability of the hydrogen atom.

\bigskip
\noindent{\bf\large 4.5. Pauli equation}

In \eqref{4.19} we have discussed the introduction of quantum and electromagnetic degrees of freedom of a classical particle. Spin internal degrees of freedom are introduced similarly. To do this, we expand the phase space \eqref{4.19} to a manifold
\begin{equation}\label{4.23}
T^*\R^3\times (\R^2, J)_\hbar\times (\R^2, -J)_{\rm{em}}\times \CPP^1\ ,
\end{equation}
where $\CPP^1$ is the Riemann sphere. Moreover, we assume that rotations of the group SO(3) acting on $\R^3$ are accompanied by rotations of the sphere $\CPP^1\cong\sSU(2)/\sU(1)$ with the help of the group $\sSU(2)\cong\,$Spin(3) twice covering the group SO(3)$\,\cong\sSU(2)/\Z_2$. In other words, these internal degrees of freedom are associated with symmetries of space $\R^3$. It is well known that quantization of the extended phase space \eqref{4.23} (with $(\R^2, J)_\hbar =\C , (\R^2, -J)_{\rm{em}}=\bar\C$ and $\CO (1)$ bundle over $\CPP^1$) produces two-component charged Pauli spinors \cite{Sni, Wood}, functions $\Psi_+$ on $\R^3$ with values in $\C^2$, sections of the bundle
\begin{equation}\label{4.24}
S\otimes L^+_\C\otimes E^-_\C\ \longrightarrow\ T^*\R^3\ ,
\end{equation}
depending only on $x^a$ and $t$, and describing electron with spin $s=1/2$ in the electromagnetic field. For this $\C^2$-spinor, the Pauli equation is introduced,
\begin{equation}\label{4.25}
\im\hbar\,{\dpar_t}\Psi_+ = \hat H^+_{\rm{Pauli}}\Psi_+\ ,
\end{equation}
and for electric field of electron-proton system its solutions give a more detailed and exact description of the hydrogen atom than the Schr\"odinger equation.

As a general rule, if we consider quantization of a manifold
\begin{equation}\label{4.26}
T^*\R^3\times (\R^2, -J)_\hbar\times (\R^2, J)_{\rm{em}}\times \overline{\CPP}^1=T^*\R^3\times\bar\C_\hbar\times\C_{\rm{em}}\times \overline{\CPP}^1
\end{equation}
with a complex conjugate internal space  of \eqref{4.23}, then we obtain a section of the bundle
\begin{equation}\label{4.27}
\bar S\otimes L^-_\C\otimes E^+_\C\ \longrightarrow\ T^*\R^3\ ,
\end{equation}
a $\bar \C^2$-spinor $\Psi_-$ with charges opposite to those of the spinor $\Psi_+$. It will satisfy the conjugate Pauli equation,
\begin{equation}\label{4.28}
-\im\hbar{\dpar_t}\,\Psi_- = \hat H^-_{\rm{Pauli}}\Psi_-\ ,
\end{equation}
which can be thought as the equation describing positrons. Let us repeat once again that the statement that antiparticles appear only after transition to a relativistic case is erroneous from a mathematical point of view.

It is well known that the Dirac equation with electromagnetic field for positive frequency bispinors in non-relativistic limit is reduced to the Pauli equation \eqref{4.25} for $\Psi_+$. Similarly, for negative frequency bispinors it is reduced to the conjugate equation \eqref{4.28} for $\Psi_-$. Note that $\Psi_+$ and $\Psi_-$ belong to the complex conjugate bundles $L^+_\C$ and $L^-_\C$ on which the group U(1)$_\hbar$ act conjugately, $\Psi_\pm\to e^{\pm\im\theta}\Psi_\pm$, and similarly for bundles $E^\mp_\C$ and the group U(1)$_{\rm{em}}$. Equations \eqref{3.36}, \eqref{3.38} as well as \eqref{4.25}, \eqref{4.28}  are invariant under the action of the group U(1)$_\hbar$ and we want to keep this invariance and orthogonality of $\Psi_+$ and $\Psi_-$ in the relativistic case.

\section{Bundles $L^\pm_\C$ and harmonic oscillator}

{\bf\large 5.1. Covariant Laplacian}

In this section, we will return to quantum internal degrees of freedom $(\R^2, \pm J)_\hbar$ and their relationship to the harmonic oscillator. To clarify this connection, we consider a system of interacting particles and antiparticles. To simplify, we will consider them moving in one-dimensional space, since for the case of three dimensions everything is similar.

Instead of the words particle and antiparticle we will often use the words particle with quantum charge $q=+1$ and $q=-1$. Thus, we consider the phase space $T^*\R=\R^2$ with the metric (cf. \eqref{3.5})
\begin{equation}\label{5.1}
g_{\R^2}=\dd x^2 + w^4\dd p^2\ ,\quad g_{\R^2}^{-1}=\dpar_ {x}\otimes\dpar_ {x} + w^{-4}\dpar_{p}\otimes\dpar_{p}\ .
\end{equation}
We have the quantum bundle $L_\C^\pm$ over $T^*\R$ for  $q=\pm 1$ with covariant derivatives
\begin{equation}\label{5.2}
\nabla_1=\dpar_ {x}\quad\mathrm{and}\quad \nabla^1=\frac{\dpar}{\dpar p} - \frac{x}{\hbar}J\ .
\end{equation}
Their action on sections $\Psi_\pm =\psi_\pm v_\pm$ of the bundles $L^\pm_\C$ are
\begin{equation}\label{5.3}
\nabla_1\Psi_\pm = \dpar_x\Psi_\pm\quad\mathrm{and}\quad \nabla^1\Psi_\pm =\mp\frac{\im}{\hbar}x\Psi_\pm\ .
\end{equation}
We now introduce covariant Laplacians on these bundles,
\begin{equation}\label{5.4}
\Delta_{L_\C^\pm} := \nabla^2_{x} + w^{-4}\nabla^2_{p}\ ,
\end{equation}
which on polarized sections \eqref{5.3} take the form
\begin{equation}\label{5.5}
\Delta_{L_\C^\pm}\Psi_\pm (x, t) = \bigl(\dpar^2_{x} - \frac{1}{\hbar^2w^4} x^2\bigr)\Psi_\pm (x, t)\ .
\end{equation}
We see that these operators are equivalent to the Hamiltonian of the quantum harmonic oscillator.

For $q=1$ we consider the equation
\begin{equation}\label{5.6}
\im\hbar\,\frac{\dpar\Psi_+ (x_1, t_1)}{\dpar t_1}=-\frac{\hbar^2}{2m}\nabla_{\R^2}^{(1)}\Psi_+ (x_1, t_1)\ ,
\end{equation}
\begin{equation}\label{5.7}
\Psi_+=\exp(-\frac{\im}{\hbar} E_1t_1)\psi_+(x_1)v_+\ \Rightarrow\  -\frac{\hbar^2}{2m} \bigl(\dpar^2_{x^1} - \frac{1}{\hbar^2w^4} x_1^2\bigr)\psi_+(x_1)=E_1\psi_+(x_1)    \ ,
\end{equation}
Similarly, for $q=-1$ we have
\begin{equation}\label{5.8}
-\im\hbar\frac{\dpar\Psi_- (x_2, t_2)}{\dpar t_2}=-\frac{\hbar^2}{2m}\nabla_{\R^2}^{(2)}\Psi_- (x_2, t_2)\ ,
\end{equation}
\begin{equation}\label{5.9}
\Psi_-=\exp(\frac{\im}{\hbar} E_2t_2)\psi_-(x_2)v_-\ \Rightarrow\  -\frac{\hbar^2}{2m} \bigl(\dpar^2_{x^2} - \frac{1}{\hbar^2w^4} x_2^2\bigr)\psi_-(x_2)=E_2\psi_-(x_2)    \ ,
\end{equation}
From \eqref{5.7} and \eqref{5.9} we see that  both cases are equivalent after choosing $t_1=t=-t_2$.

\bigskip
\noindent{\bf\large 5.2. Interacting particles}

We consider two particles with three combinations of the quantum charges $q$: $(1,-1), (1,1)$ and $(-1,-1)$. For noninteracting particles the phase space is $T^*\R\times T^*\R = \R^2\times\R^2$ with the metric
\begin{equation}\label{5.10}
g^{}_{\R^4}=\dd x^2_1 + \dd x^2_2 + w^4\dd p_1^2 + w^4\dd p_2^2\ .
\end{equation}
So, we consider two particles with the same mass $m$. The difference between the cases $(q_1, q_2)=(1,-1), (1,1)$ and $(-1,-1)$ arises after the introduction of interaction.

The two-particle state is described by the cross section of one of the three bundles,
\begin{equation}\label{5.11}
\Psi_{+-}\in L^+_\C\otimes L^-_\C\ ,\quad \Psi_{++}\in L^+_\C\otimes L^+_\C\quad\mathrm{and}\quad\Psi_{--}\in L^-_\C\otimes L^-_\C\ .
\end{equation}
If particles do not interact, then the equations for states \eqref{5.11} breaks down into equations \eqref{5.6} and \eqref{5.8} for single-particle states. To introduce the interaction, we add the term $2w^4\dd p_1\dd p_2$ to the metric \eqref{5.10}, obtaining
\begin{equation}\label{5.12}
g^{\rm{int}}=\dd x^2_1 + \dd x^2_2 + w^4(\dd p_1 + \dd p_2)^2\ .
\end{equation}
To describe the transition from \eqref{5.10} to \eqref{5.12}, we introduce the center-of-mass coordinates
\begin{equation}\label{5.13}
x=x_1-x_2\ ,\quad X=\sfrac12 (x_1+x_2)\ ,\quad p=\sfrac12(p_1-p_2) \quad\mathrm{and}\quad P=p_1+p_2\ ,
\end{equation}
in which the metric \eqref{5.12} takes the form
\begin{equation}\label{5.14}
g^{\rm{int}}=\sfrac12\dd x^2 + 2\dd X^2 + w^4\dd P^2\ ,\quad 
g^{-1}_{\rm{int}}=2\dpar_x\otimes\dpar_x + \sfrac12\dpar_X\otimes\dpar_X + w^{-4}\dpar_P\otimes\dpar_P\ ,
\end{equation}
where $\dpar_P=\sfrac12 (\dpar_{p_1}+\dpar_{p_2}).$

Covariant derivatives in all three bundles \eqref{5.11} have the form
\begin{equation}\label{5.15}
\nabla_x=\dpar_x\ ,\quad \nabla_X=\dpar_X\quad\mathrm{and}\quad \nabla_P=\sfrac12\bigl (\dpar_{p_1} +\dpar_{p_2} - \frac{x_1}{\hbar}J\otimes \unit_2 - \unit_2\otimes\frac{x_2}{\hbar}J\bigr )\ ,
\end{equation}
where $J$ is the generator of the group U(1)$_\hbar$. For polarized sections of the bundles \eqref{5.11} we have
\begin{equation}\label{5.16}
\nabla_P\Psi_{+-}=\nabla_P(\psi_{+-}(x,X,t)\,v_+\otimes v_-) =-\frac{\im}{2\hbar}x\Psi_{+-}\ ,
\end{equation}
\begin{equation}\label{5.17}
\nabla_P\Psi_{\pm\pm}=\nabla_P(\psi_{\pm\pm}(x,X,t)\,v_\pm\otimes v_\pm) =\mp\frac{\im}{\hbar}X\Psi_{\pm\pm}\ ,
\end{equation}
where $t=t_1=\pm t_2$.

Substituting for $\Psi_{+-}$ the time dependence as in  \eqref{5.7} and \eqref{5.9}, we obtain equation
\begin{equation}\label{5.18}
-\frac{\hbar^2}{2m}\bigl (2\dpar_x^2 + \sfrac12\dpar_X^2 + \frac{1}{w^4}\nabla_P^2\bigr )\Psi_{+-}(x, X)=(E_1+E_2)\Psi_{+-}(x, X)\ ,
\end{equation}
which admits factorized solutions of the form
\begin{equation}\label{5.19}
\Psi_{+-}(x, X)=\psi_{+-}(x, X)\,v_+\otimes v_-=\psi(x)\Phi(X)\, v_+\otimes v_-  \ .
\end{equation}
Substituting \eqref{5.19} into \eqref{5.18}, we obtain for $\Phi$ an equation with a solution
\begin{equation}\label{5.20}
\Phi (X, t_2)=\exp\left (\frac{\im}{\hbar}(E_0t_2 - P_0X)\right )\ ,\quad E_0=\frac{P_0^2}{2M}\ ,\quad M=2m,\ \  t_2=-t\ ,
\end{equation}
describing the free motion of the center of mass of this two-particle system, and for $\psi$ we obtain the equation of a quantum oscillator,
\begin{equation}\label{5.21}
\left (-\frac{\hbar^2}{2m_0}\dpar_x^2 + \frac{m_0\omega^2}{2}x^2\right )\psi (x)=(E_1+E_2-E_0)\psi (x)\ ,
\end{equation}
solutions of which describe the bound states of a particle and antiparticle,
\begin{equation}\label{5.22}
\psi (x,t) = \exp \left (-\frac{\im}{\hbar}(E_1+E_2-E_0)t\right )\psi (x)\ ,\quad m_0=\frac{m}{2}\ ,\quad \omega^2=\frac{1}{2m^2w^4}\ ,
\end{equation}
with $\psi(x)$ being a solution of \eqref{5.21}.

Now we will consider $\Psi_{\pm\pm}$ with time dependence as in \eqref{5.7} and \eqref{5.9} and factorization as in \eqref{5.19} with $v_\pm\otimes v_\pm$. Substituting the factorized $\Psi_{\pm\pm}$ into equation \eqref{5.18} with $\Psi_{\pm\pm}$ instead of $\Psi_{+-}$, we obtain
\begin{equation}\label{5.23}
\Phi_\pm(X,t)=\exp\left (\mp\frac{\im}{\hbar}(E_1+E_2-E_0)t\right )\Phi (X)\ ,
\end{equation}
where $\Phi(X)$ satisfy the quantum oscillator equation
\begin{equation}\label{5.24}
\bigl(-\frac{\hbar^2}{2M}\dpar^2_X + \frac{M\omega^2}{2}X^2\bigr)\Phi (X) = (E_1+E_2-E_0)\Phi (X)\ ,
\end{equation}
and the functions
\begin{equation}\label{5.25}
\psi_{\pm}(x,t)=\exp\left (\mp\frac{\im}{\hbar}(E_0t - p_0x)\right )\ ,\quad E_0=\frac{p_0^2}{2m_0}\ ,
\end{equation}
which describes the plane-wave solutions with the wave front at
\begin{equation}\label{5.26}
x=x_1-x_2=\frac{p_0}{m}t=v_0t\ .
\end{equation}
From \eqref{5.26} it follows that the distance between particles with the same quantum charge $q=\pm 1$ increases linearly with time, and their center of mass  oscillates around the point $X=0$. For the three-dimensional case, the solutions have the form similar to \eqref{5.20}-\eqref{5.22} for $(1, -1)$ and similar to \eqref{5.23}-\eqref{5.25} for $(\pm 1, \pm 1)$.

\bigskip
\noindent{\bf\large 5.3. Harmonic oscillator}

Note that we did not introduce equations \eqref{5.5}, \eqref{5.7}, \eqref{5.9}, \eqref{5.21} and \eqref{5.24} of the harmonic oscillator in one dimension by introducing potential energy. We simply used the covariant Laplacian on the bundles over the phase spaces of one or two particles with fixed metrics and quantum charges.

In three dimensions the covariant Laplacian on bundles $L_\C^\pm$ over $T^*\R^3$ has the form
\begin{equation}\label{5.27}
\Delta^{}_{\R^6}=\delta^{ab}\nabla_a\nabla_b + g_{a+3\, b+3}\nabla^{a+3}\nabla^{b+3}\ .
\end{equation}
The metric on $T^*\R^3$ and covariant derivatives are written out in Section 3, see \eqref{3.5}-\eqref{3.7} and \eqref{3.9}. On polarized sections $\Psi_\pm$ of $L^\pm_\C$ operator \eqref{5.27} take the form
\begin{equation}\label{5.28}
\Delta^{}_{\R^6}\Psi_\pm=\bigl (\delta^{ab}\dpar_a\dpar_b - \frac{1}{\hbar^2w^4}\delta_{ab}x^ax^b\bigr)\Psi_\pm\ .
\end{equation}
We will define the equation of a quantum isotropic harmonic oscillator
\begin{equation}\label{5.29}
-\frac{\hbar^2}{2m_0}\Delta^{}_{\R^6}\psi (x)=E \psi (x)\quad\mathrm{for}\quad\Psi=\exp\bigl ( -\frac{\im}{\hbar} Et \bigr) \psi(x)
\end{equation}
with
\begin{equation}\label{5.30}
\frac{1}{w^4}=m_0^2\omega^2
\end{equation}
and interpret it similarly to equation \eqref{5.21} as describing a coupled system of particle and antiparticle.

For classical system we have
\begin{equation}\label{5.31}
x^a=x^a_1-x^a_2\ ,\quad p_a=\delta_{ab}(m_0\dot x^b_1-m_0\dot x^b_2) =\sfrac12\,(p^1_a-p^2_a)
\end{equation}
since $m=2m_0$ as it was shown in one-dimensional case. The equations of motion have the form
\begin{equation}\label{5.32}
\ddot x^a + \omega^2x^a=0\ ,
\end{equation}
and we choose their solution in the form
\begin{equation}\label{5.33}
x^a=v^a_0\,\frac{1}{\omega}\sin\omega t + x^a_0\cos\omega t= \sfrac12\bigl(x^a_0+\frac{\im}{\omega}v^a_0\bigr)e^{-\im\omega t}+
\sfrac12\bigl(x^a_0-\frac{\im}{\omega}v^a_0\bigr)e^{\im\omega t}\ ,
\end{equation}
so that in the limit $\omega\to 0$ ($w^2\to\infty$) we get free motion
\begin{equation}\label{5.34}
x^a_1=\sfrac12 v_0^at+\sfrac12x^a_0\quad\mathrm{and}\quad x^a_2=-\sfrac12 v_0^at-\sfrac12x^a_0
\end{equation}
of particle and antiparticle in opposite direction.

The solutions for the quantum case \eqref{5.29} are standard and have the form
\begin{equation}\label{5.35}
|n_1, n_2, n_3\rangle =\frac{(a_1^\+)^{n_1}(a_2^\+)^{n_2}(a_3^\+)^{n_3}}{\sqrt{n_1!n_2!n_3!}}|0,0,0\rangle\ ,\quad
E_n=\hbar\omega (n+\sfrac32)
\end{equation}
with $n=n_1+n_2+n_3$, where
\begin{equation}\label{5.36}
a_c^\+=-\sqrt{\frac{\hbar w^2}{2}}\left (\dpar_c - \frac{1}{\hbar w^2} \delta_{cb}x^b\right )\quad\mathrm{and}\quad
a_c=\sqrt{\frac{\hbar w^2}{2}}\left (\dpar_c + \frac{1}{\hbar w^2} \delta_{cb}x^b\right )
\end{equation}
are creation and annihilation operators, $c=1,2,3$. In \eqref{5.35} we have broken the spherical symmetry and in \eqref{5.36} we can replace
$w^2\to (w_1^2, w_2^2, w_3^2)$. So, particles with opposite charges $q=1$ and $q=-1$ in the vacuum field $A_\hbar$ form a coupled system described by a harmonic oscillator both at the classical and quantum levels.

\bigskip
\noindent{\bf\large 5.4. Heisenberg picture}

Continuing the discussion of the connection between the gauge potential $A_\hbar$ on quantum bundles $L_\C^\pm$ and their various tensor products, we note the following. The connection $A_\hbar$ in $L_\C^\pm$ is fixed by the canonical commutation relations, and has the components
\begin{equation}\label{5.37}
A_a=0,\quad A^{a+3}=-\frac{1}{\hbar}x^a\quad\Leftrightarrow\quad A_\hbar=-\frac{1}{\hbar}\,x^a\dd p_a J
\end{equation}
where $J$ is the generator of the group U(1)$_\hbar$ acting on the sections of the bundles $L_\C^\pm$. When passing to the components to these sections, $J$ is replaced by $+\im$ or $-\im$.

The connection \eqref{5.37} enters into the covariant derivatives on $L_\C^\pm$,
\begin{equation}\label{5.38}
\nabla_a=\frac{\im}{\hbar}\,\ph_a=\dpar_a\quad\mathrm{and}\quad 
\nabla^{a+3}=-\frac{\im}{\hbar}\,\xh^a=\frac{\dpar}{\dpar p_a}-\frac{1}{\hbar}\,x^aJ\ ,
\end{equation}
which are equivalent to quantum operators $\ph_a$ and $\xh^a$. Therefore, if we want to see the dependence of $A_\hbar$ on time, we should go to the Heisenberg representation, where the wave function does not depend on time, and the operators $\xh^a(t)$ and $\ph_a(t)$ take the form
\begin{equation}\label{5.39}
\begin{split}
\xh^a(t)&=\xh^a(0)\cos\omega t+\delta^{ab}\ph_b(0)\frac{1}{m_0\omega}\sin\omega t\ ,\\
\ph_a(t)&=\ph_a(0)\cos\omega t-\delta_{ab}\xh^b(0){m_0\omega}\sin\omega t\ .
\end{split}
\end{equation}
Comparing \eqref{5.38} and \eqref{5.39}, we obtain
\begin{equation}\label{5.40}
\begin{pmatrix}\nabla_a(t)\\w^{-2}\delta_{ab}\nabla^{b+3}(t)\end{pmatrix}=\begin{pmatrix}\cos\omega t&\sin\omega t\\ -\sin\omega t&\cos\omega t\end{pmatrix}\begin{pmatrix}\nabla_a\\w^{-2}\delta_{ab}\nabla^{b+3}\end{pmatrix}
\end{equation}
and
\begin{equation}\label{5.41}
\nabla_c(t)+\im w^{-2}\delta_{cb}\nabla^{b+3}(t)=e^{-\im\omega t}(\nabla_c+\im w^{-2}\delta_{cb}\nabla^{b+3})=\gamma e^{-\im\omega t} a_c=\gamma a_c(t)\ ,
\end{equation}
\begin{equation}\label{5.42}
\nabla_c(t)-\im w^{-2}\delta_{cb}\nabla^{b+3}(t)=e^{\im\omega t}(\nabla_c-\im w^{-2}\delta_{cb}\nabla^{b+3})=\gamma e^{\im\omega t} a_c^\+=\gamma a_c^\+(t)\ ,
\end{equation}
for annihilation and creation operators \eqref{5.36} with $\gamma=\sqrt{2/\hbar w^2}$ and $J\to\im$ on $L^+_\C$. From formula
\begin{equation}\label{5.43}
\nabla^{a+3}(t)=\frac{\dpar}{\dpar p_a(t)}-\frac{\im}{\hbar}\, x^a(t)\quad\mathrm{for}\quad L_\C^+
\end{equation}
we conclude that
\begin{equation}\label{5.44}
-\hbar A^{a+3}(t)=x^a(t)
\end{equation}
is exactly the coordinates of harmonic oscillator \eqref{5.33} subject to the equations \eqref{5.32}.

Note that the curvature components $F_\hbar$ of the connection $A_\hbar$ in \eqref{3.20}-\eqref{3.23} are constant and satisfy Maxwell type equations
\begin{equation}\label{5.45}
\dpar_a F^{a\,b+3} + \dpar^a F_a^{\ \,b+3}=0\quad\mathrm{for}\quad F^{a\,b+3}:=\delta^{ac}F_c^{\ \,b+3}\ .
\end{equation}
The transition to the Heisenberg picture formally means the addition of components $F^{\ \, b+3}_t :=\dpar_t A^{b+3}$ and transition from equations \eqref{5.45} to equations
\begin{equation}\label{5.46}
\dpar_t F^{t\,b+3}+ \dpar_a F^{a\,b+3} + \dpar^a F_a^{\ \, b+3}=\omega^2 A^{b+3}\ ,
\end{equation}
which are equivalent to equations \eqref{5.32} for a harmonic oscillator. Equations \eqref{5.46} describe massive Abelian field $A_\hbar$ and are not invariant under the gauge transformations $A_\hbar\mapsto A_\hbar^\varphi =A_\hbar +\im\dd\varphi$. This invariance is restored only when $\omega\to 0$ ($w^2\to\infty$) and the field $A_\hbar$ becomes massless, and instead of oscillation in \eqref{5.33} we get free motion \eqref{5.34}. Thus, $w^{-2}$  plays the role of effective ``coupling" constant such that interaction of $q=\pm 1$ particles with vacuum field $A_\hbar$ disappears at $w^{-2}\to 0$ and they become free particles.


\section{Classical field theory and first quantization}

{\bf\large 6.1. Nonrelativistic case}

We started our consideration with a classical particle as a point in phase space $T^*\R^3$. We have consistently expanded this space by  adding internal degrees of freedom:
\begin{equation}\label{6.1}
T^*\R^3\ \longrightarrow\ T^*\R^3\times \R^2_\hbar\ \longrightarrow\ T^*\R^3 \times \R^2_\hbar\times \R^2_{\rm{em}}\  \longrightarrow         \ T^*\R^3\times \R^2_\hbar\times \R^2_{\rm{em}}\times S^2\ .
\end{equation}
Each such extension is equivalent to introducing a new complex vector bundle with a connection which is a force field acting on these internal degrees of freedom. The combination of charges of quantum particles corresponding to \eqref{6.1} is determined by the combination of matrices of complex structures specified in internal spaces in \eqref{6.1}. In particular, the addition of an internal space $(\R^2, J)_\hbar\cong\C$ leads to a quantum bundle $(L_\C^+, A_\hbar )$ describing particles, and an internal space $(\R^2, -J)_\hbar\cong\bar\C$  with the opposite complex structure leads to a quantum bundle $(L_\C^-, - A_\hbar)$ describing antiparticles.

To introduce an electric charge $\pm e$, an additional expansion of the phase space with internal spaces $(\R^2, \pm J)_{\rm{em}}$ is necessary, leading to bundles $E^\pm_\C$ for positively and negatively charged quantum particles. These bundles must be tensor multiplied by bundles $L_\C^\pm$, leading to four possible cases \eqref{4.13} of charged particles and antiparticles. The Schr\"odinger equation becomes two-component,
\begin{equation}\label{6.2}
\im\hbar\, Q\,\dpar_t\Psi = \hat H\Psi\quad\Rightarrow\quad\im\hbar\,\dpar_t\Psi_+=\hat H\Psi_+\quad\mathrm{and}\quad-\im\hbar\,\dpar_t\Psi_-=\hat H\Psi_-\ ,
\end{equation}
where $Q=-\im J$ is the quantum charge operator, and
\begin{equation}\label{6.3}
\Psi =\Psi_+ +\Psi_- =\psi_+v_++\psi_-v_-=\frac{1}{\sqrt{2}}\begin{pmatrix}\psi_++\psi_-\\-\im(\psi_+-\psi_-)\end{pmatrix}\in\C^2
\end{equation}
is a section of the bundle $L_{\C^2}^{}=L_\C^+\oplus L_\C^-$. For electrically charged quantum particles, one should pass to $\Psi$ from \eqref{4.14} with \eqref{4.15}-\eqref{4.18}. In \eqref{6.2}, $\hat H$ is the Hamiltonian operator and for free particles we have
\begin{equation}\label{6.4}
\hat H_0=-\frac{\hbar^2}{2m}\,\delta^{ab}\dpar_a\dpar_b
\end{equation}
where $m$ is a mass.

To introduce spin, the extension of the phase space \eqref{6.1} by a two-dimensional sphere $S^2$ with a complex structure $I$ on it is used
\cite{Sni, Wood}. Using the Riemann sphere $(S^2, I)\cong\CPP^1$ leads to a bundle $S\to T^*\R^3$ of two-component spinors (pulled-back from $\R^3$) and the Pauli equation describing particles, and using a sphere $(S^2, -I)\cong\overline{\CPP}^1$ with the opposite complex structure $-I$ leads to a complex conjugate bundle $\bar S$ describing antiparticles, and we have, for example, bundles 
$L^+_\C\otimes E_\C^\pm\otimes S$ and $L^-_\C\otimes E_\C^\mp\otimes \bar S$.

\bigskip
\noindent{\bf\large 6.2. Klein-Gordon field $\Psi$}

Let us now consider relativistic mechanics of a particle of mass $m$ moving in a manifold $M$. Its phase manifold is $T^*M$ and we start with the case of Minkowski space $M=\R^{3,1}$. In relativistic case we use the natural units with $\hbar =c=1$. Let us extend the phase space $T^*\R^{3,1}$ by internal degrees of freedom $(\R^2, \pm J)_\hbar$ in the same way as we did for the nonrelativistic case $T^*\R^{3}$, i.e. we introduce bundles $L_\C^\pm$,
\begin{equation}\label{6.5}
L_\C^\pm\ \stackrel{\C}{\longrightarrow}T^*\R^{3,1}
\end{equation}
which, when restricted to $T^*\R^{3}$, coincide with those considered in the previous sections. It is well known that quantization of the extended phase space $T^*\R^{3,1}\times (\R^2, \pm J)_\hbar$ for {\it free} relativistic particles in a way, invariant under the action of the Poincar\'e group, leads to the restriction
\begin{equation}\label{6.6}
\eta^{\mu\nu}p_\mu p_\nu + m^2 =0\quad\Rightarrow\quad (\eta^{\mu\nu}\dpar_\mu \dpar_\nu -m^2)\Psi (x)=0\ ,
\end{equation}
where $x=(x^\mu )$, $\mu =0,...,3$. The phase space of a free particle of mass $m$ is given by the orbit of the Poincar\'e group, 
\begin{equation}\label{6.7}
\CO_{m,s=0}=TH^3=TH_+^3\cup TH^3_-\ ,
\end{equation}
where $H^3=H_+^3\cup H^3_-$ is the two-sheeted hyperboloid in the momentum space, given in \eqref{6.6},
\begin{equation}\label{6.8}
H^3_+: \ p_0=\sqrt{\delta^{ab}p_ap_b+m^2}\ ,\quad H^3_-: \ p_0=-\sqrt{\delta^{ab}p_ap_b+m^2}\ ,
\end{equation}
and $TH^3$ is the tangent bundle of $H^3$. 
We see that the phase space \eqref{6.7} of free spinless particles consists of two connected components and, accordingly, two complex line bundles $\tilde L_\C^\pm \to TH^3_\pm$. Unconstrained cross sections of the bundles $\tilde L_\C^\pm$ correspond to particles with quantum charge $q=\pm 1$ and $\tilde L_\C^\pm$ are restrictions of bundles $ L_\C^\pm$ from \eqref{6.5} with respectively positive and negative frequency sections $\Psi_\pm$ satisfying the KG equation \eqref{6.6}. Thus, quantum bundle over the disconnected phase space \eqref{6.7} with free sections is equivalent to the quantum bundle $L_\C^+\oplus L_\C^-$ with section of the form \eqref{6.3} constrained by the KG equation.

Note that in relativistic quantum mechanics (RQM), complex-valued functions $\Phi =\psi_++\psi_-$ are considered, where $\psi_+$ and $\psi_-$ from  \eqref{6.3} have positive and negative frequencies, and in the momentum representation they are given on the manifold $H_+^3$ and $H_-^3$, respectively. In the previous sections, we discussed that in the nonrelativistic limit KG equations \eqref{6.6} for $\psi_+$ is reduced to the Schr\"odinger equation \eqref{3.36} for free particles with $\hat H=\hat H_0$, and KG equations \eqref{6.6} for $\psi_-$ is reduced to the
Schr\"odinger equation \eqref{3.38} for antiparticles also with $\hat H=\hat H_0$. In fact, only the first component of the two-component vector 
 \eqref{6.3} is considered in RQM, which violates the correspondence principle between relativistic and non-relativistic quantum mechanics.

In this paper, we consider the Klein-Gordon equations for sections $\Psi$ of the bundle
\begin{equation}\label{6.9}
L_\C^+\oplus L_\C^-\ \longrightarrow\ T^*\R^{3,1}\ ,
\end{equation}
with $\Psi$ of the form  \eqref{6.3} and $\psi_\pm^{}$ depending only on $x\in \R^{3,1}$ (polarization condition). Also, we are interested in considering particles ($q=1$) and antiparticles ($q=-1$) interacting with the vacuum field $A_\hbar$, similarly to how we considered system of two particles with $q=\pm 1$ in \eqref{5.10}-\eqref{5.36}. Such systems are not Lorentz invariant and bound states of particles and antiparticles arise in them. In particular, generalizing the consideration in \eqref{5.10}-\eqref{5.36}, one can introduce equations
\begin{equation}\label{6.10}
\left (\eta^{\mu\nu}\frac{\dpar}{\dpar x^\mu}\frac{\dpar}{\dpar x^\nu} - m^2 -\frac{1}{w^4}\delta_{ab}x^ax^b\right )\Psi (x)=0\ ,
\end{equation}
\begin{equation}\label{6.11}
\left (\eta^{\mu\nu}\frac{\dpar}{\dpar X^\mu}\frac{\dpar}{\dpar X^\nu} - M^2 \right )\Phi (X)=0\ ,
\end{equation}
where the Klein-Gordon oscillator equation \eqref{6.10} describes the bound states, and the equation  \eqref{6.11} describes  free motion of the center of mass of this system with $\sqrt{2}x^\mu = x^\mu_1- x^\mu_2$ and   $\sqrt{2}X^\mu = x^\mu_1+ x^\mu_2$.

\bigskip
\noindent{\bf\large 6.3. Bispinors as quantum particles}

We discussed the extension \eqref{6.1} of the phase space $T^*\R^{3}$ of a non-relativistic particle leading to two-component spinors having quantum charge $q=\pm 1$ and electric charge $\pm e$. In the relativistic case the Poincar\'e non-invariant phase space $T^*\R^{3}\times S^2$ in \eqref{6.1} is replaced by the orbit of the Poincar\'e group in the space dual to its Lie algebra, 
\begin{equation}\label{6.12}
\CO_{m,s} = TH_+^3\times \CPP^1\ \mathop{\cup}\ TH_-^3\times \overline{\CPP}^1\ ,
\end{equation}
corresponding to mass $m$ and spin $s=1/2$ (for $\CO (1)$ and $\bar\CO (1)$ bundles over $\CPP^1$ and $\overline{\CPP}^1$). Here 
$\overline{\CPP}^1$ is the Riemann sphere $\CPP^1$ with conjugate complex structure. Geometric quantization of the phase space  \eqref{6.12} of  {\it free} relativistic particle with spin degrees of freedom gives a bispinor $\Psi$ satisfying the Dirac equation
\begin{equation}\label{6.13}
(\im\gamma^\mu\dpar_\mu -m)\Psi (x) =0\ ,
\end{equation}
where $\gamma^\mu$ are Dirac $4\times 4$ matrices, generators of the Clifford algebra Cl$^\C (4)$ (for more details see e.g. \cite{Sni, Wood, Bona}) and references therein). Thus, Dirac spinors in Minkowski space $\R^{3,1}$ are introduced as quantization of a free relativistic particle after adding spin degrees of freedom associated with the group SO(3,1). Bispinors $\Psi$ are sections $\Psi =\Psi^++\Psi^-$ of the complex vector bundle
\begin{equation}\label{6.14}
W=W^+\cup W^-\ \stackrel{S}{\longrightarrow}\ \R^{3,1}\quad\mathrm{for}\quad S\cong \C^4\ ,
\end{equation}
where the bundles $W^\pm$ correspond to the sheets $H^3_\pm$ of the hyperboloid $H^3$, parametrizing solutions of the equation \eqref{6.13}  (positive and negative frequency).

As is known, free Klein-Gordon equation follows from the Dirac equation \eqref{6.13}. At the same time, we have shown that from the consideration of the non-relativistic limit of KG equation it follows that the positive frequency solutions are sections of the quantum bundle $L_\C^+$, and negative frequency solutions are sections of the bundle $L_\C^-$. Therefore, the Dirac spinors should also take values in the quantum bundle \eqref{6.9}, i.e. they should have additional indices $\pm$ of this bundle. Indeed, it is known that the Dirac equation with electromagnetic  fields in $\gamma^\mu\nabla_\mu$ in non-relativistic limit is reduced to the Pauli equation \eqref{4.25} if we start from positive frequency Dirac spinors $\Psi^+$. Similarly, for negative frequency bispinor $\Psi^-$, it is reduced to the conjugate equation \eqref{4.28} with opposite electric and quantum charges. Below we will look at all this from the point of view of plane-wave solutions of the Dirac equation \eqref{6.13}.

\bigskip
\noindent{\bf\large 6.4. Plane-wave solutions}

The simplest positive frequency solution of the Dirac equation \eqref{6.13} in the rest frame have the form:
\begin{equation}\label{6.15}
\Psi_s^+=e^{-\im Et}\begin{pmatrix}\xi_s^y\\0\end{pmatrix}\ ,\quad\xi_s=\xi_s^y\dpar_y\in L_\C^+\ ,\quad E^2=m^2\ .
\end{equation}
Here $\xi_s$ is a coordinate-independent two-components spinor with values in the bundle $L_\C^+$, as indicated by the index ``$y$" in $\xi^y$. We choose generators of the Clifford algebra Cl$^\C(4)$ as matrices
\begin{equation}\label{6.16}
\gamma^0=\begin{pmatrix}\unit_2&0\\0&-\unit_2\end{pmatrix}\ ,\quad
\gamma^a=\begin{pmatrix}0&\sigma_a\\-\sigma_a&0\end{pmatrix}\ ,\quad
 \gamma^5:=\im\gamma^0\gamma^1\gamma^2\gamma^3=\begin{pmatrix}0&\unit_2\\\unit_2&0\end{pmatrix}\ ,
\end{equation}
where $\sigma_a$ are Pauli matrices. Recall that a charge conjugated Dirac spinor is defined by the formula
\begin{equation}\label{6.17}
(\Psi_s^+)_c:=-\im\gamma^2(\Psi_s^+)^*=e^{\im Et}\begin{pmatrix}0\\\chi^{\bar y}\end{pmatrix}=:\Psi_s^-\ ,\quad
\chi_s=\chi^{\bar y}_s\dpar_{\bar y}\in L_\C^-\ ,
\end{equation}
where $\chi^{\bar y}_s:=\im\sigma_2\bar\xi^{\bar y}$ (star and bar denote complex conjugation). Negative frequency spinor $\Psi_s^-$ in \eqref{6.17} satisfy the same Dirac equation \eqref{6.13} but has quantum charge $q=-1$ while $\Psi_s^+$ has quantum charge $q=+1$.

Spinors $\Psi_s^+$ and $\Psi_s^-$ have in the general frame four components, but they belong to different one-dimensional complex vector spaces, as indicated by the indices $y$ and $\bar y$. This means, that instead of the spinor bundle \eqref{6.14}, we should consider the bundle
\begin{equation}\label{6.18}
\CW = W^+\otimes L_\C^+\oplus W^-\otimes L_\C^- =\CW^+\oplus\CW^-\ ,
\end{equation}
since all fundamental spinors have quantum charge that distinguishes particles and antiparticles. Note that for massive spinors the sign of frequency coincides with the sign of the Dirac norm, ${\overline\Psi}\Psi = \Psi^\+\gamma^0\Psi$, and the fundamental spinors are all massive. Replacing indices $y, \bar y$ with $\pm$ indices and $\dpar_y\to v_+$, $\dpar_{\yb}\to v_-$, the above solutions can be written in the form of 8-component columns
\begin{equation}\label{6.19}
\Psi = a_s\Psi_s^+\otimes v_+ + b_s\Psi_s^-\otimes v_- = \begin{pmatrix}a_s\Psi_s^+  + b_s\Psi_s^-\\-\im( a_s\Psi_s^+  - b_s\Psi_s^-  ) \end{pmatrix}
\end{equation}
which are sections of the bundle \eqref{6.18}. The new norm for $\Psi$ is ${\overline\Psi}^q\Psi =\Psi^+\gamma^0\otimes Q\Psi$, where $Q=-\sigma_2$. To introduce electrically charged fermions, one should introduce tensor product of bundles,
\begin{equation}\label{6.20}
\CW\otimes (E_\C^+\oplus E_\C^-)= W^+{\otimes}L_\C^+{\otimes}E_\C^+\ \oplus\  W^+{\otimes}L_\C^+{\otimes}E_\C^-\ \oplus\ W^-{\otimes}L_\C^-{\otimes}E_\C^+\ \oplus\  W^-{\otimes}L_\C^-{\otimes}E_\C^-\ ,
\end{equation}
obtaining four complex vector bundles to describe charged first quantized $s=\sfrac12$ particles.

\bigskip
\noindent{\bf\large 6.5. Internal spaces of particles}

The bundle \eqref{6.14} can be generalized to the case of a curved Lorentzian four-manifold $M$ with metric $g_M$ and Levi-Civita connection $\Gamma_\mu =(\Gamma_{\mu\hat a\hat b})$, where $\hat a, \hat b=0,...,3$ are Lorentz indices. One can introduce spinor bundle
\begin{equation}\label{6.21}
W=E(M,S)=P(M, \mbox{Spin}(3,1))\times^{}_{ \rm{Spin}(3,1)}S=\mathop{\bigcup}_{x\in M}\left\{(x,\Psi_x)\mid \Psi_x\in S_x\right\}\ ,
\end{equation}
associated with the principal bundle $P(M, \rm{Spin}(3,1))$. The bundle $P$ and $W$ are endowed with the Levi-Civita connection entering into the covariant derivative
\begin{equation}\label{6.22}
\nabla_\Gamma =\dd + \Gamma^{}_{ \rm{Spin}(3,1)}=\dd + \sfrac14\,\dd x^\mu\Gamma_{\mu\ah\bh}[\gamma^\ah , \gamma^\bh ]
\end{equation}
acting on spinors $\Psi$. Thus, spin degrees of freedom are related with space-time symmetries and the connection \eqref{6.22} specify the action of gravitational forces on $\Psi$.

Internal degrees of freedom of particles can also be related to groups $G$ not acting on the manifold $M$ and its cotangent bundle $T^*M$, as we saw for the electromagnetic forces with $G=\sU(1)_{\rm{em}}$. All these degrees of freedom can also be itroduced by geometric (=canonical) quantization. To do this, the phase space $T^*M$ of a classical particle should be extended to the phase space $T^*P$, where
\begin{equation}\label{6.23}
P(M, G)\stackrel{G}{\longrightarrow} M
\end{equation}
is the principal $G$-bundle over $M$ with Lie groups $G_x$ as fibres over $x\in M$ (see e.g. \cite{Stern, Wein, Mont}). Instead of $T^*G$
as fibres in the bundle $T^*P\to T^*M$ one can also use an orbit of the group $G$ in the dual Lie algebra $\frg^*$ which is a coset space $G/H$ with a symplectic structure \cite{Stern, Wein, Mont}. In particular, for the groups U($N$) we are considering, we should take the complex projective space $\C P^N=\sU(N)/\sU(1)\times\sU(N-1)$ and the hyperplane bundle $\CO(1)$ over it.
Then the quantization of the extended phase space produces an associated vector bundle
\begin{equation}\label{6.24}
E(M,V)=P\times_GV\ ,
\end{equation}
where $V$ is the space of some irreducible representation of $G$. Quantum particle will be a section $(x,\psi (x))$ of the bundle \eqref{6.24},
where $\psi (x)\in V_x$, $x\in M$. As usual, quantum particle is a point $x$ in space-time $M$ plus internal space $V_x$ of particle at this point.

Proceeding in the above way for the groups $G= \sU(1)$, SU(2) and SU(3), we get the internal space $\C_x$ for electromagnetic interactions, $\C^2_x$ for weak interactions and $\C_x^3$ for strong interactions. The corresponding quantum particles/antiparticles are described by complex vector bundles
\begin{equation}\label{6.25}
E_\C^+ =\mathop{\bigsqcup}_{x\in M}\C_x=\mathop{\bigcup}_{x\in M}\left\{(x, \psi (x))\mid \psi(x)\in\C_x\right\} = E(M,\C )\ ,
\end{equation}
\begin{equation}\label{6.26}
E_{\C^2}^+ =\mathop{\bigsqcup}_{x\in M}\C_x^2=\mathop{\bigcup}_{x\in M}\left\{(x, \psi (x))\mid \psi(x)\in\C_x^2\right\} = E(M,\C^2 )\ ,
\end{equation}
\begin{equation}\label{6.27}
E_{\C^3}^+ =\mathop{\bigsqcup}_{x\in M}\C_x^3=\mathop{\bigcup}_{x\in M}\left\{(x, \psi (x))\mid \psi(x)\in\C_x^3\right\} = E(M,\C^3 )\ ,
\end{equation}
and antiparticles/particles with opposite charges are related with complex conjugate bundles $E^-_\C =\overline{E^+_\C}, E^-_{\C^2} =\overline{E^+_{\C^2}}$ and $E^-_{\C^3}=\overline{E^+_{\C^3}}$. As a result, all quantum particles and antiparticles of the Standard Model (quarks and leptons) are described as sections of vector bundles of the form
\begin{equation}\label{6.28}
W^+\otimes L_\C^+\otimes\CE_\C\otimes\CE_{\C^2}\otimes\CE_{\C^3}\quad\mathrm{and}\quad
W^-\otimes L_\C^-\otimes\bar\CE_\C\otimes\bar\CE_{\C^2}\otimes\bar\CE_{\C^3}\ ,
\end{equation}
respectively. Here, $\CE_{\C^N}$ is either $E_{\C^N}^+$ or $E_{\C^N}^-$, and sections $\Psi$ of \eqref{6.28} may not have indices in one or two of the bundles \eqref{6.25}-\eqref{6.27} or their complex conjugate. Fractional charges will be discussed later and $\CE_\C$ may correspond to fractional electric charge for quarks.

\bigskip
\noindent{\bf\large 6.6. Five forces}

Thus, the matter fields $\Psi$ are quantum particle $(q=1)$ or antiparticles $(q=-1)$ and they can be introduced via the canonical quantization scheme. This is their difference from the five gauge fields,
\begin{equation}\label{6.29}
A^\hbar_{\sU(1)},\quad A^{\rm{em}}_{\sU(1)},\quad A^{\rm{weak}}_{\sSU(2)},\quad A^{\rm{strong}}_{\sSU(3)}\quad\mathrm{and}\quad
\Gamma^{\rm{gravity}}_{\rm{Spin}(3,1)}\ ,
\end{equation}
which together define a connection on the bundles \eqref{6.28}. All of them are part of the geometry of the total space of bundles \eqref{6.28}
\cite{Popov2, Popov3}. Before the second quantization, the carriers of four known interactions from \eqref{6.29}  do not have the property of particles. Note that the Higgs fields are also part of the geometry of vector bundles \eqref{6.25}-\eqref{6.27}, which was described in detail in \cite{Popov2, Popov3}.

\section{Classical field theory: gauge fields}

All fundamental fermions and antifermions take values in the complex vector spaces $\C^N$, $\bar\C^N$ and their dual. The gauge fields are given by matrices acting on these vector spaces, and charges of matter fields are determined by the type of these spaces. That is why, we recall the description of complex vector spaces \cite{KobNom}.

{\bf\large 7.1. Complex vector fields}

We consider the space $\R^{2N}$ with the Euclidean metric
\begin{equation}\label{7.1}
g_{\R^{2N}}^{}=\delta_{AB}\,\dd x^A\dd x^B\quad\mathrm{for}\quad A,B=1,...,2N\ .
\end{equation}
The tangent bundle to $\R^{2N}$ is $T\R^{2N}\cong\R^{2N}\times\R^{2N}$ and the basis of tangent space is given by vector 
fields $\{\dpar/\dpar x^A\}=\{\dpar/\dpar x^i, \dpar/\dpar x^{i+N}\}$ for $i=1,...,N$. Similarly, $\{\dd x^A\}=\{\dd x^i, \dd x^{i+N}\}$
form a basis for the cotangent bundle $T^*\R^{2N}\cong\R^{2N}\times\R^{2N}$ of one-form. On tangent space we define the matrix $J$ of complex structure
\begin{equation}\label{7.2}
J=\begin{pmatrix}0&-\unit_N\\\unit_N&0\end{pmatrix}\ ,\quad J\left (\frac{\dpar}{\dpar x^{i}}\right)=\frac{\dpar}{\dpar x^{i +N}}\quad\mathrm{and}\quad  J\left (\frac{\dpar}{\dpar x^{i+N}}\right )=-\frac{\dpar}{\dpar x^i }
\end{equation}
so that
\begin{equation}\label{7.3}
\dpar_j:=\frac{\dpar}{\dpar z^j}=\sfrac12\, \left (\frac{\dpar}{\dpar x^{j}}-\im \frac{\dpar}{\dpar x^{j+N}}\right )
\quad\mathrm{and}\quad
\dpar_{\bar\jmath}:=\frac{\dpar}{\dpar \bar z^{\bar\jmath}}=
\sfrac12\, \left (\frac{\dpar}{\dpar x^{j}}+\im \frac{\dpar}{\dpar x^{j+N}}\right )
\end{equation}
form a basis for the bundle $T^{1,0}\R^{2N}\cong\R^{2N}\times\C^N$ and $T^{0,1}\R^{2N}\cong\R^{2N}\times\bar\C^N$, respectively.
The direct sum of these complex vector bundles forms the complexified tangent bundle of $\R^{2N}: T^\C\R^{2N}\cong\R^{2N}\times \C^{2N}=\R^{2N}\times \C^{N}\oplus\R^{2N}\times \bar\C^{N}=:T^{1,0}\R^{2N}\oplus T^{0,1}\R^{2N}$.
Now the metric \eqref{7.1} can be rewritten as
\begin{equation}\label{7.4}
g_{\R^{2N}}^{}=\delta_{i\bar\jmath}\,\dd z^i\dd \bar z^{\bar\jmath} = \delta_{\bar\imath j}\,\dd \bar z^{\bar\imath}\dd z^j\ ,
\end{equation}
where $z^j = x^j + \im x^{j+N}$ and $\bar z^{\bar\jmath}=x^j - \im x^{j+N}$.

For complex vector fields $\psi_\C\in T^\C\R^{2N}=T^{1,0}\R^{2N}\oplus T^{0,1}\R^{2N}$ we have
\begin{equation}\label{7.5}
\psi_\C = \psi^j\dpar_j + \psi^{\bar\jmath}\dpar_{\bar\jmath}\quad\mathrm{with}\quad
J(\dpar_j)=\im\dpar_j\quad\mathrm{and}\quad J(\dpar_{\bar\jmath})=-\im\dpar_{\bar\jmath}\ .
\end{equation}
We can replace vector fields with columns by substitution
\begin{equation}\label{7.6}
\dpar_j\mapsto \frac{1}{\sqrt{2}}\, v_j \quad\mathrm{and}\quad \dpar_{\bar\jmath}\mapsto \frac{1}{\sqrt{2}}\, 
v_{\bar\jmath}=\frac{1}{\sqrt{2}}\,\overline{v_j}\ ,
\end{equation}
where $v_j$ has 1 at the $j$-th place and minus the imaginary unit $-\im$ at the $(j+N)$-th place. Then we have columns
\begin{equation}\label{7.7}
\psi_\C = \psi_+^j v_j + \psi_-^{\bar\jmath}v_{\bar\jmath}\quad\mathrm{for}\quad\psi_+^j=\frac{1}{\sqrt{2}}\,\psi^j
\quad\mathrm{and}\quad \psi_-^{\bar\jmath}=\frac{1}{\sqrt{2}}\psi^{\bar\jmath}\ ,
\end{equation}
\begin{equation}\label{7.8}
J\psi_\pm = \pm\im\psi_\pm\quad\mathrm{for}\quad\psi_+=\psi_+^jv_j
\quad\mathrm{and}\quad \psi_-=\psi_-^{\bar\jmath}v_{\bar\jmath}\ .
\end{equation}
The sesquilinear forms on $\C^N$ and $\bar\C^N$ columns are
\begin{equation}\label{7.9}
\langle \psi_1, \psi_2\rangle ={\bar\psi}_1^{\bar\imath}\delta_{\bar\imath j}\psi_2^j=\psi_1^\+\psi_2
\quad\mathrm{and}\quad
\langle \phi_1, \phi_2\rangle =\phi_1^i\delta_{i\bar\jmath}\phi_2^{\bar\jmath}=\phi_1^\+\phi_2
\end{equation}
for $\psi_{1,2}\in\C^N$ and $\phi_{1,2}\in\bar\C^N$. Note that $v_j$ and $v_{\bar\jmath}$ are $2N$-columns from $\C^{2N}=\C^N\oplus\bar\C^N$ and therefore both $\psi$ and $\phi$ are embedded into $\C^{2N}$. They belong to orthogonal subspaces $\C^N\subset\C^{2N}$ and $\bar\C^N\subset\C^{2N}$ since $v^\+_jv_{\bar k}=0$.

\bigskip
\noindent{\bf\large 7.2. Group $\sU(N)$}

  Consider the group $G=\sU(N)$ in the defining representation of $N{\times}N$ unitary matrices
\begin{equation}\label{7.10}
\sU(N)=\left\{g\in \sGL(N, \C)\mid g^\+g=gg^\+=\unit_N  \right\}\ ,
\end{equation}
where
\begin{equation}\label{7.11}
g=\left(g^i_j\right),\quad g^\+=\left(\bar g^{\bar\jmath}_{\bar\imath}\right)\quad\mbox{and}\quad \unit_N=\left(\delta_{i\bar\jmath}\right)=\left(\delta_{\bar\jmath i}\right)
\end{equation}
for $i, j, \bar\imath, \bar\jmath,... = 1,...,N$. In components, the definition \eqref{7.10} reads
\begin{equation}\label{7.12}
\bar g^{\bar k}_{\bar\imath}\,\delta_{\bar k l}\, g^{ l}_{j}=\delta_{\bar\imath j}\ .
\end{equation}
Matrices $g$ act on vectors $\psi=\{\psi^i\}\in\C^N$ of defining representation and $g\in \sSU(N)$ if $\det g=1$.

 Let $\{I_a\}$ with $a=1,...,N^2-1$ be the generators of SU($N$) chosen as $N{\times}N$ antihermitian matrices $I_a=\{I_{aj}^{\ \,i}\}$ 
with the structure constants $f_{ab}^c$ given by the commutation relations
\begin{equation}\label{7.13}
[I_a,I_b]=f^c_{ab}\,I_c\ .
\end{equation}
They form a basis for the Lie algebra $\frg =\rsu(N)=\,$Lie$\,\sSU(N)$. For SO(2)$\cong$U(1) we have $g=\exp(\theta J)$ and as the generator we choose $I_1=J$ from \eqref{7.2}. On complex vectors $\psi_+$ from \eqref{7.8} we have $J\to\im$.
From \eqref{7.12} we obtain
\begin{equation}\label{7.14}
I^{\+\, i}_{a\, j}:=\bar I^{\ \bar k}_{a\, \bar l}\,\delta^{\bar l i}\delta_{\bar k j}=-I^{\ \, i}_{a\, j}\ ,
\end{equation}
where $I_a=(I^{\ \, i}_{a\, j})$ act on the left on $\C^N$, matrices $\bar I_a=(\bar I^{\ \bar k}_{a\, \bar l})$ act on the left on the complex conjugate space $\bar\C^N$, and matrices $I_a^\+$ defined in \eqref{7.14} act on the right on $(\C^N)^\+ =(\bar\C^N)^{T}$. Note that $\bar I_a$ have the same commutation relations as $I_a$ in \eqref{7.13}.

\bigskip
\noindent{\bf\large 7.3. Vector fields and algebra u($N$)}

Consider the space $\C^N=(\R^{2N}, J)$ with complex coordinates $z^i$, $i=1,...,N$, and their complex conjugate $\bar z^{\bar\imath}$.  We introduce holomorphic vector fields
\begin{equation}\label{7.15}
I_a^v:= - I_{aj}^{\ \,i}\, z^j\,\partial_i \quad\mbox{with}\quad \dpar_i=\frac{\dpar}{\dpar z^i} \ .
\end{equation}
Recall that \eqref{7.15} are sections of the complex vector bundle $T^{1,0}\R^{2N}$ over $(\R^{2N}, J)$. There are also sections of the complex conjugate vector bundle $T^{0,1}\R^{2N}$,
\begin{equation}\label{7.16}
\bar I_a^v=-\bar I_{a\bar\jmath}^{\ \,\bar\imath}\bar z^{\bar\jmath}\dpar_{\bar\imath}\ ,
\end{equation}
where matrices $\bar I_a$ are complex conjugate to $I_a$. Notice that
\begin{equation}\label{7.17}
[I_a^v, I_b^v]=f^c_{ab}\,I_c^v\ ,\quad [\bar I_a^v, \bar I_b^v]=f^c_{ab}\,\bar I_c^v
\quad\mbox{and}\quad  [I_a^v, \bar I_b^v]=0\ , 
\end{equation}
so that $I_a^v$ are the image of the generators $I_a\in\rsu(N)$ under the embedding of the algebra $\rsu(N)$ into sl($N,\C$)-algebra of holomorphic vector fields on $\C^N$, $\bar I_a^v$ are their complex conjugate defined on $\bar\C^N$, and $I_a^v + \bar I_a^v$ are real vector fields on $\R^{2N}$ representing the algebra $\rsu(N)$.

Consider vector fields
\begin{equation}\label{7.18}
\psi_+ =\psi^i\dpar_i\ ,
\end{equation}
where the components $\psi^i$'s do not depend on $z=\{z^i\}\in\C^N$. 
For commutators of $I_a^v$'s with vector fields \eqref{7.18} we have
\begin{equation}\label{7.19}
\delta_a\psi_+=(\delta_a\psi^i)\dpar_i =\CL_a\psi_+=[I_a^v, \psi_+]=  I_{aj}^{\ \,i}\psi^j\dpar_i\ \Rightarrow\   \delta_a\psi^i=I_{aj}^{\ \,i}\psi^j\ ,
\end{equation}
where $\CL_a$ is the Lie derivative along the vector field $I_a^v$.
It follows from \eqref{7.19} that the vectors \eqref{7.18} belong to the defining representation $\psi_+=\{\psi^i\}\in\C^N$ of the group U$(N)$.
Similarly, for $\psi_-=\psi^{\bar\imath}\dpar_{\bar\imath}$ we have 
\begin{equation}\label{7.20}
\delta_a\psi_-= \CL_a\psi_-=[\bar I_a^v, \psi_-]=\bar I_{a\bar\jmath}^{\ \,\bar\imath}\psi^{\bar\jmath}\dpar_{\bar\imath}\ \Rightarrow\   \delta_a\psi^{\bar\imath}=\bar I_{a\bar\jmath}^{\ \,\bar\imath}\psi^{\bar\jmath}
\end{equation}
and analogous formula can be written for 
\begin{equation}\label{7.21}
(\psi^i)^\+=({\bar\psi}^{\bar\jmath}\delta_{\bar\jmath i})=(\psi_i)\in T_{1,0}\R^{2N}\quad\mbox{and}\quad
(\psi^{\bar\imath})^\+=(\bar\psi^{j}\delta_{j\bar\imath })=(\psi_{\bar\imath })\in T_{0,1}\R^{2N}\ .
\end{equation}
Namely, we have
\begin{equation}\nonumber
\delta_a\psi_+^\vee=\CL_a(\psi_i\dd z^i) =- I_{aj}^{\ \,i}\psi_i\dd z^j\ \Rightarrow\   \delta_a\psi_i
=-\psi_jI_{ai}^{\ \,j}\ ,
\end{equation}
\begin{equation}\nonumber
\delta_a\psi_-^\vee=\CL_a(\psi_{\bar\imath}\dd z^{\bar\imath}) =- \bar I_{a\bar\jmath}^{\ \,\bar\imath}\psi_{\bar\imath}\dd z^{\bar\jmath}\ \Rightarrow\   \delta_a\psi_{\bar\imath} =-\psi_{\bar\jmath} \bar I_{a\bar\imath}^{\ \,\bar\jmath}\ ,
\end{equation}
where $I_a$ and $\bar I_a$ in these formulae are related with $I_a^\+$ and $\bar I_a^\+$ by the formula \eqref{7.14}.

\bigskip
\noindent{\bf\large 7.4. Vector bundle $E^+_{\C^N}$}

Let us consider Minkowski space $M=\R^{3,1}$ at each point of which the internal space $\C_x^N$ of a particle is given. The union of all these spaces gives the rank-$N$ complex vector bundle
\begin{equation}\label{7.22}
E^+_{\C^N}=\mathop{\bigsqcup}_{x\in M}\C_x^N=\mathop{\bigcup}_{x\in M}\{(x, \psi_+(x))\mid \psi_+(x)\in \C_x^N\}\ ,
\end{equation}
where $\psi_+=\psi^i(x)\dpar_i$. Topologically, this space is the direct product
\begin{equation}\label{7.23}
E^+_{\C^N}\cong M\times \C^N\ ,
\end{equation}
but this is not so from the differential geometry point of view, since the space \eqref{7.23} carries the metric of the bundle fibred over $M$ and not the metric of the direct product of spaces. The bundle $E^-_{\C^N}=\overline{E^+_{\C^N}}$ with fibres $\bar\C_x^N$ (internal spaces of antiparticles) and its sections $\psi_-=\psi^{\bar\imath}\dpar_{\bar\imath}$ are introduced similarly and all constructions are similar.

\bigskip
\noindent{\bf\large 7.5. Frames and coframes on $E^+_{\C^N}$}

On the space \eqref{7.23} we introduce vector fields
\begin{equation}\label{7.24}
\nabla^+_\mu =\dpar_\mu + \CA^a_\mu I_a^v\ ,\quad \nabla_i^+ =\dpar_i
\end{equation}
and one-forms
\begin{equation}\label{7.25}
\theta_+^\mu =\dd x^\mu \quad\mbox{and}\quad \theta_+^i=\dd z^i + \CA_\mu^a\dd x^\mu I^{\ \,i}_{aj}z^j\ ,
\end{equation}
where $\CA_\mu^a(x)$, $x\in M$, are components of a gauge potential. In mathematical terms, $\CA =\CA_\mu^a\dd x^\mu I_a^v$ is a connection one-form on the bundle $E^+_{\C^N}$ with values in the Lie algebra $\frg =\rsu(N)$ or $\frg =\,$u(1) for $N=1$. Recall that connection $\CA$ on $E^+_{\C^N}$ defines a splitting of the tangent spaces $T_u E^+_{\C^N}$ at $u\in E^+_{\C^N}$ into horisontal $\CH_u$ and vertical $\CV_u$ subspaces, $T_u E^+_{\C^N}=\CH_u\oplus \CV_u$. To see this, we should simply rewrite the exterior derivative $\dd^{}_{E^+_{\C^N}}$ on $E^+_{\C^N}$ as follows:
\begin{equation}\label{7.26}
\dd^{}_{E^+_{\C^N}}=\dd_M + \dd_{\C^N}=\dd x^\mu\dpar_\mu + \dd z^i\dpar_i = 
\theta_+^\mu\nabla_\mu^+ + \theta_+^i\nabla_i^+\ .
\end{equation}
This splitting defines vertical subspaces $\CV_u\cong E_x$ in $E^+_{\C^N}$. Note that, generally speaking, there are no restrictions on the connection $\CA$, as was extensively discussed in \cite{Popov2, Popov3}. In particular, the Yang-Mills equations, masslessness of $\CA$ etc. has nothing to do with geometry. All properties of $\CA$, besides a smoothness class, can only be justified by experimental data.

\bigskip
\noindent{\bf\large 7.6. Metric on $E^+_{\C^N}$}

Vector fields \eqref{7.24} form a frame on the tangent bundle of $E^+_{\C^N}$ and one-forms \eqref{7.25} form a coframe. Hence, on $E^+_{\C^N}$ we can introduce the metric 
\begin{equation}\label{7.27}
g_{E^+_{\C^N}}=g_M+\delta_{i\bar\jmath}\,\theta_+^i\bar\theta_+^{\bar\jmath}\ ,
\end{equation}
where $\theta_+^i$ are given in \eqref{7.25}.  Thus, components $\CA_\mu^a$ of the gauge potential $\CA$ are part of the frame, coframe and the metric on the bundle $E^+_{\C^N}$, i.e. $\CA$ defines geometry. For $E^-_{\C^N}$ everything is similar.

\bigskip

\noindent{\bf\large 7.7. Curvature of connection on $E^+_{\C^N}$}

Calculating commutators of vector fields \eqref{7.24}, we obtain
\begin{equation}\label{7.28}
\CF_{\mu\nu}=[\nabla_\mu^+, \nabla_\nu^+] = \CF_{\mu\nu}^a I_a^v\ ,\quad \CF_{ij}=[\nabla_i^+, \nabla_j^+] =0\ ,
\end{equation}
\begin{equation}\label{7.29}
\CF_{\mu i}=[\nabla_\mu^+, \nabla_i^+] = \CA_{\mu}^a I_{a\,i}^{\ j}\nabla_j^+\ ,
\end{equation}
where 
\begin{equation}\label{7.30}
\CF_{\mu\nu}^a=\dpar_\mu\CA_\nu^a - \dpar_\nu\CA_\mu^a + f^a_{bc}\CA_{\mu}^b\CA_{\nu}^c
\end{equation}
is the field strength of $\CA$ (its curvature). Thus, the two-form on $M$,
\begin{equation}\label{7.31}
\CF = \sfrac12\, \CF_{\mu\nu}^a \dd x^\mu\wedge\dd x^\nu I_a^v
\end{equation}
is the curvature of the connection
\begin{equation}\label{7.32}
\CA = \CA_\mu^a\dd x^\mu  I_a^v
\end{equation}
on the complex vector bundle $E^+_{\C^N}$ over $M$.

Gauge fields \eqref{7.32} interact with sections $\psi_+=\psi^i(x)\dpar_i$ of $E^+_{\C^N}$ via the covariant derivatives \eqref{7.24}, i.e. they act as the force carrier for elementary fermions with values in $E^+_{\C^N}$. To see this, consider a section $\psi_+$ of the bundle $E^+_{\C^N}\to M$. If follows from \eqref{7.24} and \eqref{7.29} that 
\begin{equation}\label{7.33}
[\nabla_\mu^+, \psi_+] = [\dpar_\mu + \CA_\mu , \psi_+]=(\dpar_\mu\psi^i + \CA^a_\mu I_{a\,j}^{\ i}\psi^j)\dpar_i =(\nabla_\mu^+ \psi_+)^i\dpar_i\ .
\end{equation}
From \eqref{7.33} it follows that in \eqref{7.28}-\eqref{7.32} we can substitute $I_a^v\to I_a=(I_{a\,j}^{\ i})$ and consider $\psi_+$ as a column $\psi_+=\{\psi^i(x)\}$ on which matrices $I_a$  act from the left. The same formulae can be used for $\psi_-\in\bar\C^N$, 
$\psi_+^\+\in (\C^N)^\+$ and $\psi_-^\+\in (\bar\C^N)^\+$.

\bigskip
\noindent{\bf\large 7.8. Automorphisms of $E^+_{\C^N}$}

On each fibre $E_x\cong {\C^N}$ of the bundle $E^+_{\C^N}\to M$ the group $G_x$ ($\cong\sSU(N)$ or $\sU(1)$) acts by rotation of the basis in $E_x$, $x\in M$. There is a one-to-one correspondence between the group $G_x$ and all oriented bases, or {\it frames}, on the fibres $E_x$. Thus, $G$-frames on $E^+_{\C^N}$ are parametrized by the infinite-dimensional group
\begin{equation}\label{7.34}
\CG = C^\infty (M, G)
\end{equation}
of smooth $G$-valued functions on $M=\R^{3,1}$. This is the group of base-preserving automorphisms, Aut$_G E^+_{\C^N}$, of the bundle $E^+_{\C^N}$ and its Lie algebra is Lie$\,\CG = C^\infty (M, \frg )$. 

If we denote by $\Abb$ the space of all smooth connections on $E^+_{\C^N}$ then the group \eqref{7.34} acts on $\CA\in\Abb$ by the standard formula
\begin{equation}\label{7.35}
\CA\mapsto\CA^g = g^{-1}\CA g +g^{-1}\dd g
\end{equation}
for $g\in \CG$ and $\dd = \dd x^\mu\dpar_\mu$. Whether $\CG$ (or its subgroup) is the group of gauge transformations (i.e. one can claim that $\CA$ and $\CA^g$ in \eqref{7.35} are equivalent) or a dynamical (i.e. $\CA$ and $\CA^g$ in \eqref{7.35} are not equivalent) depends on the choice of equations of motion for $\CA$ and boundary conditions. For more detailed discussion see e.g. \cite{Popov3, Popov4} and references  therein.

\section{Classical field theory: fermions $+$ gauge fields}

{\bf\large 8.1. Free fermions and antifermions}

We will now describe interaction of spin $s=\sfrac12$ particles and their antiparticles with gauge fields introduced in the previous section. But first we will complete the description of solutions to the Dirac equation \eqref{6.13}, begun in \eqref{6.15}-\eqref{6.19}. These solutions were written for zero momenta for greater clarity. The general solution of the Dirac equation \eqref{6.13} for {\it free} (not interacting with anything) fermions can be written  as 
\begin{equation}\label{8.1}
\Psi (x)=\int \frac{\dd^3p}{\sqrt{2E_p}}\, \bigl (a_s(p)u_s(p)e^{-\im p_\mu x^\mu} v_+ + b_s(p)v_s(p)e^{\im p_\mu x^\mu}v_-\bigr )= \Psi_+(x)v_+ + \Psi_-(x)v_-
\end{equation}
where $a_s(p)$ and $b_s(p)$ are arbitrary function of 3-momenta $p_a$ and
\begin{equation}\label{8.2}
u_s(p)=\begin{pmatrix}\xi^+_s\\[4pt] -\frac{\sigma_a p_a}{(E_p+m)}\xi^+_s\end{pmatrix}\ ,\quad
v_s(p)=\begin{pmatrix} \frac{\sigma_a p_a}{(E_p+m)}\xi^-_s\\[4pt]\xi^-_s\end{pmatrix}\ ,\quad
E_p=\omega_p=\sqrt{\delta^{ab}p_a p_b+m^2}\ ,
\end{equation}
\begin{equation}\label{8.3}
\xi_s^\pm = \sqrt{E_p+m}\,\chi_s^\pm ,\
\chi_1^+=\begin{pmatrix}1\\0\end{pmatrix} ,\
\chi_2^+=\begin{pmatrix}0\\1\end{pmatrix},\
\chi_1^-=\chi_2^+=\begin{pmatrix}0\\1\end{pmatrix},\
\chi_2^-=-\chi_1^+=\begin{pmatrix}-1\\0\end{pmatrix} ,
\end{equation}
\begin{equation}\label{8.4}
v_+=\frac{1}{\sqrt 2}\begin{pmatrix}1\\-\im\end{pmatrix} ,\
v_-=\frac{1}{\sqrt 2}\begin{pmatrix}1\\\im\end{pmatrix}\quad\mbox{and}\quad
Q v_\pm=(-\im J)v_\pm =\pm v_\pm\ .
\end{equation}
Recall that $v_+$ and $v_-$ are basis vectors in the bundles $L_\C^+$ and $L_\C^-$ and formula \eqref{8.1} says that $\Psi_+$ and $\Psi_-$ belong to different spaces.

Note that for fermions, the metric \eqref{4.8} on fibres of the bundle $L_{\C^2}=L_\C^+\oplus L_\C^-$ must be included in the definition of the scalar product (summation by additional index),
\begin{equation}\label{8.5}
\overline{\Psi}^q:=\Psi^\+\gamma^0\otimes Q\quad\Rightarrow\quad\overline{\Psi}^q\Psi = \Psi_+^\+\gamma^0\Psi_+ - \Psi_-^\+\gamma^0\Psi_-=
\overline{\Psi}_+\Psi_+ - \overline{\Psi}_-\Psi_-\ ,
\end{equation}
and this leads to the fact that the energy of both positive frequency and negative frequency states is positive. In addition, after second quantization with annihilation operators of fermions $\ah_s$ and antifermions $\bh_s$, for the Hamiltonian operator we obtain the expression
\begin{equation}\label{8.6}
\hat H =\int\dd^3p\, E_p\, (\ah_s^\+\ah_s + \bh_s^\+\bh_s)\ ,
\end{equation}
without substraction of infinite vacuum energy. For charge density $\rho$ in RQM we have
\begin{equation}\label{8.7}
\rho = j^0 = \overline{\Psi}^q\gamma^0\Psi  =  \Psi_+^\+\Psi_+ - \Psi_-^\+\Psi_-\ .
\end{equation}
Correspondingly we have
\begin{equation}\label{8.8}
q=\int\dd^3x\, \rho = \left\{\begin{array}{ccl}+1&\rm{for}&\Psi_-=0\\-1&\rm{for}&\Psi_+=0\\0&\rm{for}&\Psi_-=\Psi_+^*, \ \Psi_-=\Psi_+\ \rm{or}\ \Psi_-=-\Psi_+\end{array}\right .
\end{equation}
After second quantization, for the charge operator we obtain
\begin{equation}\label{8.9}
\hat q =\int\dd^3x\, {\overline\Psi}^q\gamma^0\Psi = \int\dd^3p\,(\ah_s^\+\ah_s - \bh_s^\+\bh_s)\ .
\end{equation}
We emphasize that formulae \eqref{8.7}-\eqref{8.9} define the {\it quantum charge} associated with the bundles $L_\C^\pm$ and their structure group U(1)$_\hbar$, which defines frames on their fibres. For particles with a non-zero electric charge, we will assume that its sign is always opposite to the sign of the quantum charge. Then multiplying the expressions \eqref{8.7}-\eqref{8.9} by $e^-$ we will also get the expressions for electric charge.

\bigskip
\noindent{\bf\large 8.2. Interaction with gauge fields}

In the previous section we described in detail the geometry of the bundle $E^+_{\C^N}$ associated with the group $\sU(1)$ or $\sSU(N)$ for $N>1$. Recall that if $E\to M$ is a complex vector bundle, then the conjugate bundle $\bar E$ is obtained by having complex numbers acting through their complex conjugate. If $E$ has a Hermitian metric, then the complex conjugate bundle $\bar E$ is isomorphic to the dual bundle $E^\vee =\mbox{Hom}(E, \CO)$ through the metric, where $\CO$ is the trivial complex line bundle. Hence, similar to the four bundles \eqref{3.26}-\eqref{3.29} discussed in Section 3, for non-Abelian internal degrees of freedom we also have four bundles,
\begin{equation}\label{8.10}
\begin{array}{lcccl}
\psi_+=\psi^i\dpar_i\in &E^+_{\C^N}&\longleftrightarrow&(E^+_{\C^N})^\vee&\ni \psi_+^\vee =\psi_i\dd z^i\\[5pt]
&\downarrow\uparrow&&\downarrow\uparrow&\\[5pt]
\psi_-^\vee=\psi_{\bar\imath}\dd \bar z^{\bar\imath}\in &(E^-_{\C^N})^\vee&\longleftrightarrow&E^-_{\C^N}&\ni \psi_- =\psi^{\bar\imath}\dpar_{\bar\imath}
\end{array}
\end{equation}
where the horisontal arrows correspond to the transition to dual bundles, and the vertical arrows correspond to maps defining isomorphisms of the bundles. All charges of bundles $E^+_{\C^N}$ and $(E^-_{\C^N})^\vee$ are equal and opposite to the charges of bundles $E^-_{\C^N}$
and $(E^+_{\C^N})^\vee$. Due to isomorphism of bundles defined by the Hermitian metric $\delta_{i\bar\jmath}$ on fibres (see \eqref{7.27}), we have an equivalence (transposition operator) $\psi^i\sim\psi_{\bar\imath}=\delta_{{\bar\imath} j}\psi^j$ and $\psi_i\sim\psi^{\bar\imath}=\delta^{{\bar\imath} j}\psi_j$. A change in the signs of charges to the opposite occurs while complex conjugation, $\psi^i\to\psi^{\bar\imath}$ and 
$\psi_i\to\psi_{\bar\imath}$, and not while transposition.

When passing to the matrix basis \eqref{7.6}-\eqref{7.9} in \eqref{8.10}, we obtain
\begin{equation}\label{8.11}
E^+_{\C^N}:\ {\rm{columns}}\ \psi^iv_i\ ,\quad E^-_{\C^N}:\ {\rm{columns}}\ \psi^{\bar\imath} v_{\bar\imath}\ ,
\end{equation}
\begin{equation}\label{8.12}
(E^+_{\C^N})^\vee :\ {\rm{rows}}\ \psi_iv^{\+ i}\quad{\rm{and}}\quad (E^-_{\C^N})^\vee :\ {\rm{rows}}\ \psi_{\bar\imath} v^{\+\bar\imath}\ ,
\end{equation}
where $v^{\+ i}=(v_i)^\+$ and $v^{\+ \bar\imath}=(v_{\bar\imath})^\+$. In this case, the generators of the group SU($N$) have the form 
\begin{equation}\label{8.13}
I_a=I_{aj}^{\ \,i}v_iv^{\+j} + \bar I_{a\bar\jmath}^{\ \,\bar\imath}v_{\bar\imath}v^{\+\bar\jmath}\ ,
\end{equation}
and for sections $\phi$ of the bundle $E^+_{\C^N}\oplus E^-_{\C^N}$ we have
\begin{equation}\label{8.14}
\phi=\phi^iv_i + \phi^{\bar\imath}v_{\bar\imath}\ ,
\end{equation}
where $\phi^i$ and $\phi^{\bar\imath}$ are independent, and we can set any of them equal to zero. The covariant derivatives act on $\phi$ as follows:
\begin{equation}\label{8.15}
\begin{array}{ll}
[\dpar_\mu + \CA_\mu^a I_a, \phi ]&=\bigl(\dpar_\mu\phi^i + \CA_\mu^aI_{aj}^{\ \,i}\phi^j\bigr)v_i + 
\bigl(\dpar_\mu\phi^{\bar\imath}+ \CA_\mu^a\bar I_{a{\bar\jmath}}^{\ \,{\bar\imath}}\phi^{\bar\jmath}\bigr)v_{\bar\imath}\\[5pt]
&=: \bigl(\nabla_\mu^+\phi\bigr)^iv_i +\bigl(\nabla_\mu^-\phi\bigr)^{\bar\imath}v_{\bar\imath}\ ,
\end{array}
\end{equation}
where
\begin{equation}\label{8.16}
\nabla_\mu^+ = \dpar_\mu + \CA_\mu^aI_a\quad\mbox{and}\quad \nabla_\mu^- = \dpar_\mu + \CA_\mu^a\bar I_a\ .
\end{equation}
Due to isomorphism of bundles mentioned above, we have
\begin{equation}\label{8.17}
(\nabla_\mu^-\phi )^{\bar\imath}v_{\bar\imath}=\bigl(\dpar_\mu\phi_i - \phi_j I_{ai}^{\ \,j}\CA_\mu^a\bigr)v^{\+ i} = 
\bigl(\nabla_\mu^-\phi\bigr)_iv^{\+ i}\ ,
\end{equation}
which again confirms the equivalence of considering $\{\phi^{\bar\imath}\}$ and $\{\phi_{i}\}$.

Now we can introduce the tensor product of the fermionic bundle \eqref{6.18} with the bundle $E^+_{\C^N}$ or $E^-_{\C^N}$ and use the covariant derivatives \eqref{8.16}. We should also generalize the derivatives \eqref{8.16} to the tensor product \eqref{6.28} of bundles with different $N$, and also introduce bundles with fractional electric and quantum charges.

\bigskip
\noindent{\bf\large 8.3. Tensor products of bundles}

Having vector bundles $\CW^\pm$, $E^\pm_{\C^{N_1}}$ and $E^\pm_{\C^{N_2}}$, we can consider different combinations \eqref{6.28} of their tensor products, and we should specify covariant derivatives on them. For example, consider the bundle $E^+_{\C^{N_1}}\otimes E^+_{\C^{N_2}}$. The covariant derivatives in it have the form
\begin{equation}\label{8.18}
\nabla_\mu =\dpar_\mu + \CA_\mu^{a_1} I_{a_1}\otimes \unit_{N_2} + \unit_{N_1}\otimes \CA_\mu^{a_2} I_{a_2}\ ,
\end{equation}
where $I_a$'s are given by formula \eqref{8.13}. Sections $\phi_{++}$ of $E^+_{\C^{N_1}}\otimes E^+_{\C^{N_2}}$ have two indices,
\begin{equation}\label{8.19}
\phi_{++}=\phi^{i_1i_2}\,v_{i_1}\otimes v_{i_2}=:\phi^{i_1i_2}v_{i_1i_2}\ ,
\end{equation}
where $i_1=1,...,N_1$ and $i_2=1,...,N_2$. From \eqref{8.19} we see that only the first term of \eqref{8.13} will act on $\phi_{++}$.

To better understand the geometry of the bundle  $E^+_{\C^{N_1}}\otimes E^+_{\C^{N_2}}$, we return to the original geometric description of complex vector bundles in Section 7. On the bundle  $E^+_{\C^{N_1}}\otimes E^+_{\C^{N_2}}$ the basis in fibres is given by the field $\dpar_{i_1}\otimes\dpar_{i_2}$, so that section \eqref{8.19} is
\begin{equation}\label{8.20}
\phi_{++}=\phi^{i_1i_2}\,\dpar_{i_1}\otimes\dpar_{i_2}\ .
\end{equation}
On $\C^{N_1}$ and $\C^{N_2}$ we have coordinates $z^{i_1}$ and $z^{i_2}$, respectively. On their tensor product $\C^{N_1}\otimes\C^{N_2}$ we can introduce coordinates $z^{i_1 i_2}$ and write the frame and coframe  on $E^+_{\C^{N_1}}\otimes E^+_{\C^{N_2}}$ as follows:
\begin{equation}\label{8.21}
\nabla_\mu^{++}=\dpar_\mu - \CA_\mu^{a_1}\,I_{a_1j_1}^{\ \ l_1}\,z^{j_1l_2}\,\dpar^{}_{l_1l_2} -  \CA_\mu^{a_2}\,I_{a_2j_2}^{\ \ l_2}\,z^{l_1j_2}\,\dpar^{}_{l_1l_2}\ ,
\end{equation}
\hspace{1mm}
\begin{equation}\label{8.22}
\nabla^{++}_{i_1i_2} = \dpar_{i_1i_2}=\frac{\dpar}{\dpar z^{i_1i_2}} \ ,
\end{equation}
\hspace{2mm}
\begin{equation}\label{8.23}
\theta^\mu = \dd x^\mu,\ \ \theta_{++}^{i_1i_2}=\dd z^{i_1i_2} + \CA_\mu^{a_1}\,\dd x^\mu\,I_{a_1j_1}^{\ \ i_1}\,z^{j_1i_2}+\CA_\mu^{a_2}\,\dd x^\mu\,I_{a_2j_2}^{\ \ i_2}\,z^{i_1j_2}\ .
\end{equation}
Vector fields  \eqref{8.21}  and \eqref{8.22} act on the vector field $\phi_{++}$ in \eqref{8.20} with $\dpar_{i_1i_2}$ similar to 
\eqref{7.33}, so that the covariant derivative \eqref{8.21} can be rewritten in the matrix form \eqref{8.18}. If instead of the bundle  
$E^+_{\C^{N_1}}\otimes E^+_{\C^{N_2}}$ we want to consider e.g. the bundle  $E^+_{\C^{N_1}}\otimes E^-_{\C^{N_2}}$, then in all formulae we just need to change index $i_2$ to $\bar i_2$, etc.

\bigskip
\noindent{\bf\large 8.4. Fractional charges}

We will consider charged leptons with electric charge $e^-=-e<0$ as spinors from $\CW^+=W^+\otimes L^+_\C$ (particles) taking values in the bundle $E_\C^-$ (this is a matter of convention) and their antiparticles with charge $e^+=e>0$ as sections of the bundle $W^-\otimes L_\C^-\otimes E_\C^+$, i.e. $\Psi_{e^-}\in W^+\otimes L_\C^+\otimes E_\C^-$ and 
$\Psi_{e^+}\in W^-\otimes L_\C^-\otimes E_\C^+$. It is also generally accepted that quarks and antiquarks have fractional electric charge equal to $\pm\sfrac13\,e$ or $\pm\sfrac23\,e$. Similarly, their quantum charge can be $q=\mp\sfrac13$ or $q=\mp\sfrac23$ (with sign opposite to electric charge). We consider below description of bundles with fractional electric charge, for bundles with quantum charges everything is derived similarly.

Note that the generator of the group U(1)$_{\rm{em}}$ acting on sections $\phi_+=\phi^z\dpar_z$ and $\phi_-=\phi^{\bar z}\dpar_{\bar z}$
of bundles $E^+_\C$  and $E^-_\C$ has the form
\begin{equation}\label{8.24}
I^v =-\im (z\dpar_z - {\bar z}\dpar_{\bar z})\ .
\end{equation}
For the charge operator $Q_e^v =-\im\, e\, I^v$ we have
\begin{equation}\label{8.25}
[Q_e^v , \phi_\pm ]=\pm e \phi_\pm\ .
\end{equation}
To get charge $+\sfrac13\,e$ we should consider fibres as 3-sheeted coverings $\C_{1/3}\cong\C\times\Z_3$ of the complex plane $\C$, where $\Z_3$ is the cyclic group generated by $\zeta$ which is the 3rd root of unity, $\zeta^3=1$. If we denote by $z$ the local coordinate on $\C_{1/3}$, then in terms of it the operators \eqref{8.24} and \eqref{8.25} will take the form
\begin{equation}\label{8.26}
I^v=-\frac{\im}{3}\, (z\dpar_z - \bar z\dpar_{\bar z}) \quad\mbox{and}\quad Q_e^v = - \frac{e}{3}\, (z\dpar_z - \bar z\dpar_{\bar z}) \ ,
\end{equation}
where $z=z^{1/3}_{\rm{old}}$ on $\C_{1/3}$ and $z_{\rm{old}}$ on $\C\cong\C_{1/3}/\Z_3$.

We denote the complex line bundles with fibres $\C_{1/3}$ and its complex conjugate by $E^{1/3}_\C$ and $E^{-1/3}_\C$. On their sections $\phi_+=\phi^z\dpar_z$ and $\phi_-=\phi^{\zb}\dpar_\zb$ we have
\begin{equation}\label{8.27}
[Q_e^v , \phi_\pm ]=\pm \frac{e}{3} \phi_\pm\ .
\end{equation}
Fields $\phi_\pm$ with charges $\pm\sfrac23 e$ are obtained by passing to the tensor products of these bundles,
\begin{equation}\label{8.28}
E_\C^{2/3}=E_\C^{1/3}\otimes E_\C^{1/3}: \ Q_e^v =-\frac{2e}{3}\, (z\dpar_z - \zb\dpar_\zb), \ [Q_e^v, \phi_+]=\frac{2e}{3}\, \phi_+\ ,
\end{equation}
\begin{equation}\label{8.29}
E_\C^{-2/3}=E_\C^{-1/3}\otimes E_\C^{-1/3}: \ Q_e^v =-\frac{2e}{3}\, (z\dpar_z - \zb\dpar_\zb), \ [Q_e^v, \phi_-]=-\frac{2e}{3}\, \phi_-\ ,
\end{equation}
where $z$ and $\zb$ are now coordinates on fibres of $E_\C^{2/3}$ and $E_\C^{-2/3}$,  $z=z^{2/3}_{\rm{old}}$, $\phi_+=\phi^z\dpar_z$,  
$\phi_-=\phi^\zb\dpar_\zb$.

\bigskip
\noindent{\bf\large 8.5. Quarks}

For quantum bundles $L^{\pm 1/3}_\C$ and $L^{\pm 2/3}_\C$, the entire discussion is identical, because these are also complex line bundles, associated with the group U(1)$_\hbar$ instead of U(1)$_{\rm{em}}$. The quantum charges of sections of these bundles are $q=\pm\sfrac13$ and $q=\pm\sfrac23$. Currently the existence of six quarks (six flavors) is known, all of them are Dirac spinors with values in the bundle
$E_{\C^{3}}^+$ (color) and six antiquarks with values in $E_{\C^{3}}^-$. Their electric charges are as follows
\begin{equation}\label{8.30}
\begin{array}{cccccll}
\frac{2}{3}\,e\ :&u&c&t&\sim&q=-\frac{2}{3}&\uparrow\\
&&&&&&\sSU(2)_{\rm{weak}}\\
-\sfrac13\,e :&d&s&b&\sim&q=\sfrac13&\downarrow
\end{array}
\end{equation}
In this table, quarks are devided into three families defined by three columns, and the group $\sSU(2)_{\rm{weak}}$ of weak interactions maps quarks from the top row to the bottom row and vice versa. We assume that the quantum charge of quarks has a sign opposite to the sign of their electric charge, which is already reflected in table \eqref{8.30}. Since $q\ge 0$ corresponds to ``particles", and $q\le 0$ corresponds to ``antiparticles", it follows from this assumption that the total quantum charge of the universe is equal to zero if the total electric charge is equal to zero. Note that the use of the prefix ``anti-" is misleading and it is mathematically more correct to simply talk about a quantum charge $q$, similar to an electric charge.

From table \eqref{8.30} it follows that quarks are sections of the bundles
\begin{equation}\label{8.31}
d, s, b\ \in\ W^+\otimes L^{1/3}_\C\otimes E^{- 1/3}_\C\otimes E_{\C^{3}}^+
\end{equation}
\begin{equation}\label{8.32}
u, c, t\ \in\ W^-\otimes L^{-2/3}_\C\otimes E^{2/3}_\C\otimes E_{\C^{3}}^+
\end{equation}
The Dirac wave functions for them will have the form
\begin{equation}\label{8.33}
\Psi = \Psi_+^{y\zb i_3} v_+\otimes v_-\otimes v_{i_3}+
\Psi_-^{\yb z \bar\imath_3} v_-\otimes v_+ \otimes v_{\bar\imath_3}\ ,
\end{equation}
where the first term in \eqref{8.33} corresponds to quarks $d, s, b$, and the second term in \eqref{8.33} corresponds to $\bar d, \bar s, \bar b$ antiquarks. Similar formulae can be written for $u, c, t$ quarks and leptons. In the proposed scheme, the proton $p^+=(u u d)$ has $q=-1$, and the electron $e^-$ has $q=+1$, but the proton, consisting of three quarks, is not ``antiparticle" for the electron and forms a stable bound state 1s with it in {\bf{H}}. In conclusion of this subsection, we note  that for an example of the action of the group $\sSU(2)_{\rm{weak}}$ in \eqref{8.30}, we can consider the mapping
\begin{equation}\label{8.34}
\begin{pmatrix}u\\d\end{pmatrix}\ \longrightarrow\ \begin{pmatrix}0&\W^+\\\W^-&0\end{pmatrix}\begin{pmatrix}u\\d\end{pmatrix}=
\begin{pmatrix}\W^+ d\\ \W^-u\end{pmatrix}\ \longleftrightarrow\ \begin{array}{ccl}u&\to&d+\W^+\\d&\to&u+\W^-\end{array}\ ,
\end{equation}
where $\W^\pm$ are gauge bosons of the group SU(2)$^{}_{\rm{weak}}$. Note that for the charges of $\W^\pm$ we have $q=\mp 1$, and for neutrinos $q=0$. Gluons, photons and $Z^0$-bosons also have $q=0$ with $\psi_-=\psi_+^*$ for gauge bosons.

\section{Virtual fermions in two dimensions}

{\bf\large 9.1. Preliminary remarks}

We have described fundamental fermions as sections of the tensor products of complex vector bundles of type \eqref{6.28}, \eqref{8.31} and \eqref{8.32} with covariant derivatives on tensor products of type \eqref{8.18} included in the Dirac equation on the four-manifold $M$. All four known interactions, carried by the fields
\begin{equation}\label{9.1}
A^{\rm em}_{\sU(1)}\ ,\quad A^{\rm weak}_{\sSU(2)}\ ,\quad A^{\rm strong}_{\sSU(3)}\quad\mbox{and}\quad\Gamma^{\rm gravity}_{\Spin(3,1)} \ ,
\end{equation}
can enter into this equation. At the same time, there are no terms of interaction with the quantum field $A_\hbar\in\,$u(1)$_\hbar$ in the Dirac equation, since $A_\hbar$ does not have components along the coordinate space $M=\R^{3,1}$. To see the effect of these vacuum fields $A_\hbar$ on fermions it is necessary to extend the Dirac equation from $\R^{3,1}$ to phase space $T^*\R^{3,1}$ with the condition that spinors depend only on space-time coordinates $x^\mu\in\R^{3,1}$.

Dirac spinors on the space $T^*\R^{3,1}$ belong to the representation space $\C^{16}$ of the Clifford algebra 
\begin{equation}\label{9.2} 
\rCl^\C (8) = \rCl(6,2)\otimes\C\cong \rm{Mat}(16, \C)\ .
\end{equation}
When describing solutions to the corresponding Dirac equation with all gauge fields except the field $A_\hbar$ turned off, the complicated form of various matrix differential operators makes it difficult to compare solutions with the standard case. We will identify solutions of equations for fermions $\Psi$, interacting with fields $A_\hbar$, with virtual particles. We emphasize that they have nothing to do with those fictious virtual particles that are used in Feynman diagrams. The difference of these fermions $\Psi$ from free particles is only that they interact with the vacuum field $A_\hbar$, and they are no less real than free particles. Rather, on the contrary, free (bare) particles are a mathematical abstraction, as Bogoliubov and Shirkov insist \cite{BogShir}. 

For greater clarity, we will first consider solutions to the Dirac equations on space-time $\R^{1,1}$ and on the phase space $T^*\R^{1,1}$, and only then return to the case $T^*\R^{3,1}$. Recall that we consider spinors $\Psi$ as sections of the bundle $L_\C^+\oplus L_\C^-$ over $T^*\R^{1,1}$, i.e.
\begin{equation}\label{9.3} 
\Psi(x,t)=\psi_+v_+ + \psi_-v_-\ ,
\end{equation}
where $v_+$ is the basis in the quantum bundle $L_\C^+$, and $v_-$ is the basis in $L_\C^-$.

\bigskip

\noindent{\bf\large 9.2. Plane-wave in $\R^{1,1}$}

Consider  two-dimensional space-time $\R^{1,1}$ with the metric $\eta =(\eta_{AB})=\diag (1,-1)$, $A, B=0,1$. The Clifford 
algebra $\rCl(1,1)$ of this space has a real Majorana representation by matrices Mat(2,$\R$) and a complex Dirac representation of the algebra $\rCl^\C (2) = \rCl(1,1)\otimes \C$ by matrices Mat(2,$\C$). As generators of this algebra we take the matrices
\begin{equation}\label{9.4} 
\gamma^0 =\sigma_3\quad\mbox{and}\quad \gamma^1 =-\im\sigma_1\ .
\end{equation}
The Dirac equation for a free fermion of mass $m$ has the form
\begin{equation}\label{9.5}
(\im\gamma^A\dpar_A - m)\Psi = (\im\sigma_3\dpar_t +\sigma_1\dpar_1 - m)\Psi =0\ ,
\end{equation}
where $\dpar_A=\dpar/\dpar x^A , x^0 =t, x^1=x$. Positive frequency solution of equation \eqref{9.5} is
\begin{equation}\label{9.6}
\psi_+=\frac{1}{\sqrt{2\omega_p}}e^{-\im\omega_pt + \im px}u(p)\ ,\quad u(p)=\begin{pmatrix}\sqrt{\omega_p+m}\\[2pt]-\im\sqrt{\omega_p-m}\end{pmatrix}\ ,\quad \omega_p=\sqrt{p^2+m^2}\ .
\end{equation}
Negative frequency soution has the form
\begin{equation}\label{9.7}
\psi_-=\frac{1}{\sqrt{2\omega_p}}e^{\im\omega_pt - \im px}v(p)\ ,\quad v(p)=\begin{pmatrix}\im\sqrt{\omega_p-m}\\[2pt]
\sqrt{\omega_p+m}\end{pmatrix}\ ,\quad \omega_p=\sqrt{p^2+m^2}\ .
\end{equation}
Hence, we have the plane-wave solution
\begin{equation}\label{9.8}
\Psi = a_p\psi_+v_+ + b_p\psi_-v_-\quad\mbox{with}\quad\overline{\Psi}^q=\Psi^\+\gamma^0\otimes Q=\overline{\Psi}Q=a_p^\+\bar\psi_+v_+^\+ - b_p^\+\bar\psi_-v_-^\+
\end{equation}
for which
\begin{equation}\label{9.9}
\overline{\Psi}^q\Psi =\frac{m}{\omega_p}(a_p^\+ a_p + b_p^\+ b_p)>0\ .
\end{equation}
Here $a_p:=a(p)$ and $b_p:=b(p)$ are arbitrary complex-valued functions of momentum $p$ which become annihilation operators after the second quantization. From the solution  \eqref{9.8}, a general form of a wave packet solution is 
\begin{equation}\label{9.10}
\begin{split}
\Psi (x,t) &= \frac{1}{2\pi}\mathop{\int}_{-\infty}^{\infty}\dd p\, (a_p\psi_+v_++b_p\psi_-v_-)\\
&=\frac{1}{2\pi}\mathop{\int}_{-\infty}^{\infty}\dd p\frac{1}{\sqrt{2\omega_p}}\left (a_pu(p)e^{-\im\omega_pt+\im px}v_+ +
b_pv(p)e^{\im\omega_pt-\im px}v_-\right )\ .
\end{split}
\end{equation}
From \eqref{9.6}-\eqref{9.10} it is easy to deduce that energy of all solutions is positive.

For the quantum charge density for \eqref{9.8} we get
\begin{equation}\label{9.11}
\rho =\overline{\Psi}^q\gamma^0\Psi = a_p^\+a_p-b_p^\+b_p\ ,
\end{equation}
where $\rho\ge 0$  for ``particles" $a_p\psi_+$ and $\rho\le 0$ for ``antiparticles" $b_p\psi_-$. Note that
\begin{equation}\label{9.12}
\Psi = a_p\psi_+ v_+ + b_p\psi_-v_- = \frac{1}{\sqrt 2}
\begin{pmatrix}a_p\psi_+ + b_p\psi_-\\ -\im (a_p\psi_+ - b_p\psi_-)\end{pmatrix}\ ,
\end{equation}
and a return to usual discussion of solutions will occur if we discard the lower component in \eqref{9.12} of the $\C^2$-valued spinor and consider only upper component in \eqref{9.12}. In this case negative energies arise and the charge density becomes positive definite, which requires introduction of Grassmann-valued functions $a_p$ and $b_p$ and the replacement $b_p\mapsto b_p^\+$ in \eqref{9.8}. When using solutions \eqref{9.6}-\eqref{9.12} for the second quantization, neither replacing $b_p$ with $b_p^\+$ nor subtracting infinities is required.

\bigskip

\noindent{\bf\large 9.3. Charge conjugation}

As we already discussed, the charge conjugated spinor $\Psi_c$ is introduced through the complex conjugation of the Dirac equation and multiplying it on the left by a certain matrix to reduce it to its original form. If gauge fields are turned on, then the form of the covariant 
derivative will change according to the change in charges.  In the case under consideration we have
\begin{equation}\label{9.13}
\Psi_c=\sigma_1\Psi^*=\sigma_1\psi^*_-v_+ + \sigma_1\psi^*_+v_-\ ,
\end{equation}
which implies that 
\begin{equation}\label{9.14}
(\psi_+)_c =\sigma_1\psi^*_-\quad\mbox{and}\quad(\psi_-)_c =\sigma_1\psi^*_+\ ,
\end{equation}
which is consisted with \eqref{9.6} and \eqref{9.7}.

\bigskip

\noindent{\bf\large 9.4. Dirac equation on $\R^{2,1}\subset T^*\R^{1,1}$}

Now we will consider a generalization of the Dirac equation from space $\R^{1,1}$ to phase space $T^*\R^{1,1}=\R^{1,1}\times \R^{1,1}=\R^{2,2}$ with the metric 
\begin{equation}\label{9.15}
g_{\R^{2,2}}=- (\dd x^0)^2 + (\dd x^1)^2 +w_1^4\dd p_1^2 - w_0^4\dd p_0^2\ ,
\end{equation}
where $w_0$ and $w_1$ are length parameters (cf. \eqref{3.5}). As symplectic structure we consider
\begin{equation}\label{9.16}
\omega_{\R^{2,2}}=\dd x^0\wedge \dd p_0 + \dd x^1\wedge \dd p_1\ .
\end{equation}
First we will consider solutions to the Dirac equation on space $\R^{2,1}\subset\R^{2,2}$ with coordinates $x^0, x^1$ and $p_1$, and then we will describe solutions on space $\R^{2,2}$ to compare and see the differences.

Note that the real Clifford algebras in three dimensions have Majorana representations
\begin{equation}\label{9.17}
\rCl(1,2)\cong\mbox{Mat}(2,\C)\quad\mbox{and}\quad \rCl(2, 1)\cong\mbox{Mat}(2,\R)\oplus\mbox{Mat}(2,\R)\ ,
\end{equation}
and over the field $\C$ we have a reducible Dirac representation
\begin{equation}\label{9.18}
\rCl^\C(3)\cong \mbox{Mat}(2,\C)\oplus\mbox{Mat}(2,\C)\ .
\end{equation}
We will take the basis matrices in $\rCl(1,2)$ as
\begin{equation}\label{9.19}
\gamma^0=\sigma_3\ ,\ \ \gamma^1=-\im\sigma_1\ ,\ \ \gamma^2=-\im\sigma_2\ \ \Rightarrow\ \ \gamma^1\gamma^2=-\im\sigma_3\ ,
\end{equation}
From \eqref{9.19} we see that $\gamma^0$ and $\gamma^1\gamma^2$ are independent over $\R$  but equivalent over the field $\C$. Therefore, we should introduce $4\times 4$ matrices
\begin{equation}\label{9.20}
\Gamma^0{=}\begin{pmatrix}\gamma^0&0\\ 0&-\gamma^0\end{pmatrix},\ 
\Gamma^1{=}\begin{pmatrix}\gamma^1&0\\ 0&-\gamma^1\end{pmatrix},\ 
\Gamma_2=\begin{pmatrix}\gamma^2&0\\ 0&-\gamma^2\end{pmatrix}\ \Rightarrow\
\Gamma^0\Gamma^1\Gamma_2{=}\im\Gamma_3{=}\im\begin{pmatrix}-\unit_2&0\\ 0&\unit_2\end{pmatrix},
\end{equation}
and we see that $\Gamma^1\Gamma_2$ and $\Gamma^0$ are not equivalent.

We use the notation
\begin{equation}\label{9.21}
\begin{split}
x_2=w_1^2p_1\quad\Rightarrow\quad \frac{\dpar}{\dpar x_2}=w_1^{-2}\frac{\dpar}{\dpar p_1}=: w_1^{-2}\dpar^1\ ,\\
A^2\dd x_2 = w_1^2A^2\dd p_1 = A^1\dd p_1\ \Rightarrow A^2=w_1^{-2}A^1\ ,\\
\dpar_0=\frac{\dpar}{\dpar x^0}=\dpar_t,\ \dpar_1=\frac{\dpar}{\dpar x^1},\ \dpar^2=\frac{\dpar}{\dpar x_2}=w_1^{-2}\frac{\dpar}{\dpar p_1}= w_1^{-2}\dpar^1\ ,\\
\nabla_0=\dpar_0, \ \nabla_1=\dpar_1,\ \nabla^2=(\dpar^2+A^2J)=w_1^{-2}(\dpar^1+A^1J)=w_1^{-2}\nabla^1,\\
A^1=-x^1, \ J=\im\,(v_+v_+^\+ - v_-v_-^\+)=\begin{pmatrix}0&-1\\1&0\end{pmatrix}\ .
\end{split}
\end{equation}
With this notation the Dirac equation in $\R^{2,1}$ has the form
\begin{equation}\label{9.22}
(\im\Gamma^0\nabla_0 + \im\Gamma^1\nabla_1+\im\Gamma_2\nabla^2 - m)\Psi =0\ .
\end{equation}
Recall that $J$ is the generator of the structure group U(1)$_\hbar$ of the bundle $L_\C^+\oplus L^-_\C$ over $T^*\R^{1,1}$ and $A_\hbar$ in \eqref{9.21} is a connection on this bundle. The field $\Psi (x,t)$ in \eqref{9.22} is a section of the bundle 
$W^+\otimes L^+_\C\oplus W^-\otimes L^-_\C$,
\begin{equation}\label{9.23}
\Psi =\Psi_+v_+ + \Psi_-v_-=\begin{pmatrix}\Psi_+^L\\\Psi_+^R\end{pmatrix}\otimes v_+ + 
\begin{pmatrix}\Psi_-^L\\\Psi_-^R\end{pmatrix}\otimes v_-
\end{equation}
and $\Psi^L$, $\Psi^R$ are the eigenvectors of the matrix $\Gamma_3$ from \eqref{9.20}.

For $\Psi^L_\pm$, $\Psi^R_\pm$, equations \eqref{9.22} are split into four equations
\begin{equation}\label{9.24}
(\im\sigma_3\dpar_t + \sigma_1\dpar_1 + w_1^{-2}\sigma_2\nabla^1_\pm - m)\Psi_\pm^L =0\ ,
\end{equation}
\begin{equation}\label{9.25}
(\im\sigma_3\dpar_t + \sigma_1\dpar_1 + w_1^{-2}\sigma_2\nabla^1_\pm + m)\Psi_\pm^R =0\ ,
\end{equation}
where $\nabla_\pm^1 = \dpar^1 \mp\im x^1$ and $\nabla_\pm^1 \Psi= \mp\im x^1\Psi$ since $\dpar^1\Psi =0$.

\bigskip

\noindent{\bf\large 9.5. Solutions on $\R^{2,1}$}

Equations \eqref{9.24} can be rewritten as
\begin{equation}\label{9.26}
\begin{pmatrix}\omega - m&-\frac{\sqrt 2}{w_1}\,a_1^\+\\ \frac{\sqrt 2}{w_1}\,a_1&-(\omega +m)\end{pmatrix}
\begin{pmatrix}\psi_+^1\\\psi_+^2\end{pmatrix}=0
\quad\mbox{for}\quad 
\Psi_+^L=e^{-\im\omega t}\begin{pmatrix}\psi_+^1\\\psi_+^2\end{pmatrix}\ ,
\end{equation}
where
\begin{equation}\label{9.27}
a_1:=\frac{w_1}{\sqrt 2}\left (\dpar_1+\frac{x^1}{w_1^2}\right )\ ,\quad 
a_1^\+:=-\frac{w_1}{\sqrt 2}\left (\dpar_1-\frac{x^1}{w_1^2}\right )\ ,\quad [a_1,a_1^\+]=1\ .
\end{equation}
Solutions of the Dirac equations \eqref{9.24} for $\Psi_+^L$ are 
\begin{equation}\label{9.28}
\Psi^L_{+,n}= \frac{e^{-\im\omega_nt}}{\sqrt{2\omega_n}}\begin{pmatrix}\sqrt{\omega_n+m} \,\,|n+1\rangle\\[3pt]
\sqrt{\omega_n-m} \,|n\rangle\end{pmatrix}\quad\Leftrightarrow\quad\psi_+=
\frac{e^{-\im\omega_pt}}{\sqrt{2\omega_p}}\,\begin{pmatrix}\sqrt{\omega_p+m}\,e^{\im px}\\[3pt]
-\sqrt{\omega_p-m}\,\im\, e^{\im px}\end{pmatrix}\ ,
\end{equation}
where
\begin{equation}\label{9.29}
\omega_n=\sqrt{2w_1^{-2}(n+1)+m^2}\quad\mbox{and}\quad\omega_p=\sqrt{p^2+m^2}\ .
\end{equation}
In \eqref{9.28} and \eqref{9.29} we wrote out for comparison the solutions of the Dirac equations for spinors $\Psi^L_{+,n}$ interacting with the field $A_\hbar$ and the solution $\psi_+$ from \eqref{9.6} of the Dirac equation in $\R^{1,1}$ for free noninteracting spinors. We see that instead of momenta $p\in (-\infty , \infty)$, we get discrete numbers $n=0,1,...$ parametrizing oscillator-type solutions
\begin{equation}\label{9.30}
\Psi^L_{+,n}(x, t)= \frac{e^{-\im\omega_nt}}{\sqrt{2\omega_n}}
\begin{pmatrix}\sqrt{\omega_n+m}  \,\langle x|n+1\rangle\\[3pt]
\sqrt{\omega_n-m} \,\langle x|n\rangle\end{pmatrix}=
\frac{e^{-\im\omega_nt}}{\sqrt{2\omega_n}}
\begin{pmatrix}\sqrt{\omega_n+m}\,\,\psi_{n+1}(x)\\[3pt]
\sqrt{\omega_n-m}\,\,\psi_n(x)\end{pmatrix}\ ,
\end{equation}
where
\begin{equation}\label{9.31}
\psi_0(x)=\frac{1}{(\pi w_1^2)^{1/4}}\exp(-\frac{x^2}{2w_1^2})
\end{equation}
and $\psi_n(x)\sim H_n(\frac{x}{w_1})\,\psi_0(x)$, where $H_n$ are Hermitean polynomials.

Equations \eqref{9.24} for $\Psi_-^L$ have the form
\begin{equation}\label{9.32}
\begin{pmatrix}-(\omega+m)&\frac{\sqrt 2}{w_1}a_1\\[3pt]
-\frac{\sqrt 2}{w_1}a_1^\+&(\omega-m)\end{pmatrix}
\begin{pmatrix}\psi_-^1\\[3pt]
\psi_-^2\end{pmatrix}=0
\quad\mbox{for}\quad
\Psi_-^L=e^{\im\omega t}\,\begin{pmatrix}\psi_-^1\\[3pt]
\psi_-^2\end{pmatrix}\ ,
\end{equation}
and their solutions are
\begin{equation}\label{9.33}
\Psi^L_{-,n}= \frac{e^{\im\omega_nt}}{\sqrt{2\omega_n}}\begin{pmatrix}\sqrt{\omega_n-m} \,|n\rangle\\[3pt]
\sqrt{\omega_n+m} \,|n+1\rangle\end{pmatrix}=\left(\Psi_{+,n}^L\right )_c\ ,
\end{equation}
where $\Psi_c$ is defined in \eqref{9.13} and \eqref{9.14}. So, for $\Psi^L$ we obtain the general solution in the form
\begin{equation}\label{9.34}
\Psi^L=\mathop{\sum}_{n=0}^{\infty}\left(a_n\Psi_{+,n}^Lv_+ + b_n\Psi_{-,n}^Lv_-\right)\ ,
\end{equation}
which can be compared with \eqref{9.10}. We see that instead of a continuous set $(a_p, b_p, p\in\R )$ of functions parametrized by $p$, solutions \eqref{9.34} are parametrized by a discrete set $(a_n, b_n, n\in\Nbb )$ of complex numbers and the descrete set of energies \eqref{9.29}.

Note that from the Dirac oscillator equation \eqref{9.24}, which has the form
\begin{equation}\label{9.35}
\left(\im\sigma_3\dpar_t + \sigma_1\dpar_1 \mp \frac{\im x^1}{w_1^2}\,\sigma_2 - m\right)\Psi_\pm^L =0\ ,
\end{equation}
 follows the Klein-Gordon oscillator equation
\begin{equation}\label{9.36}
\left(-\dpar_t^2 + \dpar_x^2 - m^2 - \frac{ x^2}{w_1^4} - \frac{\sigma_3}{w_1^2}\right)\Psi_\pm^L =0\ ,
\end{equation}
with $x\equiv x^1$. Looking at these equations for fermions interacting with the vacuum field $A_\hbar$, we propose
to interpret their solutions  \eqref{9.34} as virtual particles $a_n$ and antiparticles $b_n$. Note that these solutions are localized in space, unlike wave type solutions \eqref{9.10}. Solutions that are also localized in time will be written out bellow. After the second quantization of  \eqref{9.34} we will have $\{a_m, a_n^\+\}=\delta_{mn}$  instead of $\{a_p, a_{p^\prime}^\+\}=\delta(p-p^\prime)$ and similarly for $b_m$ and $b_p$.

For $\Psi_\pm^R$ all calculations are the same and we obtain solutions of the Dirac oscillator equations \eqref{9.25} in the form
\begin{equation}\label{9.37}
\begin{aligned}
\Psi^R_{+,n}&= \frac{e^{-\im\wt\omega_nt}}{\sqrt{2\wt\omega_n\vphantom{T^2_1}}}\begin{pmatrix}\sqrt{\wt\omega_n-\wt m\vphantom{T^2_1}} \,\,|n+1\rangle\\[3pt]
\sqrt{\wt\omega_n+\wt m\vphantom{T^2_1}} \,|n\rangle\end{pmatrix}\ ,\quad\wt\omega_n=\sqrt{2w_1^{-2}(n+1)+\wt m^2\vphantom{T^2_1}}\\
\Psi^R_{-,n}&= \frac{e^{\im\wt\omega_nt}}{\sqrt{2\wt\omega_n\vphantom{T^2_1}}}\begin{pmatrix}\sqrt{\wt\omega_n+\wt m\vphantom{T^2_1}} \,|n\rangle\\[3pt]
\sqrt{\wt\omega_n-\wt m\vphantom{T^2_1}}\, \,|n+1\rangle\end{pmatrix}=\left(\Psi^R_{+,n}\right)_c\ ,
\end{aligned}
\end{equation}
where we replaced $m$ in \eqref{9.25} with $\wt m$ since  \eqref{9.24} and \eqref{9.25} are independent. The general solution of the Dirac oscillator equation \eqref{9.25} is
\begin{equation}\label{9.38}
\Psi^R=\mathop{\sum}_{n=0}^{\infty}\left(c_n\Psi_{+,n}^R v_+ + d_n\Psi_{-,n}^R v_-\right)\ ,
\end{equation}
where $(c_n, d_n, n\in\Nbb)$ are independent of $(a_n, b_n, n\in\Nbb)$. All solutions obtained have positive energy, the quantum charge density is positive for $a_n, c_n$ and negative for $b_n, d_n$.

\bigskip 

\noindent{\bf\large 9.6. Squeezed coherent states}

Note that in addition to solutions \eqref{9.28}-\eqref{9.30} with $n\ge 0$, equations \eqref{9.26} have a solution
\begin{equation}\label{9.39}
\Psi^L_{0,+}=C_0e^{-\im\omega_0t}\begin{pmatrix}\,|0\rangle\\0\end{pmatrix}\quad\mbox{with}\quad
\Psi^L_{0,+}(x_1, t)=C_0e^{-\im\omega_0t}\begin{pmatrix}\psi_0(x_1)\\0\end{pmatrix}
\end{equation}
different from $\Psi^L_{+,0}(x,t)$ in \eqref{9.30}. Here $C_0$ is a constant and $\psi_0(x_1)=\langle x_1|0\rangle$ is given in \eqref{9.31}.
Let us introduce an operator
\begin{equation}\label{9.40}
c_1=\frac{\sqrt 2}{w_1}\,a_1 =\nabla_1+{\im}\nabla_+^2=\dpar_1 + \frac{x^1}{w_1^2}
\end{equation}
which is a combination of covariant derivatives in the bundle $L_\C^+$ annihilating the state \eqref{9.39}. Using an automorphism of the bundle $L_\C^+$ given by the element $g\in\sU(1)_\hbar$, 
\begin{equation}\label{9.41}
g=e^{\im\vph}\quad\mbox{with}\quad \vph = x^1_{(0)}p_1 -p_1^{(0)} x^1\ ,
\end{equation}
we obtain a new connection $A_\hbar^\vph$,
\begin{equation}\label{9.42}
A^\vph_\hbar:\quad A^\vph_{x_1}=\im\dpar_{x_1}\vph = -\im p_1^{(0)}\ ,\quad 
A^\vph_{p_1}=A_{p_1}+\im\dpar_{p_1}\vph = -\im (x^1 - x^1_{(0)})\ ,
\end{equation}
that is not equivalent to the initial one, since the field $A_\hbar$ is massive. In this case, we obtain
\begin{equation}\label{9.43}
c_1=\dpar_1+\frac{x^1}{w_1^2}\quad\mapsto\quad c_1^\vph =\dpar_1-\im p_1^{(0)}+\frac{(x^1-x^1_{(0)})}{w_1^2}
\end{equation}
and the new solution to the Dirac oscillator equation is a squeezed coherent state
\begin{equation}\label{9.44}
\Psi_0^{\rm{squ}}(x^1, t)=C_0\,\exp\left({-\frac{(x^1-x^1_{(0)})^2}{2w_1^2}-\im\omega_0t + \im p_1^{(0)}x^1}\right)\ ,\quad\omega_0=\sqrt{2w_1^{-2}+m^2\vphantom{T^2_1}},
\end{equation}
where $x^1_{(0)}$ is the center of the wave packet, $w_1$ is its width and $p_1^{(0)}$ is the expectation value of its momentum. 
In the limit $w_1^2\to\infty$, \eqref{9.44} becomes a plane-wave solution of free Dirac equation.

Note that \eqref{9.44} can also be obtained by acting on \eqref{9.39} by the operators
\begin{equation}\label{9.45}
D(\al)=e^{\al a_1^\+ - \al^*a_1}_{}\ ,\quad S(\rho)=e^{\sfrac12\rho(a_1^2 - a_1^{\+\,2})}_{},\ \,|\al , \rho\rangle =D(\al )S(\rho) \,|0\rangle ,
\end{equation}
where the parameters $\rho$ and $\al =\al_1+\im\al_2$ can be expressed in terms of parameters $x^1_{(0)}, p_1^{(0)}$ and $w_1$. Here $\,|0\rangle$ is the vacuum state in \eqref{9.39}, $D(\al )$ is the displacement operator and $S(\rho)$ is the squeeze operator. The state \eqref{9.44} saturates the Heisenberg uncertainty relation $\Delta x\Delta p = \frac{\hbar}{2}$.

\newpage

\noindent{\bf\large 9.7. Dirac equation on $T^*\R^{1,1}$}

Let us now consider the phase space $T^*\R^{1,1}=\R^{2,2}$ with metric \eqref{9.15} and symplectic form \eqref{9.16}.  The coordinate time $x^0$ is the time measured by a stationary clock in an inertial frame. The proper time $\tau$ of a particle is the time measured by a clock that moves with it, it can be used as the time parameter. Hence, we can promote $x^0$ to an operator and to see how this will change solutions of the Dirac oscillator equation on $\R^{2,1}$ with $A_\hbar \ne 0$.

We consider the bundle $L_\C^+\oplus L_\C^-$ over $T^*\R^{1,1}$ with covariant derivative
\begin{equation}\label{9.46}
\nabla_0=\dpar_0\ ,\quad \nabla_1=\dpar_1\ ,\quad \nabla^0=\frac{\dpar}{\dpar p_0} - x^0J\ ,\quad 
\nabla^1=\frac{\dpar}{\dpar p_1} - x^1J\ ,
\end{equation}
where
\begin{equation}\label{9.47}
J=\im\,(v_+ v_+^\+ - v_- v_-^\+)\quad\mbox{with}\quad v_\pm = \frac{1}{\sqrt 2}\begin{pmatrix}1\\\mp\im\end{pmatrix}
\end{equation}
is the generator of the group $\sU(1)_\hbar$ acting in the bundle $L_\C^+\oplus L_\C^-$. In \eqref{9.46} we see the components $x^0$ and $x^1$ of the vacuum connection $A_\hbar$ on the quantum bundle $L_\C^+\oplus L_\C^-$. For sections $\psi =\psi_+v_+ + \psi_-v_-$ of this bundle depending only on $x^0$, $x^1$, we have
\begin{equation}\label{9.48}
\nabla^0\psi = -\im x^0\psi_+ v_+ +\im x^0\psi_-v_-\quad\mbox{and}\quad\nabla^1\psi = -\im x^1\psi_+ v_+ +\im x^1\psi_-v_-\ .
\end{equation}
For nonvanishing commutators of covariant derivatives \eqref{9.46} we have
\begin{equation}\label{9.49}
[\nabla_0, \nabla^0]=[\nabla_1, \nabla^1]=-J\ \Rightarrow\ [\nabla_0, \nabla^0]\psi_\pm=[\nabla_1, \nabla^1]\psi_\pm =\mp\im\psi_\pm\ ,
\end{equation}
and we can introduce operators
\begin{equation}\label{9.50}
\ph_0=-\im\dpar_0\ ,\quad\ph_1=-\im\dpar_1\ ,\quad\xh^0_\pm = \pm\im \nabla^0\ ,\quad \xh^1_\pm =\pm\im \nabla^1\ ,
\end{equation}
where the upper sign in \eqref{9.49} and \eqref{9.50} corresponds to $L^+_\C$, and lower sign there corresponds to the action on sections of $L_\C^-$.

As generators of Clifford algebra $\rCl(2,2)\otimes\C$ we choose matrices
\begin{equation}\label{9.51}
\ga^0=\begin{pmatrix}0&\unit_2\\\unit_2&0\end{pmatrix}\ ,\quad 
\ga^1=\begin{pmatrix}0&\sigma_1\\-\sigma_1&0\end{pmatrix}\ ,\quad 
\ga_2=\begin{pmatrix}0&\sigma_2\\-\sigma_2&0\end{pmatrix}\quad \mbox{and}\quad
\ga_3=\begin{pmatrix}0&\im\sigma_3\\-\im\sigma_3&0\end{pmatrix}\ .
\end{equation}
We consider spinors $\Psi$ with values in the bundle $L_\C^+\oplus L_\C^-$,
\begin{equation}\label{9.52}
\Psi =\Psi_+ v_+ + \Psi_- v_-
\end{equation}
and their conjugate $\overline{\Psi}^q:=\Psi^\+\Gamma\otimes Q$, where 
\begin{equation}\label{9.53}
\Gamma :=\im\ga^1\ga_2 = \begin{pmatrix}\sigma_3&0\\0&\sigma_3\end{pmatrix}\ ,\quad
Q=-\im J=v_+v_+^\+ - v_-v_-^\+=-\sigma_2\ ,
\end{equation}
and the scalar product is $\overline{\Psi}^q\Psi$.

The Dirac equation for $\Psi$ on $T^*\R^{1,1}$ has the form
\begin{equation}\label{9.54}
(\im\ga^0\nabla_0+\im\ga^1\nabla_1+\im\ga_2\nabla^2+\im\ga_3\nabla^3 - m)\Psi =0\ ,
\end{equation}
where
\begin{equation}\label{9.55}
\nabla_0{=}\dpar_0,\ \nabla_1{=}\dpar_1,\ \nabla^2{=}w_1^{-2}\nabla^1{=}w_1^{-2}(\dpar^1{+}A^1J)
\ \,\mbox{and}\ \,
\nabla^3{=}w_0^{-2}\nabla^0{=}w_0^{-2}(\dpar^0{+}A^0J)
\end{equation}
with $A^0 = -x^0$ and $A^1 = -x^1$. From \eqref{9.46} we obtain
\begin{equation}\label{9.56}
\nabla^2\Psi_\pm=\mp\frac{\im x^1}{w_1^2}\Psi_\pm\quad\mbox{and}\quad\nabla^3\Psi_\pm=\mp\frac{\im x^0}{w_0^2}\Psi_\pm\ .
\end{equation}

\bigskip

\noindent{\bf\large 9.8. Quantum time and fermions}

In \eqref{9.46} and \eqref{9.50} we introduced the quantum time operator $\xh^0$ acting on the wave functions according the formulas  \eqref{9.48} and \eqref{9.56}. Recall that $\dpar_{p_0}\Psi =\dpar_{p_1}\Psi =0$ and operators \eqref{9.46} are combined into operators of creation and annihilation,
\begin{equation}\label{9.57}
c_0=\frac{\sqrt 2}{w_0}\,a_0=\dpar_0 + \frac{x^0}{w_0^2}\ ,\quad
c_0^\+=\frac{\sqrt 2}{w_0}\,a_0^\+=-\left(\dpar_0 - \frac{x^0}{w_0^2}\right )\ ,
\end{equation}
\begin{equation}\label{9.58}
c_1=\frac{\sqrt 2}{w_1}\,a_1=\dpar_1 + \frac{x^1}{w_1^2}\quad\mbox{and}\quad
c_1^\+=\frac{\sqrt 2}{w_1}\,a_1^\+=-\left(\dpar_1 - \frac{x^1}{w_1^2}\right )\ .
\end{equation}
Using these operators, equation \eqref{9.54} is rewritten as
\begin{equation}\label{9.59}
\begin{pmatrix}c_0&-c_1^\+\\c_1&-c_0^\+\end{pmatrix}\psi_+^2 + \im\,m\,\psi_+^1=0\ ,\quad 
\begin{pmatrix}-c_0^\+&c_1^\+\\-c_1&c_0\end{pmatrix}\psi_+^1 + \im\,m\,\psi_+^2=0\ ,
\end{equation}
\begin{equation}\label{9.60}
\begin{pmatrix}-c_0^\+&c_1\\-c_1^\+&c_0\end{pmatrix}\psi_-^2 + \im\,m\,\psi_-^1=0\ ,\quad 
\begin{pmatrix}c_0&-c_1\\c_1^\+&-c_0^\+\end{pmatrix}\psi_-^1 + \im\,m\,\psi_-^2=0\ .
\end{equation}
Here we used the substutution 
\begin{equation}\label{9.61}
\Psi_+=\begin{pmatrix}\psi_+^1\\\psi_+^2\end{pmatrix}\in\C^4\quad\mbox{and}\quad
\Psi_-=\begin{pmatrix}\psi_-^1\\\psi_-^2\end{pmatrix}\in\C^4
\end{equation}
in \eqref{9.52} and \eqref{9.54}.

As solutions of equations \eqref{9.59}, we get
\begin{equation}\label{9.62}
\psi_+^1{=}\begin{pmatrix}\al_+ |n_0,n_1+1\rangle\\
\al_- |n_0+1,n_1\rangle \end{pmatrix},\quad
\psi_+^2{=}\frac{\im}{m}\begin{pmatrix}(-\al^{}_+\omega_{n_0}+\al^{}_-\omega_{n_1})\,|n_0+1, n_1+1\rangle\\
(\al^{}_-\omega_{n_0}-\al^{}_+\omega_{n_1})\,|n_0, n_1\rangle \end{pmatrix} ,
\end{equation}
where
\begin{equation}\label{9.63}
 \al_\pm =\sqrt{\omega_{n_0}\pm m},\ \omega_{n_0}=\sqrt{\omega_{n_1}^2+m^2\vphantom{T^2_1}}=\sqrt{2w_0^{-2}(n_0+1)\vphantom{T^2_1}},\ 
\omega_{n_1}^2=2w_1^{-2}(n_1+1)\ .
\end{equation}
Note that here we can change the operator $c_1$ as in \eqref{9.43} and similarly
\begin{equation}\label{9.64}
c_0=\dpar_0+\frac{x^0}{w_0^2}\quad\mapsto\quad
c_0^\vph =\dpar_0-\im p_0^{(0)}+\frac{(x^0-x^0_{(0)})}{w_0^2}\ ,
\end{equation}
and write the ground state as
\begin{equation}\label{9.65}
\psi_{0,0}(x^0, x^1)=C_0(w_0) C_1(w_1)e^{-\frac{(x^0-x^0_{(0)})^2}{2w_0^2}-\frac{(x^1-x^1_{(0)})^2}{2w_1^2}+\im p_0^{(0)}x^0+\im p_1^{(0)}x^1}\sim\langle x\,|n_0=0, n_1=0\rangle\ ,
\end{equation}
where $ C_0(w_0) C_1(w_1)$ is a normalization constant. We see that obtained solutions are localized both in space and time.

Depending on the choice of normalization coefficients in \eqref{9.65} we have various limiting cases:
\begin{equation}\label{9.66}
C_1(w_1)=1:\quad \mathop{\lim}_{w_1\to\infty}e^{-\frac{(x^1-x^1_{(0)})^2}{2w_1^2}+\im p_1^{(0)}x^1}= e^{\im p_1^{(0)}x^1},
\end{equation}
\begin{equation}\label{9.67}
C_1(w_1)=\frac{1}{\sqrt{2\pi w_1^2}}:\quad \mathop{\lim}_{w_1\to 0}\frac{1}{\sqrt{2\pi w_1^2}}\,e^{-\frac{(x^1-x^1_{(0)})^2}{2w_1^2}+\im p_1^{(0)}x^1}= \delta (x^1-x^1_{(0)})\,e^{\im p_1^{(0)}x^1}\ .
\end{equation}
Similarly we have
\begin{equation}\label{9.68}
C_0(w_0)e^{-\frac{(x^0-x^0_{(0)})^2}{2w_0^2}+\im p_0^{(0)}x^0}\ \longrightarrow\ \left\{
\begin{array}{ll}e^{\im p_0^{(0)}x^0} &\mbox{for}\ w_0\to\infty\\ \delta (x^0-x^0_{(0)})\,e^{\im p_0^{(0)}x^0}&\mbox{for}\ w_0\to\ 0\end{array}\right.
\end{equation}
i.e. we can have localization either in energy or in time.

Solutions of equations \eqref{9.60} have the form
\begin{equation}\label{9.69}
\psi_-^1 = \begin{pmatrix}\al_- \,|n_0+1, n_1\rangle\\\al_+\,|n_0, n_1+1\rangle\end{pmatrix}\ ,
\psi_-^2 = \frac{\im}{m}\begin{pmatrix}(\al_-\omega_{n_0} - \al_+\omega_{n_1}) \,|n_0, n_1\rangle\\(-\al_+\omega_{n_0} + \al_-\omega_{n_1})\,|n_0+1, n_1+1\rangle\end{pmatrix}
\end{equation}
with the same formulae \eqref{9.63}. It is easy to verify that $\Psi_-$ written out in \eqref{9.69} is charge conjugate to $\Psi_+$ from \eqref{9.62},
\begin{equation}\label{9.70}
\Psi_-= \begin{pmatrix}\psi_-^1\\\psi_-^2\end{pmatrix}=C\Psi_+^*=-\ga^0\ga^1\Psi_+^*=
\begin{pmatrix}\sigma_1&0\\0&-\sigma_1\end{pmatrix}\begin{pmatrix}(\psi_+^1)^*\\(\psi_+^2)^*\end{pmatrix}\ ,
\end{equation}
where ``$*$" means complex conjugation.

\section{ Virtual fermions in four dimensions}

\noindent{\bf\large{10.1. Gamma matrices}}

Having completed the consideration of the two-dimensional case with a comparison of fermions (``coupled" or ``virtual") interacting with the vacuum field $A_\hbar$  and fermions (``free" or ``bare") not interacting with any field, we move on to the four-dimensional case. Let us consider the space $\R^{6,1}\subset T^*\R^{3,1}=\R^{6,2}$ with the metric 
\begin{equation}\label{10.1}
g^{}_{\R^{6,1}} = -(\dd x^0)^2 + g^{}_{\R^{6}}\ ,
\end{equation}
where $g^{}_{\R^{6}}$ is the metric on the phase space $T^*\R^{3}$ given in \eqref{3.5}. Minkowski space $\R^{3,1}$ with the metric
\begin{equation}\label{10.2}
g^{}_{\R^{3,1}}=\eta_{\mu\nu}\dd x^\mu \dd x^\nu\ ,\quad \eta =(\eta_{\mu\nu})=\diag(-1, 1, 1, 1)
\end{equation}
is a subspace in $\R^{6,1}$. Clifford algebras for spaces $\R^{3,1}$ and $\R^{1,3}$ have matrix representation 
\begin{equation}\label{10.3}
\rCl(3,1)\cong \mbox{Mat} (4, \R)\quad\mbox{and}\quad \rCl(1,3)\cong \mbox{Mat} (2, \Hbb)\ ,
\end{equation}
where $\Hbb$ is the associative algebra of quaternions. Hence, spinors in the spaces $\R^{3,1}$ and $\R^{1,3}$ will be columns $\R^4$ and $\Hbb^2$, respectively, and these will be Majorana spinors. One can identify $\Hbb$ with $\C^2$ and then Mat(2,$\Hbb$) will be embedded into Mat(4,$\C$) as a subalgebra defined by some reality conditions. Instead, Dirac considered the complexified Cliford algebra,
\begin{equation}\label{10.4}
\rCl^\C(4):= \rCl(3,1)\otimes\C = \rCl(1,3)\otimes\C \cong \mbox{Mat}(4, \C)
\end{equation}
for which spinors $\Psi$ are $\C^4$-valued.

We choose generators of the algebra $\rCl^\C(4)$ as matrices
\begin{equation}\label{10.5}
\ga^0=\im\begin{pmatrix}\unit_2&0\\0&-\unit_2\end{pmatrix},\quad \ga^a=\begin{pmatrix}0&\im\sigma_a\\-\im\sigma_a&0\end{pmatrix},\quad \im\ga^0\ga^1\ga^2\ga^3 =\begin{pmatrix}0&\unit_2\\\unit_2&0\end{pmatrix}=:\ga^5\ ,
\end{equation}
where $\sigma_a$ are Pauli matrices. Generators \eqref{10.5} satisfy the anticommutation relations
\begin{equation}\label{10.6}
\{\ga^\mu, \ga^\nu\}=\ga^\mu\ga^\nu + \ga^\nu\ga^\mu =2\eta^{\mu\nu}\unit_4\quad\mbox{and}\quad \{\ga^\mu, \ga^5\}=0\ .
\end{equation}
For $\R^{6,1}$ we have
\begin{equation}\label{10.7}
\rCl^\C(7)\cong \mbox{Mat}(8, \C)\oplus\mbox{Mat}(8, \C)\quad\mbox{and}\quad \rCl^\C(8)\cong \mbox{Mat}(16, \C)\ ,
\end{equation}
analogously to the cases $\R^{2,1}$ and $\R^{2,2}$. We choose generators of $\rCl^\C(7)$ in the form 
\begin{equation}\label{10.8}
\wt\Gamma^0=\begin{pmatrix}\Gamma^0&0\\0&-\Gamma^0\end{pmatrix}\quad\mbox{and}\quad 
\wt\Gamma^M=\begin{pmatrix}\Gamma^M&0\\0&-\Gamma^M\end{pmatrix},\ M=1,...,6,
\end{equation}
where $\Gamma^M\in\mbox{Mat}(8, \C)$ are generators of the algebra $\rCl^\C(6)$, $\{\Gamma^M,\Gamma^M\}=2\delta^{MN}\unit_8$, and we will choose them in the form
\begin{equation}\label{10.9}
\begin{split}
\Gamma^0=\ga^0\otimes\sigma_3\ ,\quad&\Gamma^1=\ga^1\otimes\sigma_3\ ,\quad\Gamma^2=\ga^3\otimes\sigma_3\ ,\quad\Gamma^4=\unit_4\otimes\sigma_2\\
&\Gamma^3=\ga^2\otimes\sigma_3\ ,\quad\Gamma^5=\ga^5\otimes\sigma_3\ ,\quad\Gamma^6=\unit_4\otimes\sigma_1\ .
\end{split}
\end{equation}
We have ${\overline{\wt\Psi}}:=\wt\Psi^\+\hat\Gamma^0$ for $\wt\Psi\in\C^8\oplus\C^8$, $\hat\Gamma^0:=-\im\wt\Gamma^0$.

\bigskip

\noindent{\bf\large{10.2. Dirac equation on $\R^{6,1}$}}

Over the space $\R^{6,1}$ there is given the quantum bundle $L_\C^+\oplus L_\C^-$ described in detail in the previous sections, and we consider spinors $\wt\Psi$ having an additional index of the two-dimensional space of fibres of the bundle $L_\C^+\oplus L_\C^-$,
\begin{equation}\label{10.10}
\wt\Psi = \begin{pmatrix}\Psi\\\Phi\end{pmatrix}=\begin{pmatrix}\Psi_+v_++\Psi_-v_-\\\Phi_+v_++\Phi_-v_-\end{pmatrix}\in
\begin{pmatrix}\C^{16}\\\C^{16}\end{pmatrix}\ .
\end{equation}
 The bundle $L_\C^+\oplus L_\C^-$ is endowed with the metric, quantum connection $A_\hbar$ and covariant derivatives $\wt\nabla_M$ described in previous sections.
The Dirac equation for spinors $\wt\Psi$ from \eqref{10.10} has the form 
\begin{equation}\label{10.11}
(\wt\Gamma^0\dpar_0 + \wt\Gamma^M\wt\nabla_M - \wt m)\wt\Psi =0\quad\mbox{for}\quad \wt m=\diag (m\unit_{16},\ m^\prime\unit_{16})\ .
\end{equation}
Here the covariant derivatives are of the form
\begin{equation}\label{10.12}
\wt\nabla_a=\nabla_a=\dpar_a\ ,\quad \tilde\nabla_{a+3}=w^{-2}\delta_{ab}\nabla^{b+3}=w^{-2}\delta_{ab}(\dpar^b-x^bJ)\ ,\quad
\dpar^b=\frac{\dpar}{\dpar p_b}\ ,
\end{equation}
so for commutators we have
\begin{equation}\label{10.13}
[\nabla_a, \nabla^{b+3}_+]=-\im\delta_a^b\ \mbox{on}\ L_\C^+\quad\mbox{and}\quad
[\nabla_a, \nabla^{b+3}_-]=\im\delta_a^b\ \mbox{on}\ L_\C^-\ ,
\end{equation}
as shown in  \eqref{3.22}. Note that the equations for $\Psi$ and $\Phi$ in \eqref{10.10} are independent, so we will consider only the equation for $\Psi$, for $\Phi$ everything is the same. From \eqref{10.10}-\eqref{10.12} we obtain equations for $\Psi_\pm$,
\begin{equation}\label{10.14}
(\Gamma^0\dpar_0 + \Gamma^M\wt\nabla_M^\pm - m)\Psi_\pm =0\ ,
\end{equation}
from which there follow the Klein-Gordon oscillator equations
\begin{equation}\label{10.15}
\left(-\dpar_0^2 + \delta^{ab}\dpar_a\dpar_b - m^2 - \frac{1}{w^4}\,\delta_{ab}x^ax^b + \frac{\im}{w^2}\,\delta_{ab}\, [\Gamma^a, \Gamma^{b+3}]\right)\Psi_\pm =0
\end{equation}
for components of spinors $\Psi_\pm\in\C^8$.

\bigskip

\noindent{\bf\large{10.3. Ladder operators}}

For greater generality, we replace $w^2\to (w_1^2, w_2^2, w_3^2)$ (anisotropic case) and introduce the operators
\begin{equation}\label{10.16}
\begin{split}
c_1=\dpar_1 + \frac{x^1}{w_1^2}=:\frac{\sqrt 2}{w_1}\,a_1\ ,\quad&c_1^\+=-(\dpar_1 - \frac{x^1}{w_1^2})=:\frac{\sqrt 2}{w_1}\,a_1^\+\ ,\\
c_2=\dpar_2 + \frac{x^2}{w_2^2}=:\frac{\sqrt 2}{w_2}\,a_2\ ,\quad&c_2^\+=-(\dpar_2 - \frac{x^2}{w_2^2})=:\frac{\sqrt 2}{w_2}\,a_2^\+\ ,\\
c_3=\dpar_3 + \frac{x^3}{w_3^2}=:\frac{\sqrt 2}{w_3}\,a_3\ ,\quad&c_3^\+=-(\dpar_3 - \frac{x^3}{w_3^2})
=:\frac{\sqrt 2}{w_3}\,a_3^\+\ .
\end{split}
\end{equation}
Note that 
\begin{equation}\label{10.17}
[c_a, c_b^\+]=\frac{2}{w_a^2}\,\delta_{ab}\quad\mbox{and}\quad [a_b, a_c^\+]=\delta_{bc}
\end{equation}
and when $w_a^2\to\infty$ the first commutators will go to zero, but the second ones will not.

After some calculations we obtain 
\begin{equation}\label{10.18}
\Gamma_+:=\Gamma^M\wt\nabla_M^+ = -\im\begin{pmatrix}
0&0&c_2^\+&c_1^\+&c_3&0&0&0\\
0&0&-c_1&c_2&0&c_3&0&0\\
c_2&-c_1^\+&0&0&0&0&c_3&0\\
c_1&c_2^\+&0&0&0&0&0&c_3\\
c_3^\+&0&0&0&0&0&-c_2^\+&-c_1^\+\\
0&c_3^\+&0&0&0&0&c_1&-c_2\\
0&0&c_3^\+&0&-c_2&c_1^\+&0&0\\
0&0&0&c_3^\+&-c_1&-c_2^\+&0&0
\end{pmatrix}
\end{equation}
Now we set $\Psi_+=e^{-\im\omega t}\psi$ and obtain the equation
\begin{equation}\label{10.19}
(\im\omega \Gamma^0 + m -\Gamma_+)\psi =0\ ,
\end{equation}
where $\Gamma_+$ is given in \eqref{10.18}.

\newpage

\noindent{\bf\large{10.4. Solutions}}

Solutions of equation \eqref{10.19} are
\begin{equation}\label{10.20}
\begin{pmatrix}\psi_1\\\psi_2\\\psi_7\\\psi_8\end{pmatrix}{=}
\begin{pmatrix}\al_1|n_1+1, n_2+1, n_3\rangle\\\al_2|n_1, n_2, n_3\rangle\\ \al_3|n_1+1, n_2, n_3+1\rangle\\ \al_4|n_1, n_2+1, n_3+1\rangle \end{pmatrix},\
\begin{pmatrix}\psi_3\\\psi_4\\\psi_5\\\psi_6\end{pmatrix}{=}\frac{-\im}{(\omega_{n_1n_2n_3}+m)}
\begin{pmatrix}c_2\psi_1-c^\+_1\psi_2+c_3\psi_7\\ c_1\psi_1+c^\+_2\psi_2+c_3\psi_8\\
c_3^\+\psi_1-c^\+_2\psi_7-c_1^\+\psi_8\\
c_3^\+\psi_2+c_1\psi_7-c_2\psi_8\\\end{pmatrix}\ ,
\end{equation}
where $\al_1, \al_2, \al_3$ and $\al_4$ are arbitrary constants and
\begin{equation}\label{10.21}
\omega_{n_1n_2n_3}=\left(\frac{2(n_1+1)}{w_1^2} + \frac{2(n_2+1)}{w_2^2} +\frac{2(n_3+1)}{w_3^2} + m^2\right)^\sfrac12\ .
\end{equation}
We do not write out the explicit form of the components $\psi_3, \psi_4, \psi_5, \psi_6$; they are easily calculated using \eqref{10.16}
and \eqref{10.20}. 

It is easy to show that the charge conjugation operator for equation \eqref{10.19} is given by the matrix
\begin{equation}\label{10.22}
C=\Gamma^4\Gamma^5\Gamma^6 = \begin{pmatrix}\im\sigma_2&0\\0&-\im\sigma_2\end{pmatrix}\otimes\sigma_1=
\begin{pmatrix}
0&0&\im\sigma_2&0\\
0&0&0&-\im\sigma_2\\
\im\sigma_2&0&0&0\\
0&-\im\sigma_2&0&0
\end{pmatrix}
\end{equation}
and we have the solutions of the Dirac oscillator equation \eqref{10.11} for $\Psi\in \C^{16}$ in the form
\begin{equation}\label{10.23}
\Psi_{(n)}=\Psi_+^{(n)}v_+ + \Psi_-^{(n)}v_-\quad\mbox{with}\quad
\Psi_-^{(n)}=(\Psi_+^{(n)})_c= C(\Psi_+^{(n)})^*\ ,
\end{equation}
where $\Psi_+^{(n)}=e^{-\im\omega_{(n)}t}\psi_{(n)}$ is given in \eqref{10.20}, for $(n)=(n_1 n_2 n_3)$.

From the viewpoint of Minkowski space, equations \eqref{10.14} with solutions \eqref{10.23} describe two particles $(q=1)$ and two antiparticles 
$(q=-1)$. The equations for $\Phi$ from \eqref{10.10} are solved in exactly the same way, and also describe two particles and two antiparticles. The energy of all solutions is positive and $\overline{\wt\Psi}^q\wt\Psi = \wt\Psi^\+\hat\Gamma^0\otimes Q\wt\Psi$ for $\wt\Psi$ from \eqref{10.10}. Recall that in the Klein-Gordon oscillator equation \eqref{10.15} one can interpret $x^a$ as a distance $x^a_1-x^a_2$ between a particle and an antiparticle  whose center of mass moves with a constant velocity. Then solutions \eqref{10.20}-\eqref{10.23} are bound states of the systems of virtual particles and antiparticles. If we lift the Dirac equation with $A_\hbar$ from $\R^{6,1}\subset T^*\R^{3,1}$ to $T^*\R^{3,1}$, then the results will be similar to those discussed in \eqref{9.57}-\eqref{9.70} for $T^*\R^{1,1}$, the solutions will be bound states localized not only in space, but also in time. We will not write out their explicit form. 

To summarize our discussion of fundamental spin $s=\sfrac12$ fields, it is reasonable to assume that quarks, antiquarks and gluons exist only in bound states of virtual particles and they are held in these bound states inside hadrons by vacuum quantum fields $(A_\hbar , F_\hbar )$  described in detail in this paper. This is confirmed by experimental data which indicate that hadrons contain an infinite number of virtual gluons and quark-antiquark pairs, and cloud of these virtual particles form constituent quarks with measurable masses $m$ which are much larger than those  of valence quarks. At the same time, leptons and photons, unlike quarks and gluons, can exist both in an interacting with $A_\hbar$ state (particles) and in a free state (waves). This is the meaning of wave-particle duality.

\section{~Virtual gauge bosons}

\noindent{\bf\large 11.1. {Internal degrees of freedom}}

The main idea of this paper is that particles have internal degrees of freedom described by fibres of vector bundles over the phase space of classical particle. The first quantization itself is the introduction of the bundle $L_\C^+$ for particles (quantum charge $q=1$) and the bundle $L_\C^-$ for antiparticles (quantum charge $q=-1$). These bundles have U$(1)_\hbar$ as structure group and vacuum gauge fields ($A_\hbar , F_\hbar )$ as connection and curvature. 
We call them vacuum gauge fields because the quantum charge density $\rho$ is not the source of these fixed background fields. In fact, the modulus of this density is the probability of finding  a particle (for $\rho_+$) or an antiparticle (for $\rho_-$) at point $x$. Note that the prefix ``anti-" in this paper is considered as conditional; it implies the charge $q=-1$ (leptons) or $q=-\sfrac23$ and $q=-\sfrac13$ (quarks), but does not implies annihilation. Also, I believe that proton has $q=-1$, so the total quantum charge of universe is zero, as is the total electric charge.

Quantum degrees of freedom $L_\C^\pm$ introduce interaction with the quantum gauge fields $(A_\hbar , F_\hbar )$. The spin degrees of freedom (Pauli spinors $\to$ Dirac bispinors) are related to the bundle $W$ of $\C^4$-spinors, the group Spin(3,1) and the gravitational interaction carried out by the Levi-Civita connection $\Gamma^{}_{\Spin(3,1)}$, which is not considered in this paper. In relativistic quantum mechanics, we always have both particles and antiparticles as solutions of the Dirac equation, therefore spinors $\Psi$ have an additional index that parametrizes $\C^2$-fibres of the bundle $L_\C^{+\,q}\oplus L_\C^{-\,q}$ ($q=1$ for leptons and $q=\sfrac13$ or $\sfrac23$ for quarks) plus additional indices of the vector bundles $E_\C^{\pm q}$ (electromagnetic forces), $E_{\C^2}^{\pm }$ (weak forces) and $E_{\C^3}^{\pm }$ (strong forces). In total, we have five types of gauge fields that are defined by connections and curvatures in these bundles. All these connections, except the vacuum connection $A_\hbar$, are dynamical and generated by sections $\Psi$ (matter fields) of these vector bundles. The relationship between the matter fields $\Psi$ and the geometry of the bundles specified by connections $A^{}_{\sSU(N)}$ is determined by the equations of motion imposed on them. All these connections can be considered as part of the definition of quantum particles $=$  bundles over the phase space of classical particles with sections $\Psi$ and connections $\CA$.

\bigskip

\noindent{\bf\large 11.2.  Klein-Gordon equation for free $\CA$}

Until now, connections $\CA$ were classical fields, part of the geometry of vector bundles. The next step is to assume that $A_{\sU(1)}^{\rm{em}}$, $A_{\sSU(2)}^{}$ and $A_{\sSU(3)}^{}$ (as well as $\Gamma_{\Spin(3,1)}^{}$) also have indices  that parametrizes the fibres of the bundle  $L_\C^{+}\oplus L_\C^{-}$, i.e. have properties of quantum particles. This assumption follows from the decomposition of these fields  into positive and negative frequency parts, which are then declared to be operators of annihilation and creation in QFT.

We will consider only Maxwell field $\CA_\mu = \im A_\mu\in\ru(1)=\im\R$, since for the subsector of free non-Abelian fields everything is similar. Maxwell's equations without sources in the Lorenz gauge $\dpar_\mu A^\mu =0$ reduce to the massless Klein-Gordon equation 
\begin{equation}\label{11.1}
\eta^{\mu\nu}\dpar_\mu\dpar_\nu A_\sigma =0
\end{equation}
for each component $A_\sigma$ of the gauge potential. That is why we will omit the index $\sigma$ in $A_\sigma$ and consider an arbitrary component of this potential, which we denote as $A$.

The solution to the equation \eqref{11.1} is a superposition of positive frequency $A_+$ and negative frequency $A_-$ plane-wave solutions, written as the sum
\begin{equation}\label{11.2}
A=A_+ + A_-\ .
\end{equation}
The reality condition requires $A_-=A_+^*$, where $*$ denotes complex conjugation. Using solutions $A_\pm$, we introduce the matrix
\begin{equation}\label{11.3}
A\,\unit_2\ \to\tilde A=A_+P_+ + A_-P_- = \sfrac12\, \begin{pmatrix}A_++A_-&\im (A_+-A_-)\\-\im (A_+-A_-)&A_++A_-\end{pmatrix}\ ,
\end{equation}
each element of which is a real solution to the KG equation \eqref{11.1}. This matrix takes values in the endomorphisms End($L_{\C^2}^{}$) of the bundle $L_{\C^2}^{}=L_\C^+\oplus L_\C^-$, where
\begin{equation}\nonumber
P_+=v_+\otimes v_+^\+ = \sfrac12\,\begin{pmatrix}1&\im\\-\im&1\end{pmatrix}\quad\mbox{and}\quad
P_-=v_-\otimes v_-^\+ =\sfrac12\,\begin{pmatrix}1&-\im\\\im&1\end{pmatrix}
\end{equation}
are projectors
\begin{equation}\nonumber
P_\pm :\quad L_{\C^2}\ \to\ L_{\C}^\pm\ , \quad P_\pm^2=P_\pm\ , \quad P_\pm P_\mp=0\ , \quad P_++P_-=\unit_2\ ,
\end{equation}
from the bundle $L_{\C^2}$ onto the subbundles $L_{\C}^\pm$. If we consider $A_+=A_-=A$ to be simply real solution of equation \eqref{11.1}, then \eqref{11.3} reduces to the standard case $\tilde A= A \unit_2$ of gauge theories.

\bigskip

\noindent{\bf\large 11.3.  Klein-Gordon oscillator equation for virtual $\CA$}

For free field, $\tilde A$ contains the same information as $A$ in \eqref{11.2}, but the structure group 
SO(2)$_\hbar\cong\sU(1)_\hbar$ of the vector bundles $L_{\R^2}^{}$ and $L_\C^+\oplus L_\C^-$ acts on $\tilde A$. Let us assume that the group $\sU(1)_\hbar$ acts from the left. Then for the generator $J$ of this group we have 
\begin{equation}\label{11.4}
J\tilde A=J(A_+P_++A_-P_-)=\im (A_+P_+-A_-P_-)
\quad\mbox{and}\quad J^2\tilde A=-\tilde A\ .
\end{equation}
For this action,  the field $\tilde A$ interacts with the vacuum field $A_\hbar$ entering the Klein-Gordon oscillator equation appearing from the KG equation raised to the phase space $T^*\R^{3,1}$. In the bundle $L_{\R^2}^{}$ lifted to $T^*\R^{3,1}$ we have covariant derivatives $\nabla_a$, $\nabla^{a+3}$ from \eqref{3.9}, \eqref{3.22}-\eqref{3.24} and \eqref{5.27}, and for covariant d'Alembertian we have
\begin{equation}\label{11.5}
\left(-\dpar_t^2+\Delta_{L_{\R^2}^{}}\right)\tilde A=\left(\eta^{\mu\nu}\dpar_\mu\dpar_\nu - \frac{1}{w^4}\delta_{ab}x^ax^b\right)\tilde A=0
\end{equation}
for any component $\tilde A_\sigma$. If we introduce the operator $\nabla^0$ from \eqref{9.46} and extend the metric on $T^*\R^{3,1}$ by the term $-w_0^4\dd p_0^2$, then the term $w_0^{-4}(x^0)^2$ will be added in the bracket in  \eqref{11.5} and it will be possible to generalize the solutions that we will now write out.

Substituting the expansion
\begin{equation}\label{11.6}
\tilde A = e^{-\im\omega t}\phi_+(x)P_+ + e^{\im\omega t}\phi_+^*(x)P_-
\end{equation}
into equation \eqref{11.5}, we obtain the equation \eqref{5.29} of an isotropic harmonic oscillator after replacing $2m_0E$ by $\omega^2$. Hence, solutions to \eqref{11.5} are given by formulae \eqref{5.35}. Thus, virtual photons are described by oscillating solutions  \eqref{11.6}
with real functions
\begin{equation}\label{11.7}
\phi^{}_{+(n)}(x)\ ,\quad (n)=(n_1n_2n_3)\ ,\quad \omega_n=\sqrt{2w^{-2}(n+\sfrac32)}\ , \quad n=n_1+n_2+n_3
\end{equation}
and the ground state is
\begin{equation}\label{11.8}
\tilde A_0=\sfrac12\, e^{-\im\omega_0t}e^{-\frac{r^2}{2w^2}}P_+ + \sfrac12\, e^{\im\omega_0t}e^{-\frac{r^2}{2w^2}}P_-=\begin{pmatrix}\cos\omega_0t&\sin\omega_0t\\-\sin\omega_0t&\cos\omega_0t\end{pmatrix}e^{-\frac{r^2}{2w^2}}
\end{equation}
with $\omega_0=\sqrt{3w^{-2}}$ and $r^2=\delta_{ab}x^ax^b$. If we return to the hydrogen atom, then fluctuating electromagnetic field \eqref{11.8} can generate spontaneous emission and absorption of photons by the atom \cite{Dirac}.

Oscillating solutions
\begin{equation}\label{11.9}
\tilde A_{(n)}=\begin{pmatrix}\cos\omega_nt&\sin\omega_nt\\-\sin\omega_nt&\cos\omega_nt\end{pmatrix}\phi_{+(n)}(x), \quad (n)=(n_1n_2n_3)
\end{equation}
of \eqref{11.5} are proposed to identify with virtual photons. They exist during absorption and emission, and when $w^2\to\infty$ ($\omega_n\to 0$) they become solutions of the free Klein-Gordon equation \eqref{11.1} and then propagate as waves. Despite the fact that $\Psi$ (matter) are quantum particles and $\CA_\mu$ (geometry) are not, the free fields of both behave like waves. The behavior of $\Psi$ and $\CA_\mu$ as particles occur when they interact with the vacuum field $A_\hbar$.

\bigskip

\noindent{\bf\large 11.4.  Quantum time and gauge bosons}

Solutions \eqref{11.7}-\eqref{11.9} are Gaussian in coordinates $x^a$. It is also possible to obtain solutions that are Gaussian in the coordinate $x^0$ (i.e. localized both in space and time) by introducing time and energy operators. To do this, we introduce on $T^*\R^{3,1}$ the symplectic 2-form
\begin{equation}\label{11.10}
\omega^{}_{\R^{6,2}}=\omega^{}_{\R^{6}} + \dd x^0\wedge\dd p_0=\dd x^\mu\wedge\dd p_\mu
\end{equation}
and  operators
\begin{equation}\label{11.11}
\nabla_0=\im\ph_0=\frac{\dpar}{\dpar x^0}=\dpar_0\quad\mbox{and}\quad\nabla^7=-\im\xh^0=\frac{\dpar}{\dpar p_0}-x^0J\ ,
\end{equation}
with commutation relations 
\begin{equation}\label{11.12}
[\nabla_0, \nabla^7]=-J=[\ph_0,\xh^0]\ ,
\end{equation}
in addition to the operators $\nabla_a, \nabla^{b+3}$ from \eqref{3.22}-\eqref{3.25}. These operators are covariant derivatives in the bundle $L_\C^+\oplus L_\C^-$ over $T^*\R^{3,1}$ with the structure group $\sU(1)_\hbar$ whose generator $J$ is equal to $\im$ on $L_\C^+$ and 
$-\im$ on $L_\C^-$.

After extending the metric \eqref{10.1} to
\begin{equation}\label{11.13}
g^{}_{\R^{6,2}}=-(\dd x^0)^2 + \delta_{ab}\dd x^a\dd x^b + w^4\delta^{ab}\dd p_a\dd p_b-w_0^4\dd p^2_0\ ,
\end{equation}
where $g^{}_{77}=g^{}_{p_0p_0}=-w_0^4$ with the length parameter $w_0$, and the change $w^2\to (w_1^2, w_2^2, w_3^2)$, the covariant d'Alembertian on the bundle $L_\C^+\oplus L_\C^-$ over $T^*\R^{3,1}$ will have the form
\begin{equation}\label{11.14}
\left(\eta^{\mu\nu}\dpar_\mu\dpar_\nu + \frac{1}{w_0^4}\, (x^0)^2 - \frac{1}{w_1^2}\, (x^1)^2-\frac{1}{w_2^2}\, (x^2)^2-\frac{1}{w_3^2}\, (x^3)^2\right)\tilde A=0\ .
\end{equation}
This equation can be rewritten in terms of the ladder operators \eqref{10.16} plus
\begin{equation}\label{11.15}
c_0=\dpar_0 + \frac{x^0}{w_0^2}=:\frac{\sqrt 2}{w_0}\, a_0\quad\mbox{and}\quad
c_0^\+=-(\dpar_0 - \frac{x^0}{w_0^2})=:\frac{\sqrt 2}{w_0}\, a_0^\+\ ,
\end{equation}
in the form
\begin{equation}\label{11.16}
\left(2w_0^{-2}(a_0^\+ a_0 +\sfrac12) - 2w_1^{-2}(a_1^\+ a_1 +\sfrac12)-2w_2^{-2}(a_2^\+ a_2 +\sfrac12) - 2w_3^{-2}(a_3^\+ a_3 +\sfrac12)\right)\tilde A=0\ .
\end{equation}
Solutions of equations \eqref{11.16} for each component of the matrix $\tilde A$ have the form
\begin{equation}\label{11.17}
|n_0, n_1, n_2, n_3\rangle =\frac{(a_0^\+)^{n_0}(a_1^\+)^{n_1}(a_2^\+)^{n_2}(a_3^\+)^{n_3}}{\sqrt{n_0!n_1!n_2!n_3!}}|0,0,0,0\rangle\ .
\end{equation}
The energy eigenvalues are constrained by the equation
\begin{equation}\label{11.18}
E^2_{n_0}=E^2_{n_1}+E^2_{n_2}+E^2_{n_3}\quad\mbox{for}\quad E^2_{n_\nu}=2w^{-2}_\nu (n_\nu +\sfrac12)\ .
\end{equation}
For the ground state we have
\begin{equation}\label{11.19}
\langle x|0,0,0,0\rangle\sim\exp\left(-\frac{x_0^2}{2w_0^2}-\frac{x_1^2}{2w_1^2}-\frac{x_2^2}{2w_2^2}-\frac{x_3^2}{2w_3^2}\right)\ .
\end{equation}
By automorphisms of the bundle $L_\C^+\oplus L_\C^-$ one can shift the center of the Gaussian distribution  \eqref{11.19} to any point $x_0^\mu$ of the space-time $\R^{3,1}$.

\section{~Conclusions}

The main purpose of this paper was to discuss the geometric view of quantization as the introduction of internal degrees of freedom of relativistic and non-relativistic particles. In non-relativistic quantum mechanics, an internal space $(\R^2, J)_\hbar\cong\C$ is added (with the group SO(2)$_\hbar\cong\sU(1)_\hbar$ acting on it) at each point of the phase space $T^*\R^{3}$ and hence a complex line bundle $L_\C^+\to T^*\R^{3}$ is defined. Quantization is the transition from commuting partial derivatives in $T^*\R^{3}$ to non-commuting covariant derivatives in the bundle $L_\C^+$ endowed a connection $A_\hbar$. A quantum particle is a section $\psi$ (wave function) of the bundle $L_\C^+$ over $T^*\R^{3}$, and the connection $A_\hbar$ on $L_\C^+$ specifies the parallel transport of a quantum particle $\psi$ along curves in $T^*\R^{3}$. The commutators of covariant derivatives define a constant Abelian field $F_\hbar = \dd A_\hbar \sim \im\omega$, where $\omega$ is the symplectic 2-form on $T^*\R^{3}$. Fields $A_\hbar$ and $F_\hbar$ on the quantum bundle $L_\C^+$ are not dynamical, they are background fields fixed by the canonical commutation relations. The quantum charge density $\rho (x)=\psi^*\psi$ for $L_\C^+$ is the probability density of finding a particle at a point $x$ and it can generate electromagnetic field on the bundle $L_\C^+\otimes E_\C^\pm$ but not the fields $(A_\hbar , F_\hbar)$. That is why we consider $A_\hbar$ and $F_\hbar$ to be the background vacuum fields.

Thus, we view quantum mechanics as an Abelian gauge theory with a massive u(1)$_\hbar$-valued gauge field $A_\hbar$. Note that any gauge field specifies an interaction and we know four forces of nature carried by fields
\begin{equation}\label{12.1}
A^{\rm em}_{\sU(1)}\ ,\quad A^{\rm weak}_{\sSU(2)}\ ,\quad A^{\rm strong}_{\sSU(3)}\quad\mbox{and}\quad\Gamma^{\rm gravity}_{\rm{SO}(3,1)} \ .
\end{equation}
Consequently, the field $A_\hbar$ also sets a certain force acting on quantum particles $\psi$.  Note that the field $A_\hbar$ enters only in the covariant derivatives $\nabla^{a+3}=-\im\hbar^{-1}\xh^a$, and the potential energy $V(\xh^a)$ actually defines the force acting on particles in non-relativistic quantum mechanics. In this paper, we aimed to explore the possibility of generalizing this geometric view to the relativistic case.
To achieve this goal, we restricted ourselves to quadratic potentials $V\sim \xh^2$ associated with covariant Laplacians on quantum bundles. In addition, we used the fact that if the bundle $L_\C^+$ describes particles, then the complex conjugate bundle $L_\C^-=\overline{L_\C^+}$ describes antiparticles. After that we introduced complex vector bundles
\begin{equation}\label{12.2}
E^\pm_\C\ ,\quad E^\pm_{\C^2}\ ,\quad E^\pm_{\C^3}\quad\mbox{and}\quad W^\pm
\end{equation}
for electromagnetic degrees of freedom, weak interactions, strong interactions and spin degrees of freedom, also coupling them with the quantum degrees of freedom $L_\C^+$ of particles and $L_\C^-$ of antiparticles.

Connections \eqref{12.1} are defined in complex vector bundles \eqref{12.2} and serve as carriers of four known interactions. The connection $A_\hbar$ in the quantum bundles $L_\C^\pm$, $L_\C^+\otimes L_\C^-$ etc. specifies the interaction with vacuum of first quantized theories. To see this interaction, we generalized Klein-Gordon and Dirac equations to the phase space and showed that the extended equations have infinite number of oscillator type solutions with discrete energy values. They describe various bound states of particles and antiparticles in the background vacuum field $A_\hbar$. These particles become free and are described by wave packets if the interaction with vacuum fields $A_\hbar$ disappears. The wave-particle duality lies in turning on and off the interaction with the field $A_\hbar$. Finally, we argued that the total quantum charge of universe is zero, and the confinement of quarks and gluons is due to the fact that their interaction with $A_\hbar$ is always on and they can only be in bound states in the form of hadrons.

\bigskip 

\noindent 
{\bf\large Acknowledgments}

\noindent
I am grateful to Tatiana Ivanova for stimulating discussions and remarks.



\begin{thebibliography}{99}

\bibitem{Sni}
J. Sniatycki, {\it Geometric quantization and quantum mechanics}, Springer-Verlag, Berlin, 1980.

\bibitem{Wood}
N.M.J. Woodhouse, {\it Geometric quantization}, Clarendon Press, Oxford, 1980.

\bibitem{Dirac}
P.A.M. Dirac, {\it The principles of quantum mechanics}, Clarendon Press, Oxford, 1958.

\bibitem{Ali}
S.T. Ali and M. Englis, ``Quantization methods: a guide for physicists and analysts",\\ Rev. Math. Phys. {\bf 17} (2005) 391
[arXiv:math-ph/0405065 [math.ph]].

\bibitem{KobNom}
S. Kobayashi and K. Nomizu, {\it Foundation of differential geometry}, v.II, John Wiley \& Sons, New-York, 1969.

\bibitem{Greiner1}
W. Greiner, {\it Relativistic quantum mechanics. Wave equations,} Springer, Berlin, 2000.

\bibitem{Greiner2}
W. Greiner and J. Reinhardt, {\it Field quantization,} Springer, Berlin, 1996.

\bibitem{Hurt}
N.E. Hurt, {\it Geometric quantization in action}, D.Reidel Publishing Company, Dordrecht, 1983.

\bibitem{Popov1}
A.~D.~Popov,
``On exact solvability of $\CN=4$ super Yang-Mills,''\\
Nucl. Phys. B \textbf{978} (2022) 115742
[arXiv:2106.04460 [hep-th]].

\bibitem{Bona}
A. Bona, ``The Dirac equation in geometric quantization",\\ Ann. Henri Poincare {\bf 4} (2003) 487.

\bibitem{Stern}
S. Sternberg, ``Minimal coupling and the symplectic mechanics of a classical particle in the presence of a Yang-Mills field", Proc. Natl. Acad. Sci. USA {\bf 74} (1977) 5253.

\bibitem{Wein}
A. Weinstein, ``A universal phase space for particles in Yang-Mills fields", \\ Lett. Math. Phys. {\bf 2} (1978) 417.

\bibitem{Mont}
R. Montgomery, ``Canonical formulations of a classical particle in a Yang-Mills field and Wong's equations", Lett. Math. Phys. {\bf 8} (1984) 59.

\bibitem{Popov2}
A.D.~Popov,
``Stueckelberg and Higgs mechanisms: frames and scales,''\\
Universe \textbf{8} (2022)  361
[arXiv:2204.13368 [hep-th]].

\bibitem{Popov3}
A.D.~Popov,
``Geometric confinement in gauge theories,''\\
Symmetry \textbf{15} (2023) 1054
[arXiv:2211.03096 [hep-th]].

\bibitem{Popov4}
A.D.~Popov,
``Yang-Mills-Stueckelberg theories, framing and local breaking of symmetries,'' \\Rev. Math. Phys., in press,
[arXiv:2110.00405 [hep-th]].

\bibitem{BogShir}
N.N.  Bogoliubov and D.V. Shirkov, {\it Introduction to theory of quantized fields}, John Wiley \& Sons, Canada, 1980.

\end{thebibliography}
\end{document}